\documentclass[prd,preprintnumbers,aps,preprint,amsmath,amssymb,superscriptaddress,floatfix,nofootinbib,unsortedaddress]{revtex4}
\pdfoutput=1

\usepackage{amssymb, bm, amsmath, graphicx, subfigure,physymb}
\usepackage[colorlinks=true,linktocpage=true,linkcolor=blue,citecolor=blue]{hyperref}
\usepackage[usenames,dvipsnames]{color}
\usepackage{subfigure}
\usepackage{slashed}
\usepackage{multirow,array}
\usepackage[utf8]{inputenc}
\usepackage{array}
\usepackage{enumerate}
\usepackage{amsfonts}
\usepackage{dsfont}
\usepackage{mathtools}
\usepackage{mathrsfs}
\usepackage{slashed}
\usepackage{latexsym}
\usepackage{hyperref}
\usepackage{epstopdf} 
\usepackage{bbm}

\pdfoutput=1

\newcommand{\barq}{{\bar q}}

\newcommand{\thalf}{\tfrac{1}{2}}

\newcommand{\tvec}[1]{\vec{#1}_\bot}

\newcommand{\ord}[1]{\mathcal{O} \left( #1 \right)}
\newcommand{\tr}{\mathrm{tr}}

\begin{document}

\title{
Toward Initial Conditions of Conserved Charges Part II: \\
The ICCING Monte Carlo Algorithm
}
\author{Patrick Carzon}
\email[Email: ]{pcarzon2@illinois.edu}
\affiliation{Illinois Center for Advanced Studies of the Universe, Department of Physics, University of Illinois at Urbana-Champaign, Urbana, IL 61801, USA}
\author{Mauricio Martinez}
\email[Email: ]{mmarti11@ncsu.edu}
\affiliation{North Carolina State University, Raleigh, NC 27695, USA}
\author{Matthew D. Sievert}
\email[Email: ]{msievert@nmsu.edu}
\affiliation{Department of Physics, New Mexico State University, Las Cruces, NM 88003, USA}
\author{Douglas E. Wertepny}
\affiliation{(formerly) Ben-Gurion University of the Negev, Beer-Sheva 84105, Israel}
\author{Jacquelyn Noronha-Hostler}
\email[Email: ]{jnorhos@illinois.edu}
\affiliation{Illinois Center for Advanced Studies of the Universe, Department of Physics, University of Illinois at Urbana-Champaign, Urbana, IL 61801, USA}

\begin{abstract}
At top collider energies where baryon stopping is negligible, the initial state 
of heavy ion collisions is overall charge neutral and predominantly composed of gluons.  Nevertheless, there can also be significant local fluctuations of the baryon number, strangeness, and electric charge densities about zero, perturbatively corresponding to the production of quark/antiquark pairs.  These previously ignored local charge fluctuations can permit the study of charge diffusion in the quark-gluon plasma (QGP), even at top collider energies.  In this paper we present a new model denoted ICCING (Initial Conserved Charges in Nuclear Geometry) which can reconstruct the initial conditions of conserved charges in the QGP by sampling a ($g \rightarrow q\bar{q}$) splitting probability over the initial energy density.  We find that the new charge distributions generally differ from the bulk energy density; in particular, the strangeness distribution is significantly more eccentric than standard bulk observables and appears to be associated with the geometry of hot spots in the initial state.  The new information provided by these conserved charges opens the door to studying a wealth of new charge- and flavor-dependent correlations in the initial state and ultimately the charge transport parameters of the QGP.
\end{abstract}

\date{\today}
\maketitle

\tableofcontents

%
\section{Introduction}
%

The defining feature of the exotic quark-gluon plasma (QGP) produced in ultrarelativistic collisions of heavy nuclei is its extremely small viscosity.  This ``nearly perfect fluid'' flows almost isentropically starting from an early far-from-equilibrium state, through thermalization, and across the confinement phase transition until the fluid freezes out into the measured particle spectra.  This standard paradigm of heavy-ion collisions has been well established by comparisons of data from the Large Hadron Collider (LHC) and Relativistic Heavy Ion Collider (RHIC) with simulations from event-by-event viscous relativistic hydrodynamics using realistic equations of state \cite{Gardim:2012yp, Gardim:2012im, Heinz:2013bua, Shen:2015qta, Niemi:2015qia, Niemi:2015voa, Noronha-Hostler:2015uye, Gardim:2016nrr, Gardim:2017ruc, Eskola:2017bup, Giacalone:2017dud, Gale:2012rq, Bernhard:2016tnd, Giacalone:2016afq, Zhu:2016puf, Zhao:2017yhj, Zhao:2017rgg, Alba:2017hhe, Noronha-Hostler:2019ytn, Sievert:2019zjr, Bozek:2011if, Bozek:2012gr, Bozek:2013ska, Bozek:2013uha, Kozlov:2014fqa, Zhou:2015iba, Mantysaari:2017cni, Weller:2017tsr, Pratt:2015zsa, Moreland:2015dvc, Auvinen:2018uej}.  Because of the extremely low viscosity of the QGP, the final-state bulk correlations are highly sensitive to the initial geometry of the collision, complicating the extraction of final-state transport parameters.  An appropriate description of the initial state at the time of hydrodynamization, including in particular its event-by-event fluctuations, is essential for reproducing the measured anisotropic flow and its multiparticle cumulants \cite{Renk:2014jja, Giacalone:2017uqx}.

In principle, the initial conditions of hydrodynamics consist of a complete specification of the full initial energy-momentum tensor $T^{\mu \nu}$ and the initial currents $J^\mu$ of all three conserved charges: baryon number $B$, strangeness $S$, and electric charge $Q$. However, in practice the most common approach has been to only initialize the energy density $\epsilon=T^{00}$ and set all other components zero.  More recently, progress has been made in incorporating the full $T^{\mu \nu}$ \cite{Gardim:2011qn, Gardim:2012yp, Gale:2012rq, Schenke:2019pmk, Liu:2015nwa, Kurkela:2018wud,Plumberg:2021bme,Chiu:2021muk} including initial flow and in some cases the initial shear stress as well.  Models range from phenomenological generalizations of a Monte-Carlo Glauber approach (e.g., \cite{Moreland:2014oya, Bernhard:2016tnd, Nagle:2018ybc, Romatschke:2013re}), to full microscopic calculations of $T^{\mu\nu}$ (e.g., \cite{Schenke:2012wb, Gale:2012rq}).  Regarding the currents $J^\mu$ of conserved charges, most of the focus has been on the effect of finite net baryon density \cite{Werner:1993uh, Itakura:2003jp, Shen:2017bsr, Akamatsu:2018olk, Mohs:2019iee}.  These considerations are driven primarily by efforts to locate and study the QCD critical point at finite net baryon density \cite{Karpenko:2015xea, Borsanyi:2018grb, Noronha-Hostler:2019ayj, Monnai:2019hkn, Monnai:2016kud, Bazavov:2018mes, Critelli:2017oub, Parotto:2018pwx, Demir:2008tr, Denicol:2013nua, Denicol:2018wdp, Kadam:2014cua, Stephanov:1999zu, Stephanov:2017ghc, Jiang:2015hri, Mukherjee:2016kyu, Nahrgang:2018afz, An:2019osr, Du:2019obx, Batyuk:2017sku, Shen:2017bsr}, either by decreasing the collision energy as in the RHIC Beam Energy Scan program, or by studying the QGP in a far forward regime \cite{Brewer:2018abr, Li:2018fow}.  At top collider energies, where the baryon chemical potential $\mu_B$ and total charge $Q$ deposited into the QGP vanish, the conserved currents $J^\mu$ are typically set homogeneously to zero.

This assumption -- that the vanishing of the total charge $Q = 0$ in the plasma justifies setting the charge density $\rho$ uniformly to zero -- corresponds to a mean field approximation in the charge sector.  It neglects the role of {\it{local fluctuations}} of the charge densities (and, more generally, the conserved currents $J^\mu$) around zero, even though event-by-event fluctuations in $T^{\mu\nu}$ are known to play an essential role in the initial conditions of heavy-ion collisions.  This description of the initial state with an energy-momentum distribution $T^{\mu \nu}$ but with zero conserved currents $J^\mu = 0$ is actually consistent with a leading-order picture in perturbative QCD, in which the initial state is composed entirely of soft gluons.  These gluons deposit energy into the plasma, for instance by classical Yang-Mills evolution, but carry none of the $BSQ$ charges themselves.  

%
\begin{figure}
\begin{center}
	\includegraphics[width=0.5\textwidth]{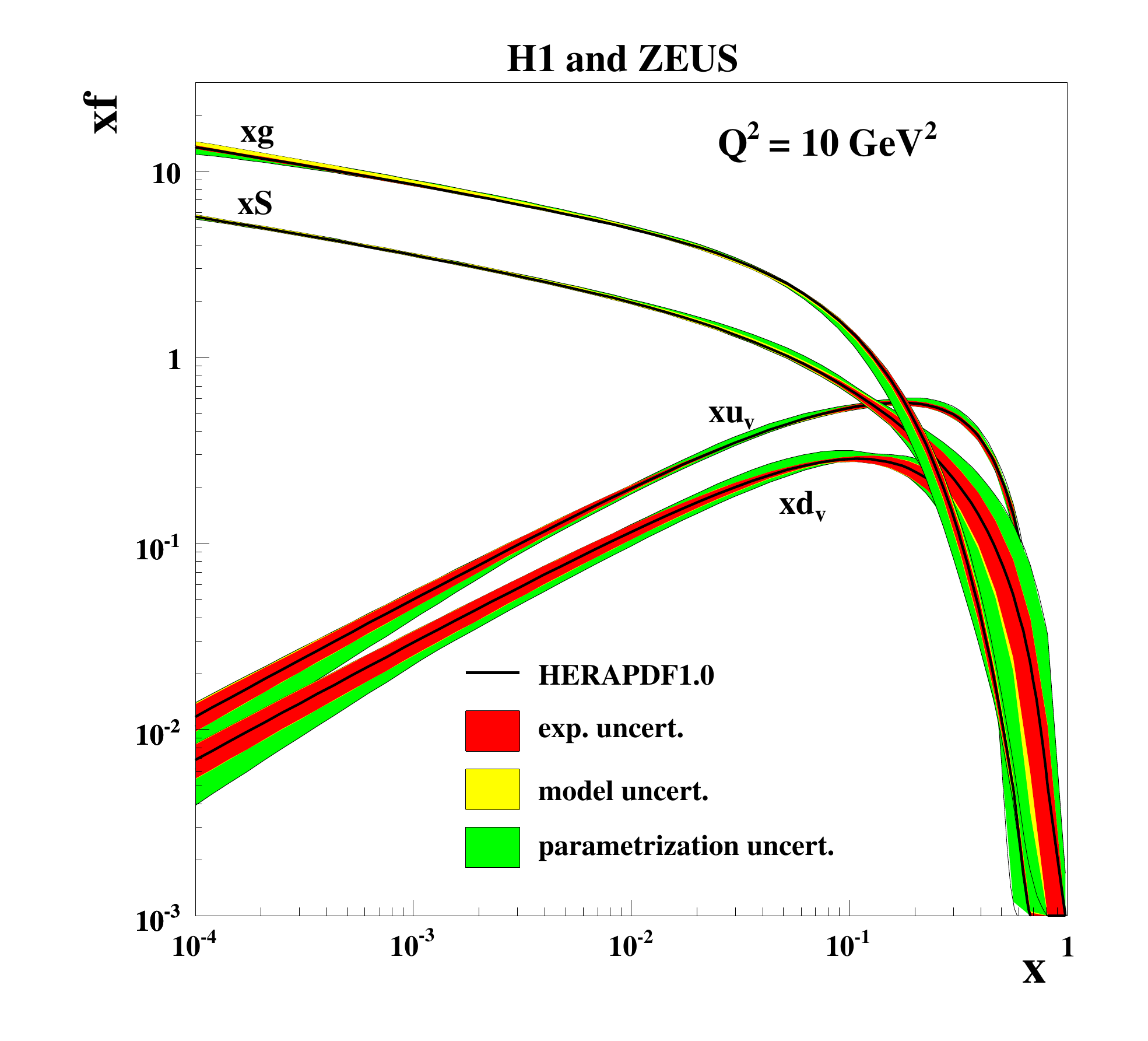}
	\caption{
	    Parton distribution functions (PDFs) extracted by the H1 and ZEUS Collaborations showing the abundance of sea quarks at small $x$ even in cold nuclear matter.  Note that the sea quarks $(xS)$ are comparable in number to gluons $(xg)$ at small $x$.  This figure reproduced following Creative Commons Attribution License guidelines from Fig.~19(b) of Ref.~\cite{Aaron:2009aa}.
	}
	\label{f:HERAPDF}
	\end{center}
\end{figure}
%

This perturbative picture of a gluon-dominated initial state, however, does not hold beyond the leading order.  Beginning at next-to-leading order, the gluons abundant in the initial state can pair produce quarks and antiquarks in equal amounts, leading to nonzero charge densities and currents in the initial state even though the total charge remains zero.  Indeed, in cold nuclear matter and even in the proton, the $q \bar{q}$ pairs in the form of sea quarks are never negligible in comparison to gluons as seen in the gluon $x g$ and sea quark $x S$ distributions extracted by the H1 and ZEUS Collaborations in Fig.~\ref{f:HERAPDF}.  At the low values of $x$ responsible for producing the initial state in heavy-ion collisions, the two distributions scale proportionately $x S \propto x g$, consistent with the perturbative expectation $x S \approx \alpha_s \, x g$.  The abundance of sea quark fluctuations in the proton is not due solely to perturbative $g \rightarrow q \bar{q}$ splitting; nonperturbative mechanisms \cite{Shuryak:2002qz} can contribute as well.  Even more $q \bar{q}$ pair production can occur in the early pre-equilibrium dynamics of heavy-ion collisions (e.g. \cite{Kurkela:2018vqr}), further contributing to the event-by-event fluctuations of the local conserved charge densities away from their zero mean-field value.

Even if the $q \bar{q}$ pair fluctuations make a comparatively small (perturbatively $\sim \ord{\alpha_s}$) contribution to the initial energy-momentum tensor $T^{\mu \nu}$ relative to the gluons, they nonetheless provide the {\it{leading}} contribution to the initial conserved currents $J^\mu$ at top collider energies.  Because of this, the inclusion of the initial-state $q \bar{q}$ fluctuations and their subsequent evolution can make it possible to study the transport coefficients (see e.g., \cite{Rougemont:2015ona, Rougemont:2017tlu, Greif:2017byw, Denicol:2018wdp}) associated with $BSQ$ charge diffusion and dissipation entirely {\it{outside}} of the low-energy / forward QGP programs.  Indeed, an extraction of charge transport properties at top energies $(\mu_B \approx 0)$ where the theory and systematics are {\it{best}} under control is a necessary baseline to study the modifications induced by the critical point as part of the Beam Energy Scan program.

To study charge transport at top collider energies, two ingredients are needed: an initial condition which accounts for the $q \bar{q}$ pairs, and an implementation of hydrodynamics which propagates the three conserved charges along with all relevant dissipative currents \cite{Rougemont:2015ona, Rougemont:2017tlu, Greif:2017byw, Denicol:2018wdp}.  Recent progress in the latter has been made in Refs.~\cite{Du:2019obx,Denicol:2018wdp,Batyuk:2017sku}, but a comprehensive treatment including all three conserved charges and their interplay is still lacking.  In this paper, we provide the former: a Monte Carlo algorithm which samples the $g \rightarrow q \bar{q}$ splitting probability to transform an initial state given only by the energy density $\epsilon = T^{0 0}$ into one accompanied by the corresponding charge densities of baryon number $B$, strangeness $S$, and electric charge $Q$.  The resampling algorithm, which we denote ``Initial Conserved Charges in Nuclear Geometry (ICCING),'' can utilize any external choice of the initial energy density and any choice of the $g \rightarrow q \bar{q}$ splitting probabilities, supplementing existing models of the initial state with new charge information that can be directly read into subsequent hydrodynamic simulations.  We illustrate the algorithm and its physical consequences on initial energy profiles produced from Trento \cite{Moreland:2014oya, Bass:2017zyn, Moreland:2018gsh} and using the $g \rightarrow q \bar{q}$ spatial correlation function derived by some of us in Ref.~\cite{Martinez:2018ygo} within the color-glass condensate (CGC) effective theory \cite{McLerran:1993ni,McLerran:1993ka,McLerran:1994vd} of QCD at high energies.

The rest of this paper is organized as follows.  In Sec.~\ref{sec:Theory} we describe the choice of $g \rightarrow q \bar{q}$ splitting probabilities which we use as external input to the ICCING algorithm, with additional details about its theoretical basis deferred to Appendix~\ref{app:Theory}.  In Sec.~\ref{sec:algorithm} we describe the main algorithm itself, explaining how we perform the Monte Carlo resampling of the initial state to produce the conserved $BSQ$ charge densities.  In Sec.~\ref{sec:Results} we analyze the results of the ICCING algorithm applied to Pb Pb collisions at $5.02~\mathrm{TeV}$, with particular emphasis on the different information about the initial state encoded in the strangeness distribution.  Finally, in Sec.~\ref{sec:Concl} we conclude by summarizing the key features of the initial charge profiles produced by ICCING and outlining future work to study the propagation of these initial charge profiles into final-state observables.  Details of the eccentricities and cumulants, including subtleties unique to their application to conserved charges, are presented in Appendix~\ref{app:Eccentricities}.  Additional heuristic estimates of certain parameters and scales are given in Appendices~\ref{app:TrentoEstimate} and \ref{app:Densities}.

%
\section{Theory Input on the Quark-Gluon Splitting Function}
\label{sec:Theory}
%

ICCING is an algorithm which takes as input external information about the probability of gluons to split into quark/antiquark pairs and uses it to reconstruct the initial condition, including the conserved charge densities of the produced quarks and antiquarks.  One of the particular strengths of ICCING is its flexibility in the choice of this external theory input; the algorithm itself is agnostic as to the choice of the quark/gluon splitting function.  While the algorithm itself can be applied to any choice of splitting function, in this paper we make the particular choice to use the quark/gluon multiplicities obtained from a color-glass condensate (CGC) calculation for this input.  The theoretical calculation underlying this choice was presented in Ref.~\cite{Martinez:2018ygo} and its application here is discussed in detail in Appendix~\ref{app:Theory}.

The result of the calculation of Ref.~\cite{Martinez:2018ygo} and its adaptation in Appendix~\ref{app:Theory} is an expression for the differential probability distribution $\frac{dP}{dr_\bot \, d\alpha}$ for a gluon to split into a $q \barq$ pair with given kinematics.  Here $\tvec{r}$ is the separation vector between the quark and antiquark, and $\alpha$ is the fraction of light-front momentum (i.e., energy) carried by the quark.  The differential probability to split into quarks with given kinematics can then be integrated over $r_\bot$ and $\alpha$ to obtain the total probability for a gluon to split into quarks of a given flavor.  We have evaluated these expressions in two different CGC models: the Golec-Biernat-Wusthoff (GBW) \cite{Golec-Biernat:1998zce} and McLerran-Venugopalan (MV) \cite{McLerran:1993ni, McLerran:1993ka, McLerran:1994vd} models.  For the GBW model, the differential probability distribution is given by~\cite{Martinez:2018ygo}
\begin{align} \label{e:prob1_first}
\frac{dP}{dr_\bot d\alpha} &= 
\frac{\alpha_s}{4 \pi}  m^2 r_\bot
\Big[ 1 - \exp\Big( - \tfrac{1}{4} [\alpha^2 + (1-\alpha)^2] \, r_\bot^2 Q_s^2 (\vec{x}_\bot) \Big) \Big]
\notag \\ &\hspace{1cm} \times
\Big[ \big(\alpha^2 + (1-\alpha)^2 \big) K_1^2 ( m r_\bot) + K_0^2 (m r_\bot) \Big] .
\end{align}
with integrated total probability
\begin{align} \label{e:multratio1_first}
P_{tot} &=
\frac{\alpha_s}{4 \pi}
\int\limits_0^1 d\alpha \, \int\limits_0^\infty d\zeta \, \zeta \,
\Big[ 1 - \exp\Big( - \tfrac{1}{4} [\alpha^2 + (1-\alpha)^2] \, \tfrac{Q_s^2 (\vec{x}_\bot)}{m^2} \, \zeta^2 \Big) \Big]\,
\notag \\ &\hspace{1cm} \times
\Big[ \big(\alpha^2 + (1-\alpha)^2 \big) K_1^2 (\zeta) + K_0^2 (\zeta) \Big] .
\end{align}
We note that both the differential and integrated probabilities depend on the position $\tvec{x}$ in the transverse plane at which the splitting occurs through the saturation scale $Q_s (\tvec{x})$.

Similarly for the MV model, the differential and integrated probabilities are given by
\begin{align} \label{e:prob2_first}
\left.\frac{dP}{dr_\bot d\alpha}\right|_{MV} &=
\frac{\alpha_s}{4 \pi}  m^2 r_\bot
\Big[ 1 - \exp\Big(
- \tfrac{1}{4} \alpha^2 \, r_\bot^2 Q_s^2 \ln\tfrac{1}{\alpha r_\bot \Lambda}
- \tfrac{1}{4} (1-\alpha)^2 \, r_\bot^2 Q_s^2 \ln\tfrac{1}{(1-\alpha) r_\bot \Lambda}
\Big) \Big]
\notag \\ &\hspace{1cm} \times
\Big[ \big(\alpha^2 + (1-\alpha)^2 \big) K_1^2 ( m r_\bot) + K_0^2 (m r_\bot) \Big]
\end{align}
and
\begin{align} \label{e:multratio2_first}
P_{tot}\Big|_{MV} &=
\frac{\alpha_s}{4 \pi}
\int\limits_0^1 d\alpha \, \int\limits_0^\infty d\zeta \, \zeta \,
\Big[ 1 - \exp\Big(
- \tfrac{1}{4} \alpha^2 \, \zeta^2 \tfrac{Q_s^2}{m^2}  \,  \ln\tfrac{1}{\alpha \zeta \Lambda/m}
- \tfrac{1}{4} (1-\alpha)^2 \, \zeta^2 \tfrac{Q_s^2}{m^2} \,  \ln\tfrac{1}{(1-\alpha) \zeta \Lambda/m}
\Big) \Big]
\notag \\ &\hspace{1cm} \times
\Big[ \big(\alpha^2 + (1-\alpha)^2 \big) K_1^2 (\zeta) + K_0^2 (\zeta) \Big] ,
\end{align}
respectively.  

%
\begin{figure}
\begin{center}
	\includegraphics[width=0.7\textwidth]{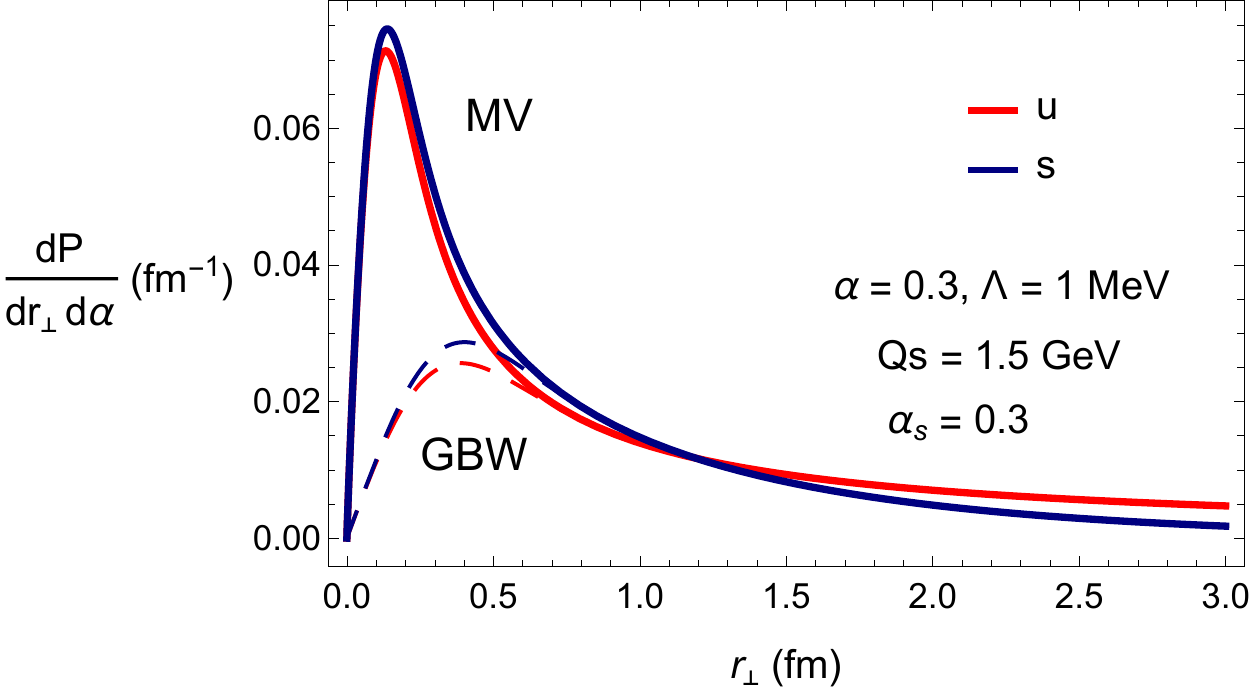}
	\caption{
	    Differential $q \barq$ splitting probability as a function of the distance $r_\bot$ between them for both the GBW~(Eq.~\eqref{e:prob1_first}, dashed curves) and MV~(Eq.~\eqref{e:prob2_first}, solid curves) models.  We plot both the up quark and strange quark probabilities for representative parameter values. The down quark distribution is indistinguishable from the up quark distribution and is not shown explicitly.
	}
	\label{f:probabilities}
	\end{center}
\end{figure}
%
%
%
%
\begin{figure}
\begin{center}
\includegraphics[width=0.49\textwidth]{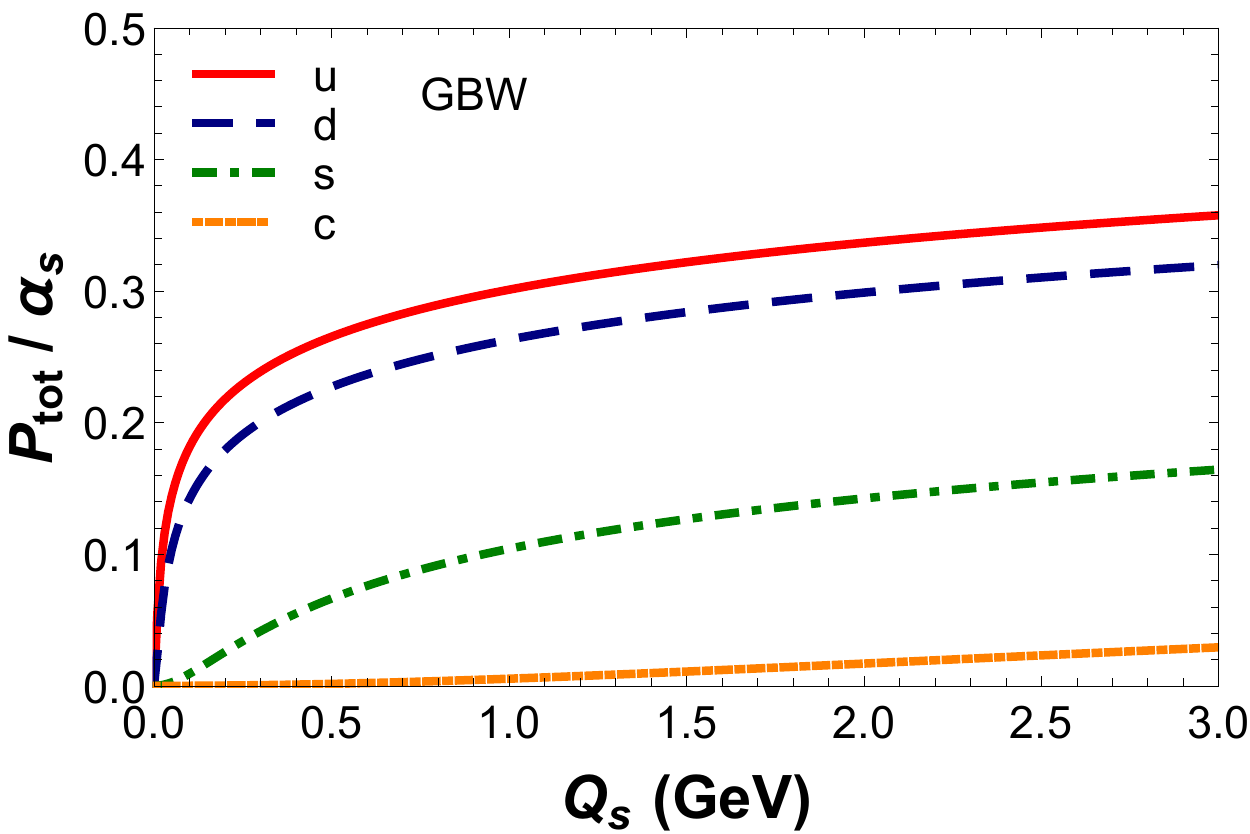}
\includegraphics[width=0.49\textwidth]{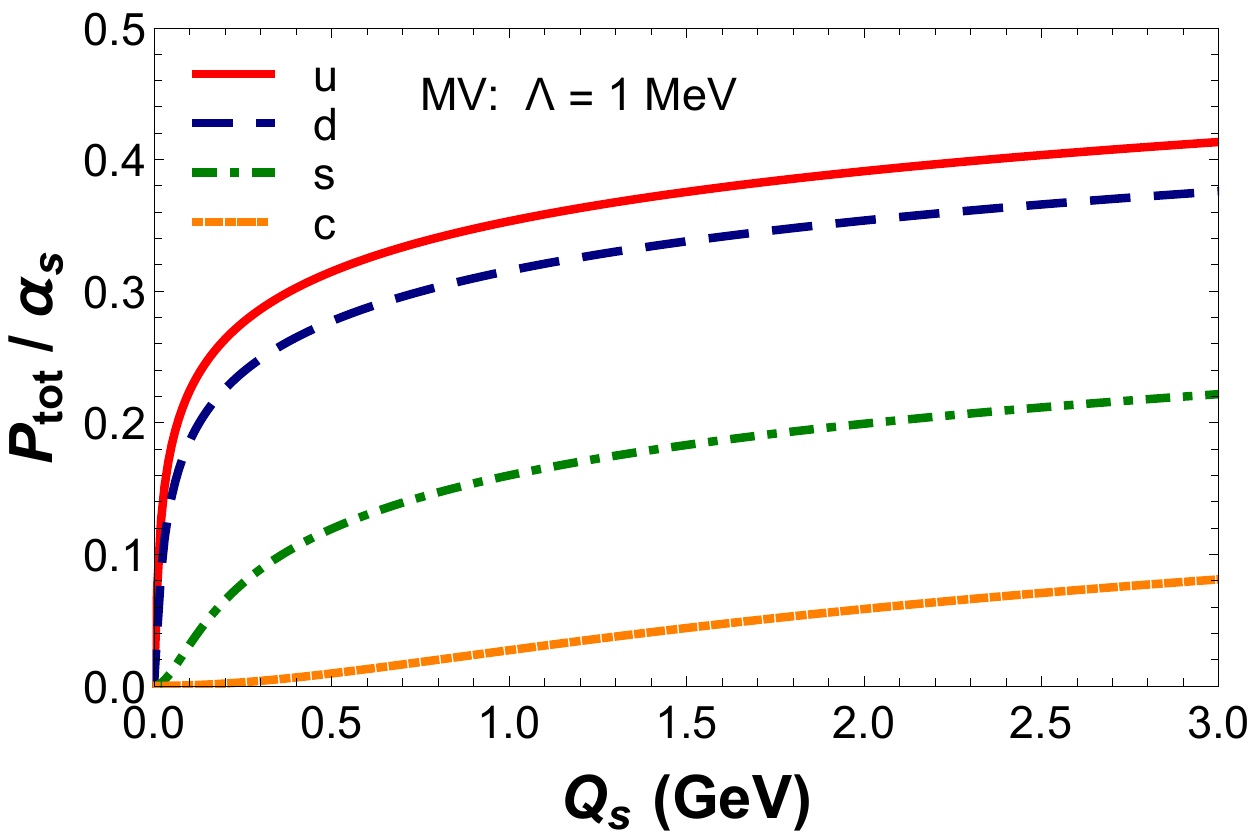}
	\caption{Total quark production probabilities (scaled by the strong coupling $\alpha_s$) as a function of the target saturation scale $Q_s$ for various flavors in the GBW (left) and MV (right) models.}
	\label{f:multratio}
	\end{center}
\end{figure}
%

The differential and integrated probabilities for both CGC models are plotted in Figs.~\ref{f:probabilities} and \ref{f:multratio}.  As can be seen clearly in Fig.~\ref{f:probabilities}, both the MV and GBW models agree in the long-distance behavior of the splitting function, corresponding to the deep saturation regime.  At short distances, however, the MV model correctly reflects the transition to a power-law behavior in the dilute regime, while the GBW model continues the simple Gaussian behavior present at large distances.  As a result, the MV model has a much more pronounced peak at short distances than the GBW model and has greater probability to produce quarks overall.  In both models, the flavor dependence arises only through the mass of the quarks, leading to a small difference in the probability distribution between the up quarks and strange quarks.  For the integrated probabilities shown in Fig.~\ref{f:multratio} which fix the overall chemistry of the initial state, we see that in both models the total probability increases as a function of the saturation scale $Q_s$.  Both models also reflect the mass ordering $P_\mathrm{tot} (u) > P_\mathrm{tot} (d) > P_\mathrm{tot} (s) > P_\mathrm{tot} (c)$ arising from the splitting functions.  More discussion about these models and their application here can be found in Appendix~\ref{app:Theory}.

%
\section{The ICCING Algorithm}
\label{sec:algorithm}
%

%
\begin{figure}
\begin{center}
	\includegraphics[width=1\textwidth]{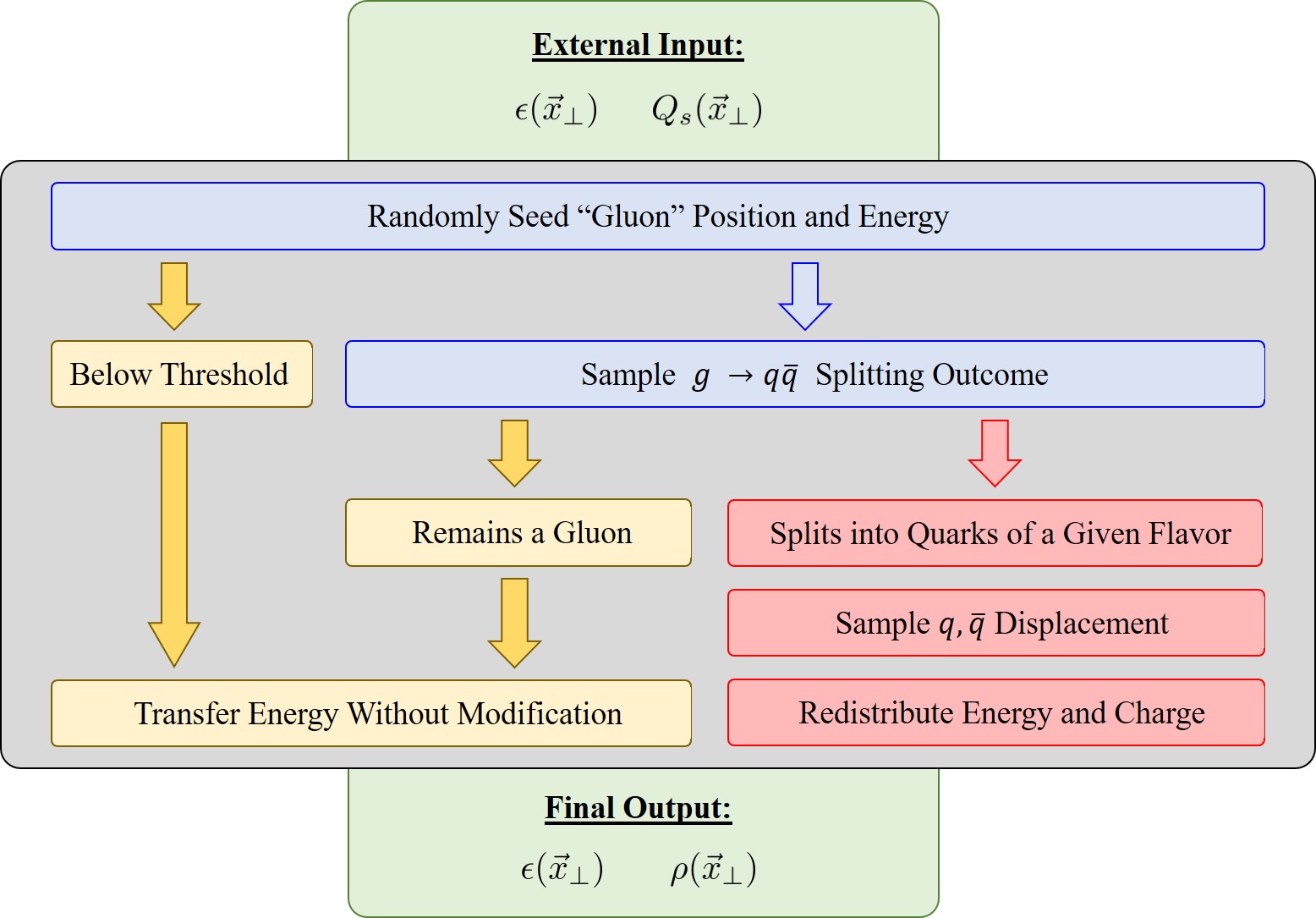}
	\caption{Decision tree of the ICCING algorithm.
	}
	\label{f:Algorithm}
	\end{center}
\end{figure}
%

%
\begin{figure}
\begin{center}
	\includegraphics[width=0.45\textwidth]{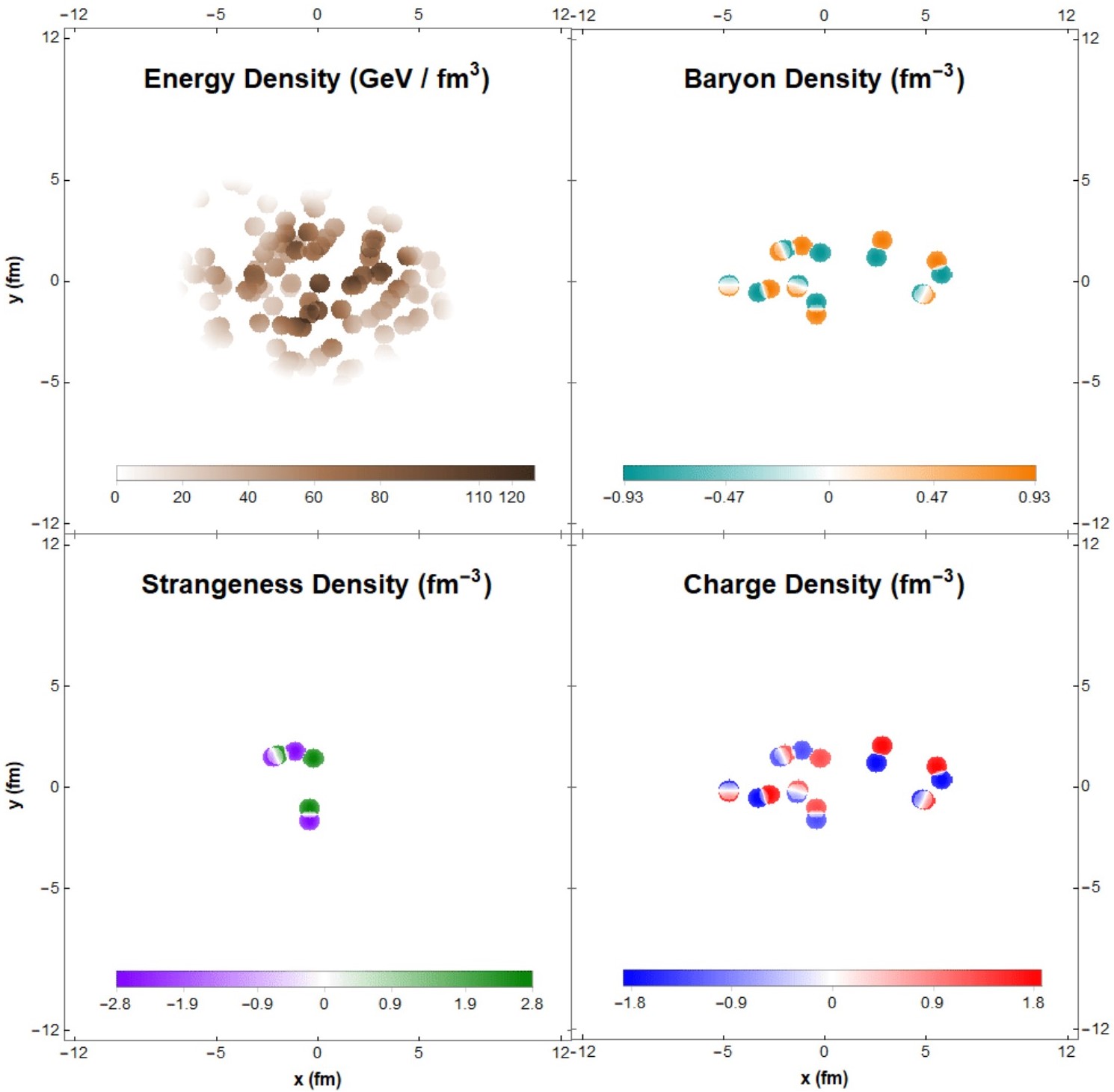}
	\includegraphics[width=0.45\textwidth]{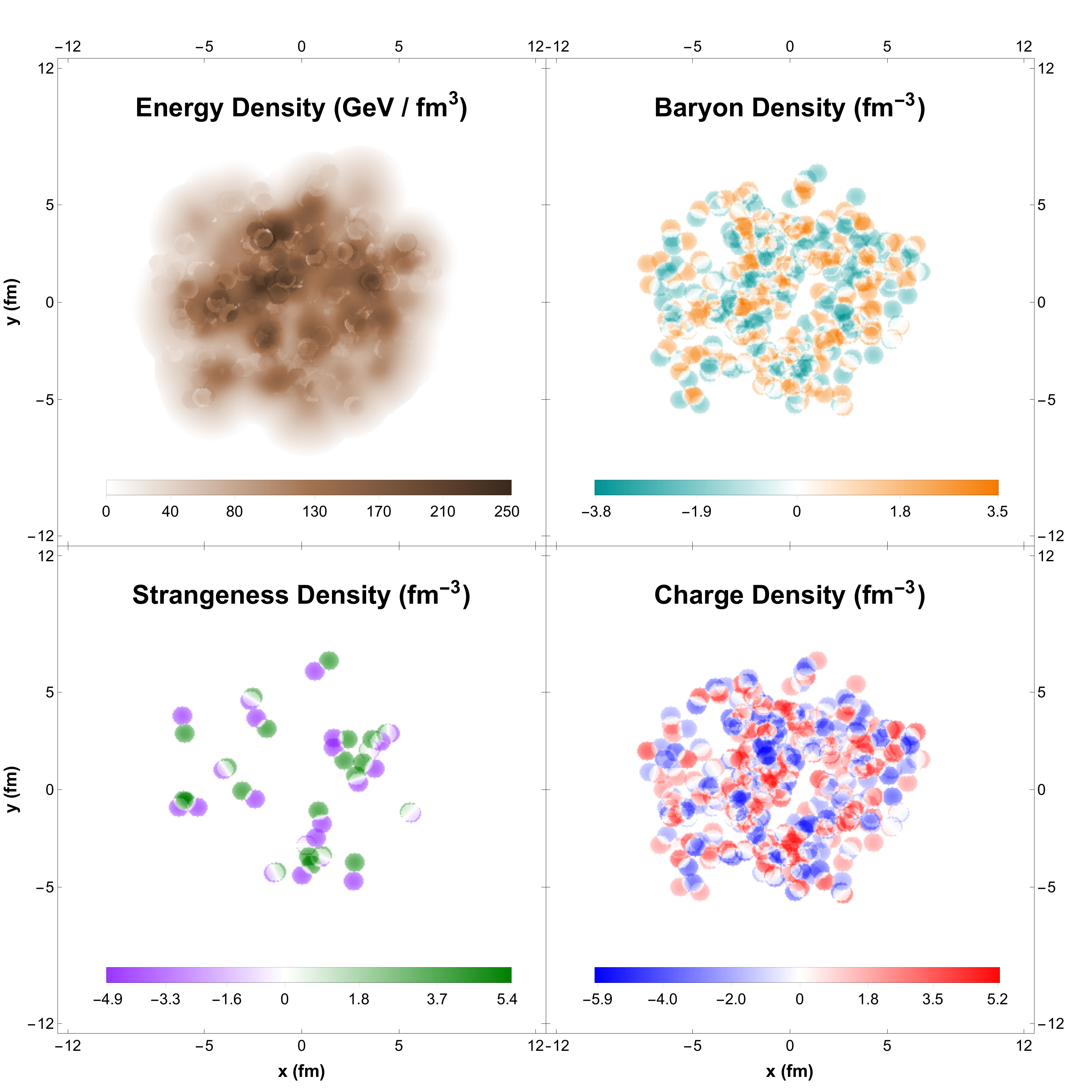}
	\caption{Left:  the ICCING algorithm as it begins to resample an event, selecting blobs of energy and giving them the chance to split into $q \barq$ pairs.  Right: a different event after being fully resampled by the ICCING algorithm, resulting in a reconstructed energy density as well as new distributions of the three conserved charges.  Note that some artifacts of the energy redistribution can be seen in the modified energy density.
	}
	\label{f:Events}
	\end{center}
\end{figure}
%

To begin, let us outline our approach to generating initial conserved charges at a high level, as visualized in Fig.~\ref{f:Algorithm}.  At its core, ICCING consists of an algorithm which uses Monte Carlo sampling of the probabilities given in Sec.~\ref{sec:Theory} as a basis for introducing $q \barq$ pairs into the initial state with the desired chemistry and spatial correlations.  The starting point for the ICCING algorithm is an externally-provided initial 2D profile of the energy density $\epsilon (\vec{x}_\bot)$.  Rather than provide our own model computation of this initial energy density, we instead keep this input arbitrary to allow for maximum generality of the approach.  Philosophically, we consider this initial energy density to consist entirely of gluons, and we perform a Monte Carlo sampling of the quark flavor ratios shown in Fig.~\ref{f:multratio} to determine the probability that a given gluon will split into a $q \barq$ pair of a particular flavor, or will remain a gluon.  If the gluon does split into a $q \barq$ pair, then we perform a second Monte Carlo sampling of the radial probability distributions shown in Fig.~\ref{f:probabilities}.  Using this sampled distance between the quark and antiquark, we redistribute the gluon energy to the new quark and antiquark positions, and we also deposit the corresponding conserved charges: baryon density, strangeness density, and electric charge density.  In this way, we resample all of the initial energy distribution and allow it the possibility to produce quarks and their conserved charges.  Typical events generated by this resampling procedure are shown in Fig.~\ref{f:Events}.  In the left-hand panel, the sampling procedure is underway for an event, allocating individual blobs of energy from the initially-provided $\epsilon(\vec{x}_\bot)$ and sampling their splitting probabilities into quarks.  In the right-hand panel, the resampling of a different event has finished, resulting in new distributions of the associated conserved charges.\footnote{For an analysis of how the absolute scales of the charge density seen in Fig.~\ref{f:Events} arise, see Appendix~\ref{app:Densities}.}

%
\subsection{External Input}
\label{subsec:Trento}
%

Now let us spell out the step-by-step procedure in greater detail.  As a start, while we emphasize that the algorithm we employ is applicable to any initial energy density distribution, in this paper we will obtain $\epsilon(\vec{x}_\bot)$ from the Trento model \cite{Moreland:2014oya}.  Trento is a parameterized generalization of a Monte-Carlo-Glauber model for the initial conditions of a nucleus-nucleus collision, ultimately providing the ``reduced thickness function'' $T_R (\vec{x}_\bot)$ which combines the fluctuating nuclear profile functions $T_A (\vec{x}_\bot)$ of one nucleus and $T_B (\vec{x}_\bot)$ of the second.  A range of parameters affect the event-by-event profile of $T_R$, but the most important of these is the exponent $p$ which defines the method of combining $T_A$ and $T_B$.  In our case we will choose $p=0$ which is the favored output of a Bayesian analysis \cite{Bernhard:2016tnd} and combines the nuclear thickness functions according to the geometric mean:
\begin{align}
    T_R (\vec{x}_\bot) \overset{(p=0)}{=} \sqrt{T_A (\vec{x}_\bot) \, T_B (\vec{x}_\bot)} .
\end{align}

The key physical assumption is that a thermodynamic variable quantifying the initial state of a heavy-ion collision can be described by a proportionality to $T_R$.  In our case, we take that quantity to be the entropy density $s(\vec{x}_\bot)$, with the proportionality factor
\begin{subequations}
\begin{align} \label{e:Trentomodel}
s(\vec{x}_\bot) &= a \, T_R (\vec{x}_\bot) , \\
a &= 119 \: \mathrm{fm}^{-1} \label{e:Trentoa}
\end{align}
\end{subequations}
for Pb Pb collisions at $5.02 \, \mathrm{TeV}$ \cite{Sievert:2019zjr}.  Note that $T_R$ is a number density per unit area, having dimensions of $\mathrm{fm}^{-2}$, such that the entropy density has the correct units of $\mathrm{fm}^{-3}$ when $a$ has dimensions of $\mathrm{fm}^{-1}$.  A useful heuristic estimate of $a$ is performed in Appendix~\ref{app:TrentoEstimate}.

For our purposes, it is necessary to specify the externally-provided initial state in terms of the energy density, rather than the entropy density.  For that, we use the equation of state provided by Ref.~\cite{Alba:2017hhe} at $\mu_B = 0$ to map the entropy density $s(\vec{x}_\bot)$ into the energy density $\epsilon(\vec{x}_\bot)$ in units of $\mathrm{GeV} / \mathrm{fm}^3$.  To speed up the resampling algorithm, we immediately discard grid points which have numerically infinitesimal entropy densities below some grooming threshold which by default we take to be $S_\mathrm{chop} = 10^{-20} \, \mathrm{fm}^{-3}$.  For the other Trento parameters, we take a nucleon width of $w = 0.51 \, \mathrm{fm}$, a multiplicity fluctuation parameter $k=1.6$, and we consider Pb Pb collisions at $\sqrt{s_{NN}} = 5.02 \, \mathrm{TeV}$.   See Ref.~\cite{Moreland:2014oya} for details of these parameters and how they are implemented in Trento.  These parameters have been used, together with a final-state hydrodynamics model, to successfully describe the anisotropic flow of bulk particle production at the LHC \cite{Moreland:2014oya, Bernhard:2016tnd, Alba:2017hhe, Giacalone:2017dud,Bernhard:2019bmu}.

The other piece of information which we extract from Trento is the initial profile of the saturation scale $Q_s (\vec{x}_\bot)$.  The square of the saturation scale is in general proportional to the nuclear thickness function, say $T_B (\vec{x}_\bot)$, with some model-dependent proportionality factor:
\begin{align}   \label{e:kappadef}
Q_s (\vec{x}_\bot) = \kappa \sqrt{T_B (\vec{x}_\bot)} .
\end{align}
The precise value of $\kappa$ varies depending on the choice of model and on the normalization used to define the saturation scale, but for a default value we take $\kappa = 1 \, \mathrm{GeV} \, \mathrm{fm}$.  To benchmark this number, let us compare with the value of $\kappa$ computed in perturbative QCD using the normalization of Ref.~\cite{Kovchegov:2012mbw}.  If one replaces the nucleons composing a nucleus with single quarks or gluons, the corresponding saturation scales are given by
\begin{align}
Q_s^2 (\vec{x}_\bot) =
\begin{cases}
\frac{4\pi \alpha_s^2 C_F}{N_c} \, T_B (\vec{x}_\bot) \qquad &\mathrm{if \, nucleon \rightarrow quark} \\
4\pi \alpha_s^2 \, T_B (\vec{x}_\bot) \qquad &\mathrm{if \, nucleon \rightarrow gluon}
\end{cases} .
\end{align}
Choosing the ballpark value $\alpha_s \approx 0.3$ we obtain for these cases
\begin{align} \label{e:kappapQCD}
\kappa =
\begin{cases}
0.140 \, \mathrm{GeV \, fm} \qquad &\mathrm{if \, nucleon \rightarrow quark} \\
0.210 \, \mathrm{GeV \, fm} \qquad &\mathrm{if \, nucleon \rightarrow gluon}
\end{cases} .
\end{align}
Thus we see that the default value $\kappa = 1 \, \mathrm{GeV \, fm}$ is roughly a factor of $5$ larger than if a nucleon were composed of a single parton, which seems a reasonable ballpark.  We will explore the quantitative dependence of our results on the value of $\kappa$ later in Sec.~\ref{sec:Results}.

We note that the (semi)dilute / dense regime of the color glass condensate effective theory has been assumed in obtaining the analytical expressions in Sec.~\ref{sec:Theory}.  This assumption inherently treats the two colliding nuclei asymmetrically, with one being considered ``dense'' (in this case, $T_B$ which is used here to generate the saturation scale $Q_s$), and the other being considered ``(semi)dilute'' (leading to a linear dependence on the analogous scale $\mu^2$ which ultimately cancels out).  Because of this inherent asymmetry of the formulas, only the energy density derived from $T_R$ and the saturation scale derived from \textit{one} of the colliding nuclei $T_B$ are used.  To provide this information, we have modified the out-of-the-box Trento code to print off the profile function $T_B$ in addition to the reduced thickness function $T_R$.  While the correlations and probabilities obtained in Sec.~\ref{sec:Theory} are strictly valid only for asymmetric collision systems such as Cu Au, we will push them beyond their range of validity to apply to various systems including Pb Pb collisions.  Ultimately, the algorithm we describe here can also be modified to sample different probability distributions than the ones outlined in Sec.~\ref{sec:Theory} which can be more rigorously applied to symmetric collisions, or may not be based on the underlying color glass condensate theory at all.

%
\subsection{Monte Carlo Sampling}
\label{subsec:MonteCarlo}
%

Next let us detail how we perform the sampling of the $g \rightarrow q \barq$ splitting probabilities summarized in Sec.~\ref{sec:Theory}.  Suppose we have identified a gluon at a particular point $\vec{x}_\bot$ in the transverse plane, which may now split into a $q \barq$ pair of various flavors, or may remain a gluon without splitting.  The first step in the sampling is to identify the outcome of the possible $q \barq$ splitting, which is controlled by the total splitting probabilities shown in Fig.~\ref{f:multratio}.  We divide up the interval $[0,1]$ into outcomes consisting of splitting into up, down, strange, and charm quarks, with the remainder leading to the gluon staying intact without splitting.  We do not consider top or bottom quarks at this time.  Then by throwing a random number between $0$ and $1$, we identify the outcome of the potential splitting for the gluon in question.

\begin{figure}
\begin{center}
	\includegraphics[width=0.32\textwidth]{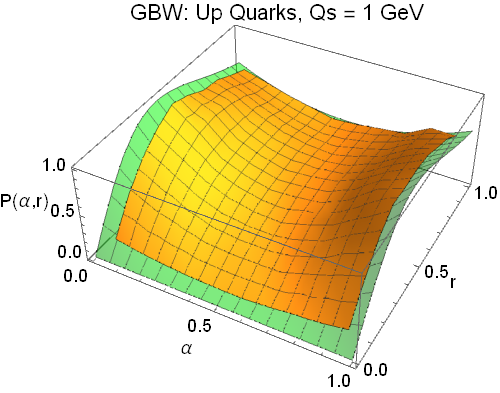}
	\includegraphics[width=0.32\textwidth]{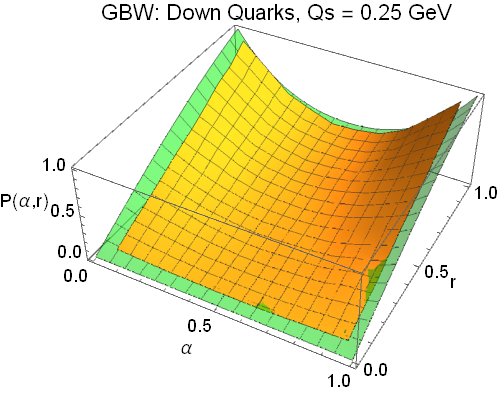}
	\includegraphics[width=0.32\textwidth]{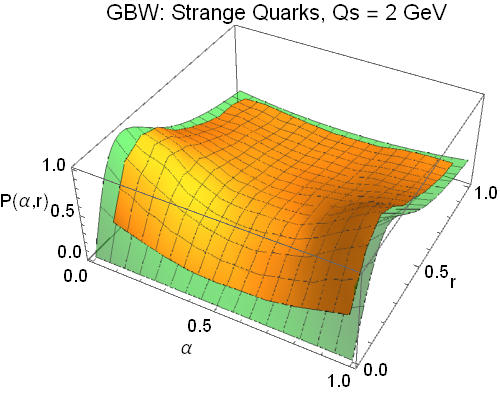}
	\caption{
	    Plots of the input and output of the Monte Carlo sampling, validating that it successfully reproduces the theoretical probability distribution \eqref{e:prob1_first} for the GBW model.  We have similarly checked that the procedure correctly reproduces the theoretical distribution \eqref{e:prob2_first} for the MV model.
	}
	\label{f:MCvalidate}
	\end{center}
\end{figure}

If the gluon splits into a $q \barq$ pair of a given flavor, then the second step is to determine the displacement and energy of the quark and antiquark relative to the gluon.  This information is controlled by the distributions in distance $r_\bot$ and light-front momentum fraction $\alpha$ given in Eqs.~\eqref{e:prob1_first} or \eqref{e:prob2_first} and illustrated in Fig.~\ref{f:probabilities}.  These distributions depend on the quark flavor through its mass, the saturation scale $Q_s (\vec{x}_\bot)$, and in the case of the MV model, a cutoff scale $\Lambda$.  To determine the positions of the produced quark and antiquark, we perform a simultaneous Monte Carlo sampling of $r_\bot$ and $\alpha$.  The splitting fraction $\alpha$ is randomly chosen from an interval $[\alpha_\mathrm{min}, 1 - \alpha_\mathrm{min}]$ which is almost the entire range from $0$ to $1$.  For consistency we must exclude the endpoints around $\alpha \approx 0 , 1$ because in this case either the quark ($\alpha$) or antiquark $(1-\alpha)$ carries so little energy that the underlying assumptions of the calculation are violated.  For this reason, we must introduce the parameter $\alpha_\mathrm{min}$ which determines this small excluded region of phase space; by default, we take $\alpha_\mathrm{min} = 0.01$.  Similarly, we randomly choose the distance $r_\bot$ from the interval $[0, d_\mathrm{max}]$, with the maximum $q \barq$ separation $d_\mathrm{max}$ limited by the breakdown of the underlying perturbative calculation at long distances.  Clearly $d_\mathrm{max}$ should be chosen to be a number on the order of $\ord{1/\Lambda_{QCD}}$, but its precise value is unspecified.  As a default, we cut this distance off at $d_\mathrm{max} = 1 \, \mathrm{fm}$.  The values of $(\alpha, r_\bot)$ for a given splitting are then determined using standard rejection sample techniques.  As seen explicitly in three illustrative cases shown in Fig.~\ref{f:MCvalidate}, this procedure accurately reproduces the input distribution (green surface) after repeated Monte Carlo sampling (orange surface).

%
%

Finally, we throw a uniformly random angle $\phi \in [0, 2\pi]$ for the transverse orientation of the displacement vector $\vec{r}_\bot$ with respect to the coordinate axes.  Then the displacement vector of the quark relative to the gluon is $\Delta \vec{x}_\bot^{q} = (1-\alpha) \vec{r}_\bot$ and the displacement vector of the antiquark relative to the gluon is $\Delta \vec{x}_\bot^{\barq} = -\alpha \vec{r}_\bot$.  This combination explicitly preserves the center of momentum of the $q \barq$ pair and is an explicit feature of the $g \rightarrow q \barq$ light front wave function \cite{Kovchegov:2006qn}.  In this way, we use the theoretical input distributions specified in Sec.~\ref{sec:Theory} to sample each gluon splitting.

%
\subsection{Energy and Charge Redistribution}
\label{subsec:Redistribute}
%

Finally, we explain in detail the primary loop of the ICCING algorithm which selects a ``gluon'' from the input energy density, performs the Monte Carlo sampling described above, and then redistributes the energy and charge as appropriate after the splitting.  The underlying assumption of this method is that the input energy density $\epsilon (\vec{x}_\bot)$ can be considered to be composed entirely of gluons, which may then be given the opportunity to split into $q \barq$ pairs.  In practice, we do this by randomly choosing a point in the energy density to act as the ``seed'' for a gluon; in our formulation, any point $\vec{x}_\bot$ containing nonzero energy density is equally likely to be chosen.  Then we draw a circle around this center point and count up the total energy enclosed.  The choice to associate a circular blob of energy with gluons is a simple model, and it depends in particular on the radius $r$ of that circle.  The width of the energy deposition profile of a gluon is a much more sophisticated quantity than we implement here, being appropriately described by generalized parton distributions (GPDs) \cite{Ji:1996ek, Radyushkin:1996nd, Ji:1996nm, Radyushkin:1996ru, Collins:1996fb, Diehl:2003ny}, but in general we expect $r$ to be some nonperturbative scale.  By default, we take $r= 0.5 \, \mathrm{fm}$ such that the diameter of the circle is $1 \, \mathrm{fm}$, and we explore the dependence of our results on this choice in Sec.~\ref{sec:Results}.

The energy contained in particular circle of radius $r$ is given by integrating the energy density, but we must consider that $\epsilon(\vec{x}_\bot)$ is in fact the {\it{three}}-dimensional energy density in units of $\mathrm{GeV} / \mathrm{fm}^3$.  To obtain the total enclosed energy $E_R$ which is to be compared with the quark mass thresholds, we have to integrate over the hypersurface defined by the initialization time $\tau_0$ at which this initial energy density is considered.  We take our default value $\tau_0 = 0.6 \, \mathrm{fm}$ from the Trento Bayesian analysis.  The overall scale for the energy density is controlled by the initialization time $\tau_0$ and by the parameter $a$ from Eq.~\eqref{e:Trentoa}.  As usual, this overall scale must be constrained by matching the total multiplicity from experiment.  Then by integrating the energy over the circle, which we denote as the region $R$, we obtain
\begin{align}
E_R &= \int\limits_{R} d^3 x \, \epsilon (\vec{x}) = \tau_0 \int\limits_{R} d\eta \, d^2 x_\bot \, \epsilon(\vec{x}_\bot) \\
\frac{dE_R}{d\eta} &= \tau_0 \int\limits_R d^2 x_\bot \, \epsilon(\vec{x}_\bot) = \tau_0 \Delta x \Delta y \sum\limits_{i \in R} \epsilon_i
\end{align}
with $\Delta x$ and $\Delta y$ the grid spacings in units of $\mathrm{fm}$.  Here we have assumed that the externally-provided energy density $\epsilon(\vec{x}_\bot)$ is a 2D, boost-invariant distribution which is intended to be the initial condition for a $2+1$D hydrodynamics simulation.  This assumption can be relaxed in future work if coupled to a more elaborate $3+1$D setup, but for now the distributions are taken to be boost invariant.  As such, the quark production rates will be boost invariant as well, and will be compared to the necessary thresholds to produce a $q \barq$ pair of a given flavor {\it{per unit rapidity}}.\footnote{As discussed in detail in Appendix~\ref{app:Theory}, the usual definition of rapidity is not applicable in coordinate space, where we have integrated over the transverse momentum in order to fix the transverse position.  Nevertheless, the distributions \eqref{e:prob1_first} and \eqref{e:prob2_first} are boost invariant, consistent with this interpretation.  It is also interesting to note that while the distributions are invariant under boosts of the total $q \barq$ momentum, they in principle contain additional information about the breaking of boost invariance with respect to the {\it{relative}} quark and antiquark rapidities through their dependence on the light-front momentum fraction $\alpha$.  Considerations of such $3+1$D information are beyond the scope of this calculation which we leave to future work.}

Having tallied the total energy $E_R$ (per unit rapidity) contained within the circular region $R$, we next assign some fraction of that enclosed energy to belong to a single ``gluon.''  Within the boost-invariant approximations used here, the gluon distribution as expressed in e.g. Eq.~\eqref{e:gluon_boost} depends inversely on its light-front momentum $q^+$, which for an ultrarelativistic particle is equivalent to its energy ($q^+ \approx \sqrt{2} E_G$):
\begin{align}
    \frac{d\sigma^G}{d q^+} \propto \frac{1}{q^+}
    \qquad \rightarrow \qquad
    \frac{dP}{dE_G} \propto \frac{1}{E_G}.
\end{align}
While the boost-invariant distribution $\sim 1 / E_G$ is the natural, consistent choice for the distributions used in Sec.~\ref{sec:Theory}, we can implement a somewhat more flexible distribution of gluon energies 
\begin{align}   \label{e:lambdaparam}
   \frac{dP}{dE_G} \propto  \left(\frac{1}{E_G} \right)^\lambda
\end{align}
by allowing the exponent $\lambda$ to deviate from $1$.  This flexibility can allow the ICCING algorithm to mimic the effects of linear small-$x$ quantum evolution by modifying the gluon distribution, with natural choices being $\lambda = 1.79$ (the ``perturbative'' or ``hard Pomeron'' \cite{Kovchegov:2012mbw} for $\alpha_s = 0.3$) and $\lambda = 1.08$ (the ``phenomenological'' or ``soft Pomeron'' \cite{Donnachie:1992ny}).  One may also contrast these results with a uniform distribution of gluon energies corresponding to $\lambda = 0$.

\begin{figure}
        \begin{centering}
    	\includegraphics[width=0.46 \textwidth]{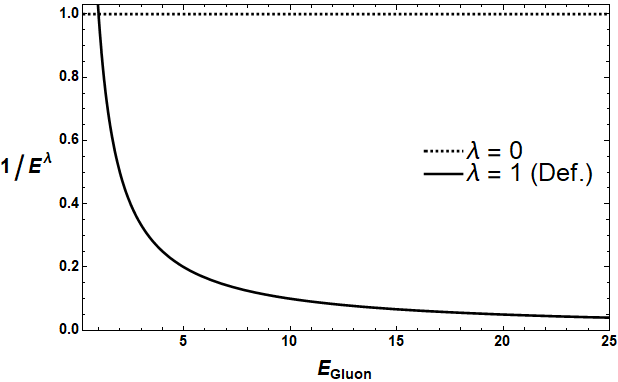} \,
    	\includegraphics[width=0.44\textwidth]{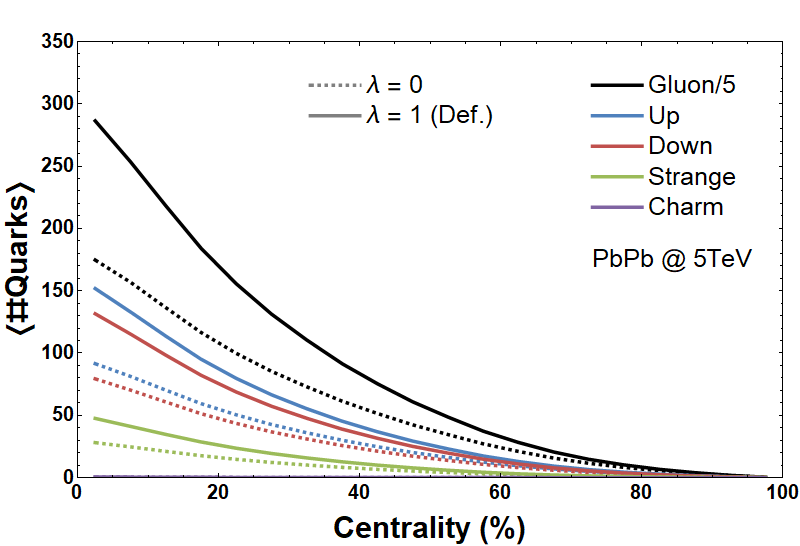}
    	\end{centering}
    	\caption{Illustration of the function \eqref{e:lambdaparam} used to sample the gluon energy (left) and its effect on the quark multiplicities (right).  Note that the gluon multiplicities have been scaled down by a factor of 5 for better comparison to the quarks.}
    	\label{f:lambdaEffect}
    \end{figure}

For a chosen value of the exponent $\lambda$ controlling the gluon energy distribution, we again perform a Monte Carlo rejection sample of the distribution to randomly choose a gluon energy $E_G \in [E_\mathrm{thresh} , E_R]$ between a minimum gluon threshold $E_\mathrm{thresh}$ and the total available energy $E_R$.  The choice of exponent $\lambda$ affects the relative probability to apportion the energy $E_R$ into a small number of harder gluons, versus a larger number of softer gluons.  Consequently, this change in the number of gluons leads to a change in the typical number of quark pairs produced.  This feature is illustrated in Fig. \ref{f:lambdaEffect} by comparison between the constant case $\lambda = 0$ and the boost-invariant case $\lambda = 1$.  While somewhat counterintuitive, the more steeply-falling energy distribution for $\lambda = 1$ leads to a significant \textit{increase} in the average multiplicity of all $q \barq$ pairs produced as a result.

    %
   %

If the total enclosed energy $E_R$ is less than the threshold value, we set $E_G = E_\mathrm{thresh}$ and prohibit these low-energy gluons from splitting.  The purpose of the threshold energy $E_\mathrm{thresh}$ is to provide a small cutoff for blobs of energy around the periphery of the fireball which are so low as to be unable or unlikely to ever produce $q \barq$ pairs.  This significantly affects the speed of the computation, and its impact on the quark multiplicities is explored in Sec.~\ref{sec:Results}.  Physically, a natural choice for gluons which are unable to pair produce quarks would be the up quark threshold $E_\mathrm{thresh} = 2  m_u = 4.6 \, \mathrm{MeV}$; as discussed in Sec.~\ref{sec:Results}, this is the safest choice to ensure that the quark multiplicities are not artificially suppressed.

If the gluon is seeded near the center of the fireball, then there will be ample energy $E_R$ enclosed to meet the minimum threshold $E_\mathrm{thresh}$, and only some fraction $\frac{E_G}{E_R} \leq 1$ of the enclosed energy will be reallocated.  If the gluon is seeded far enough out in the periphery that the enclosed energy falls below the threshold $E_\mathrm{thresh}$, then the region $R$ need not be further considered as a possible source of $q \barq$ splitting.  In this case, we subtract the energy density from the input grid, point by point, and add it directly into the output grid without modification.
\begin{align*}
    & \mathrm{If \,} (E_R < E_\mathrm{thresh}), \\
    & \qquad \Big[ \mathrm{For \,} i \in R, \\
    & \qquad \qquad \Big\{ \epsilon_i^\mathrm{(output)} += \epsilon_i^\mathrm{(input)};  \\
    & \qquad \qquad \epsilon_i^\mathrm{(input)} = 0; \Big \} \\
    & \qquad \Big]
\end{align*}
In this case, the result is that the energy distribution contained in the region $R$ is transferred directly from the input grid to the output grid {\it{without modifying its geometry}}.  This point is important to avoid applying unnecessary artifacts which could smear of the input distribution and significantly modify the geometry of the fireball.

%
\begin{figure}
\begin{center}
	\includegraphics[width=\textwidth]{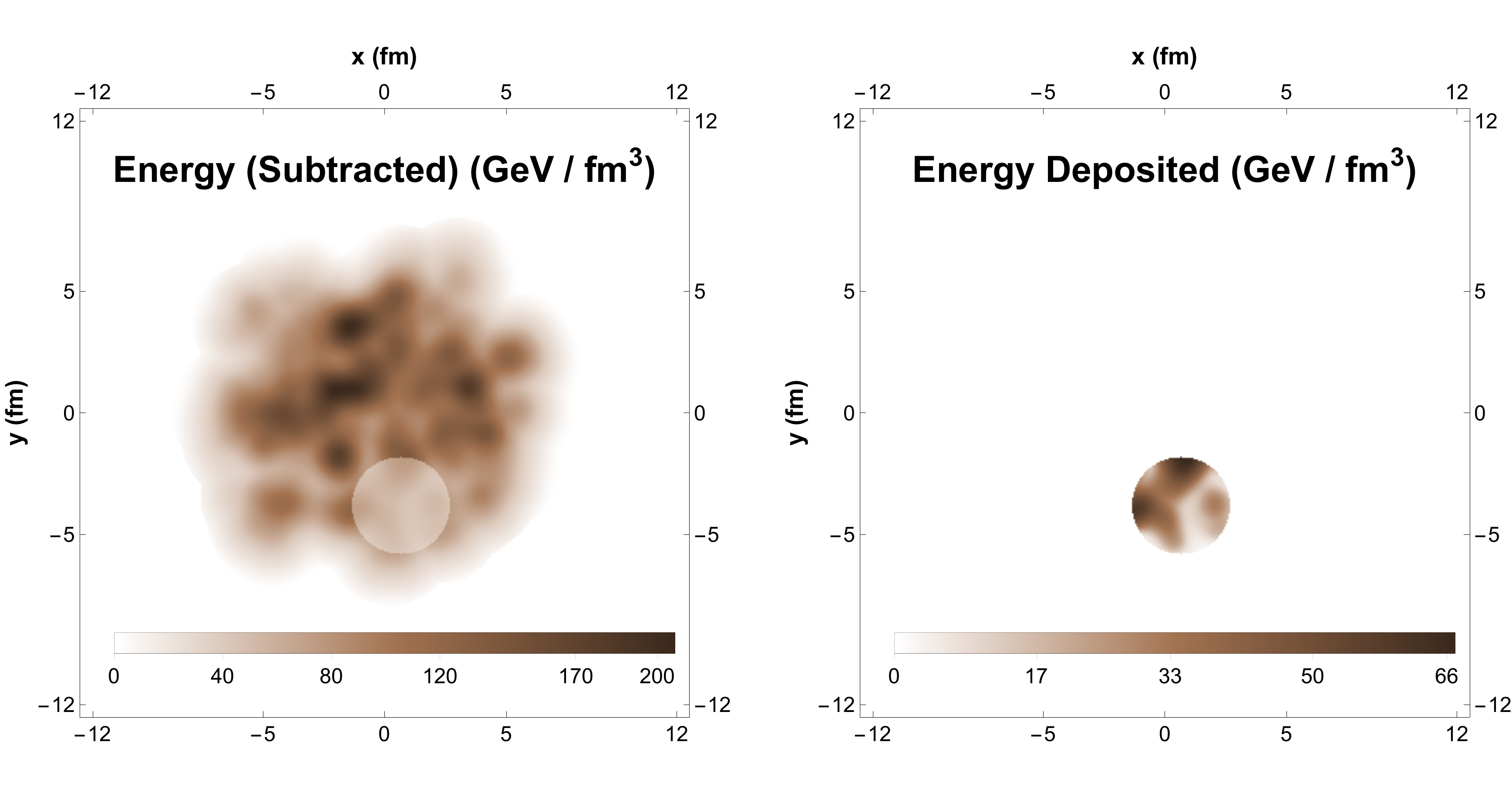}
	\caption{Illustration of how the ICCING algorithm transfers energy when the gluon {\it{does not}} split into a $q \barq$ pair.  The energy is deducted from the input grid (left) and deposited in the output grid (right) as shown.  	The energy transfer is done point by point and proportionately to the total enclosed energy.  As a result, the transferred energy retains the underlying geometric structure of the original energy density, as seen in both the input energy grid after subtraction and the output energy grid after deposition.  The gluon radius here has been greatly increased to clearly show these details.
	}
	\label{f:cutpaste}
	\end{center}
\end{figure}
%

If there is enough energy $E_R$ enclosed to produce the desired gluon $E_G \geq E_\mathrm{thresh}$, then the next step is to determine the outcome of the possible $q \barq$ splitting using the Monte Carlo procedure described in Sec.~\ref{subsec:MonteCarlo}.  If the result is to try to produce a quark flavor which is too heavy for the gluon to meet the mass threshold -- that is, if $E_G < 2 m$ -- then the procedure simply starts over, picking a new seed for a new gluon.  If the outcome is that the gluon {\it{does not}} split and remains a gluon, then again the gluon energy is subtracted from the input grid and added to the output grid in a way which preserves the underlying geometry, as illustrated in Fig.~\ref{f:cutpaste}.  In this case, the total amount of energy $E_G$ being deducted from the input grid is only a fraction $\frac{E_G}{E_R}$ of the total energy enclosed, so that fraction of the energy is transferred from each grid point $\epsilon_i \in R$, proportionately.
\begin{align*}
    & \mathrm{If \,} (\mathrm{outcome} = \mathrm{remains \, a \, gluon}), \\
    & \qquad \Big[ \mathrm{For \,} i \in R, \\
    & \qquad \qquad \Big\{ \epsilon_i^\mathrm{(output)} += \frac{E_G}{E_R} \epsilon_i^\mathrm{(input)};  \\
    & \qquad \qquad \epsilon_i^\mathrm{(input)} -= \frac{E_G}{E_R} \epsilon_i^\mathrm{(input)}; \Big \} \\
    & \qquad \Big]
\end{align*}
In this way, the energy $E_G$ is subtracted from $E_R$ without reducing any individual grid point $\epsilon_i$ to zero; instead, the grid points are depleted proportionately until they fail to meet the threshold criterion $E_R \geq E_\mathrm{thresh}$.  Again, transferring the energy to the output grid in a way which reflects the geometry in the region $R$ is important to preserve the original collision geometry as faithfully as possible.

%
\begin{figure}
\begin{center}
	\includegraphics[width=\textwidth]{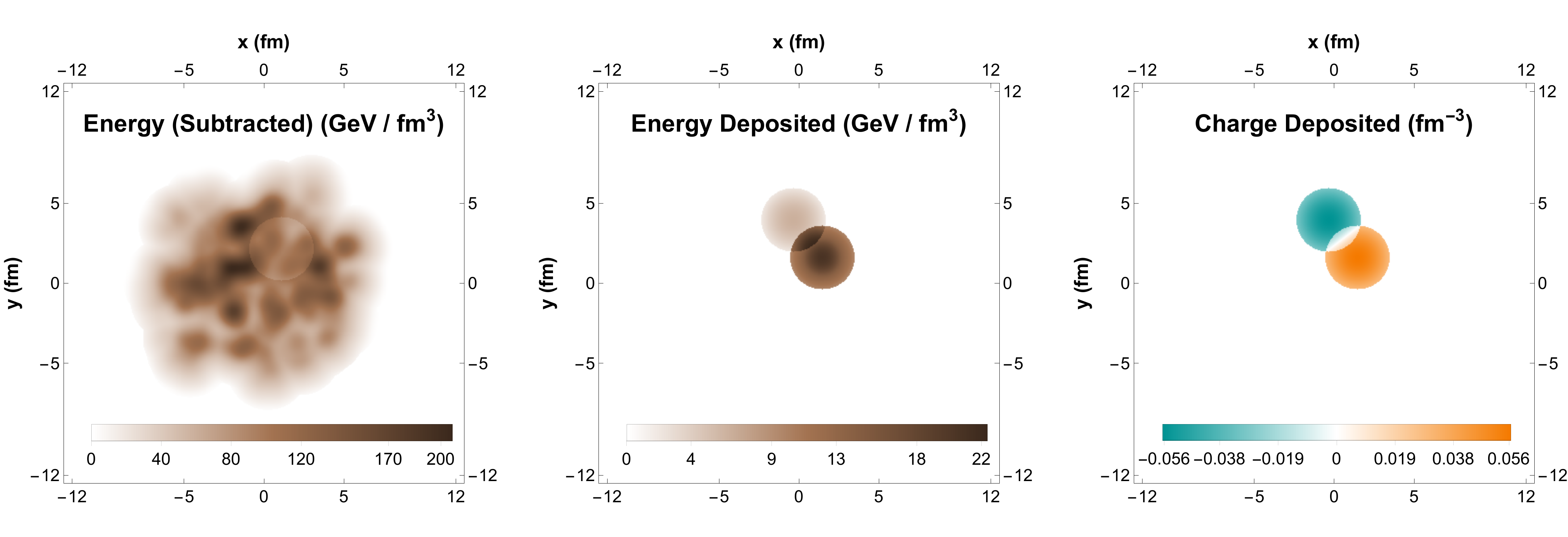}
	\caption{
	Illustration of how the ICCING algorithm transfers energy when the gluon {\it{does}} split into a $q \barq$ pair.  The energy is deducted from the input grid (left plot) proportionately, preserving the underlying geometry in the input grid.  But it is deposited in Gaussian blobs for the $q \barq$ pair, which are displaced relative to the original gluon position.  This modifies the energy density distribution (center plot) and also leads to a net displacement of positive and negative charge (here baryon density, right plot).  Note that the energy is in general shared unequally between the quark and antiquark; in this case, the quark carried about 75\% of the original gluon energy, as visible in the center plot.  Here both the radii and overall $q \barq$ displacement have been greatly increased to clearly show these details.
	}
	\label{f:cutsplit}
	\end{center}
\end{figure}
%

Most importantly, if the outcome of the splitting is to produce a quark flavor for which the gluon meets the mass threshold $(E_G \geq 2 m)$, then the energy $E_G$ will need to be redistributed in space, along with the associated conserved charges of the quarks.  First, the total energy $E_G$ is subtracted proportionately, point by point: $\epsilon_i^\mathrm{(input)} -= \frac{E_G}{E_R} \epsilon_i^\mathrm{(input)}$.  But now this total energy $E_G$ will be deposited at new locations, centered on the quark and antiquark positions, as illustrated in Fig.~\ref{f:cutsplit}.  We do this by drawing circles, with the same radius $r$ used for the gluon, around the quark position $\vec{x}_\bot + \Delta \vec{x}_\bot^q$ and the antiquark position $\vec{x}_\bot + \Delta \vec{x}_\bot^\barq$.  The total energy deposited inside the quark region is $\alpha E_G$, while the total energy deposited inside the antiquark region is $(1-\alpha) E_G$.  We choose to deposit these energies with Gaussian spatial profiles of width $r$ about their center, properly normalized for the finite number of grid points enclosed inside their respective circles (see Appendix~\ref{app:Densities}).  As with the method of subtracting the gluon energy from the input grid, the choices to use Gaussian profiles for depositing the quark energies in the output grid and the choice to use the same radius value $r$ are arbitrary.  In principle, the energy distribution around a quark in perturbative QCD is also calculable using GPDs, but we leave this extra layer of calculation for future work.  The spatial extent of the quark energy deposition is again a nonperturbative scale of $\ord{1 \, \mathrm{fm}}$, but there is no {\it{a priori}} reason for it to be equal to the gluon radius.

In addition to depositing the redistributed energy in Gaussian blobs centered on the quark and antiquark positions, we can now finally deposit similar Gaussian blobs for the charge densities of baryon number, strangeness, and electric charge (see also Appendix~\ref{app:Densities}).  We summarize the total charges for the four relevant quark flavors in Table~\ref{t:charges}.  Depending on the flavor of the produced quarks, we deposit the corresponding charges in their own output grids to track the charge densities.  The generation of these charge densities, along with a consistent modification of the energy density (such that every grid point containing charge density also has a corresponding energy density) is the primary new information provided by the ICCING algorithm.

%
\begin{table}
\centering
\begin{tabular}{ | c || c | c | c | } \hline
Flavor & B & S & Q \\ \hline
u & $\tfrac{1}{3}$ & 0 & $\tfrac{2}{3}$ \\ \hline
d & $\tfrac{1}{3}$ & 0 & $-\tfrac{1}{3}$ \\ \hline
s & $\tfrac{1}{3}$ & $-1$ & $-\tfrac{1}{3}$ \\ \hline
c & $\tfrac{1}{3}$ & 0 & $\tfrac{2}{3}$\\ \hline
\end{tabular}
\caption{The three conserved charges for the relevant quark flavors: baryon number (B), strangeness (S), and electric charge (Q).}
\label{t:charges}
\end{table}
%

The algorithm repeats in this fashion, decrementing an energy $E_G$ from the input grid and incrementing it to the output grid, until all of the energy has been transferred from input to output, generating the associated conserved charge densities in the process.  We have explicitly verified that this algorithm preserves conservation of energy and each of the three conserved charges in practice.  The decision tree for the algorithm detailed in this Section is summarized in Fig.~\ref{f:Algorithm}.  The algorithm terminates when all of the input energy has been transferred to the output grid, and it also performs error checks every step to ensure that none of the regions being drawn around the quark or gluon positions exceed the dimensions of the provided grids. In practice this is unlikely to happen since the gluon must be centered on a valued point, but in the case it does go out of bounds the algorithm terminates early and returns an ``out of bounds'' error message to prevent the propagation of unreliable data.  The final output of the ICCING algorithm are grids of the redistributed energy density $\epsilon^\mathrm{(output)}$ along with the charge densities $\rho_B$, $\rho_S$, and $\rho_Q$ of baryon number, strangeness, and electric charge, respectively.  These grids are written to file in a format which can then be read directly as the input to a subsequent hydrodynamics code for the evolution to the final observed particles.  As emphasized previously, we have designed ICCING as a universal tool which aims to be agnostic of our particular preference of initial-state and hydrodynamic models.  In principle, the procedure we have outlined here can be used to resample any initial energy density to construct one realization of the associated conserved charge densities.  This resampling procedure introduces statistical fluctuations of its own, such that one fixed input energy density can produce multiple outputs after incorporating the $g \rightarrow q \barq$ splitting.

%
\begin{table}[h]
\centering
\begin{tabular}{ | c | c | }
\hline
{\qquad \bf{External Parameters} \qquad} &  {\qquad \bf{ICCING Parameters} \qquad}                                      \\ \hline
Pb Pb 5.02 TeV                           &  GBW Model \eqref{e:prob1} - \eqref{e:multratio1}                            \\ 
$T_R \overset{(p=0)}{=} \sqrt{T_A T_B}$  &  $\alpha_s = 0.3$                                                            \\ 
$k = 1.6$                                &  $\kappa = 1 \, \mathrm{GeV} \, \mathrm{fm}$                                 \\ 
$w = 0.51 \, \mathrm{fm}$                &  $r = 0.5 \, \mathrm{fm}$                                                    \\ 
$a = 119 \, \mathrm{fm}^{-1}$            &  $d_{\max} = 1.0 \, \mathrm{fm}$                                             \\ 
$\tau_0 = 0.6 \, \mathrm{fm}$            &  $\alpha_{\mathrm{min}} = 0.01$                                              \\ 
                                         &  $S_{\mathrm{chop}} = 10^{-20} \, \mathrm{fm}^{-3}$                           \\ 
                                         &  $E_{\mathrm{thresh}} = 0.25 \, \mathrm{GeV}$                                \\ 
                                         &  $\lambda = 1$                                                               \\ \hline
\end{tabular}
\caption{The default parameter set for ICCING.  The meaning and usage of the parameters are detailed in Sec.~\ref{sec:algorithm}.}
\label{t:default}
\end{table}
%

%
\section{Results}
\label{sec:Results}
%

    %
    \begin{figure} 
    	\includegraphics[width=\textwidth]{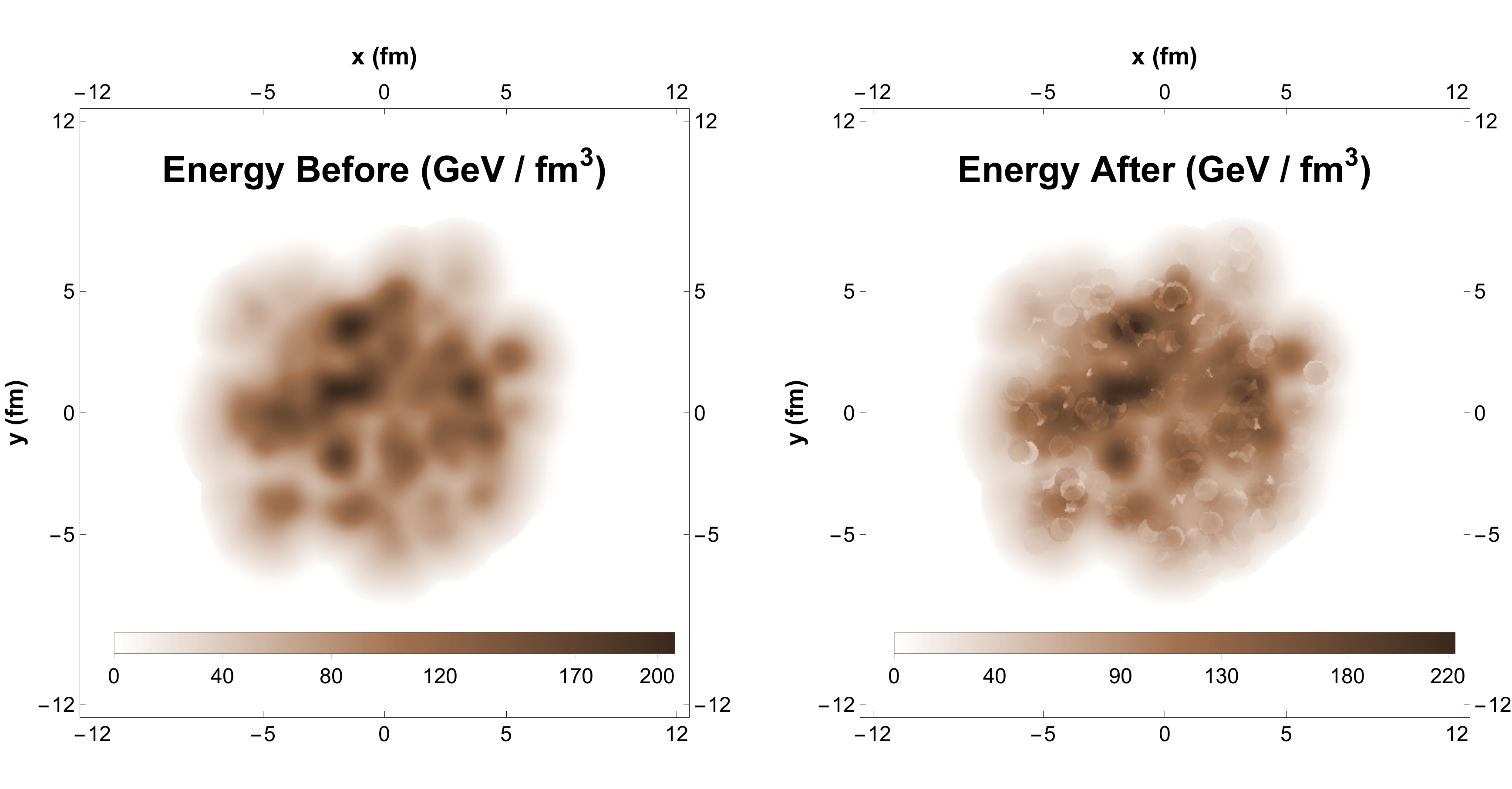}
    	\caption{
    	Comparison of the energy density before (left plot) and after (right plot) running the ICCING algorithm.  As a result of redistributing the energy density from the $g \rightarrow q \barq$ splitting, the energy density profile is somewhat modified, including visible artifacts associated with the model implementation.}
    	\label{f:BeforeAfterPlot}
    \end{figure}
    %

To begin, we summarize the default parameter set used for our simulations in Table~\ref{t:default}.  The meaning and usage of these parameters are explained in Sec.~\ref{sec:algorithm}; unless otherwise specified, these are the parameters used to generate our results.  We will quantify the geometric shape of the energy density and conserved charges by using the eccentricities $\varepsilon_2 , \varepsilon_3$ and their cumulants $\varepsilon_n \{2\} , \varepsilon_n \{4\}$ as defined in  Appendix~\ref{app:Eccentricities}, examining in particular the shape of the various charges in comparison with the bulk energy density.

We start the discussion of our results by analyzing how the ICCING algorithm modifies the bulk energy density when redistributing the energy due to the quantum mechanical $g\to q\bar{q}$ splitting process.  As clearly visible from the comparison of the energy density before and after application of ICCING in Fig.~\ref{f:BeforeAfterPlot}, the algorithm modifies the energy profile in a nontrivial way.  Clearly visible by eye are the circular ``holes'' in the energy density produced by ICCING when it subtracts a ``gluon'' blob of energy and redistributes it into quarks.  To some degree this modification reflects the addition of the partonic splitting physics and is a real effect introduced by the procedure, but the particular details of how this splitting is implemented (such as the choice of radius $r$) reflect artificial choices made in the model implementation.  It is therefore important to quantify this distortion, not least because Trento itself (before augmentation by ICCING), together with an appropriate hydrodynamic model, has been shown to accurately describe the flow harmonics and fluctuations of bulk particle production at the LHC \cite{Moreland:2014oya, Bernhard:2016tnd, Alba:2017hhe, Giacalone:2017dud}.  If ICCING distorts the energy profile too much, it could substantially disrupt existing fits to the data, rather than benignly supplying additional charge distributions. 

\begin{figure}
        \begin{centering}
    	\includegraphics[width=0.45 \textwidth]{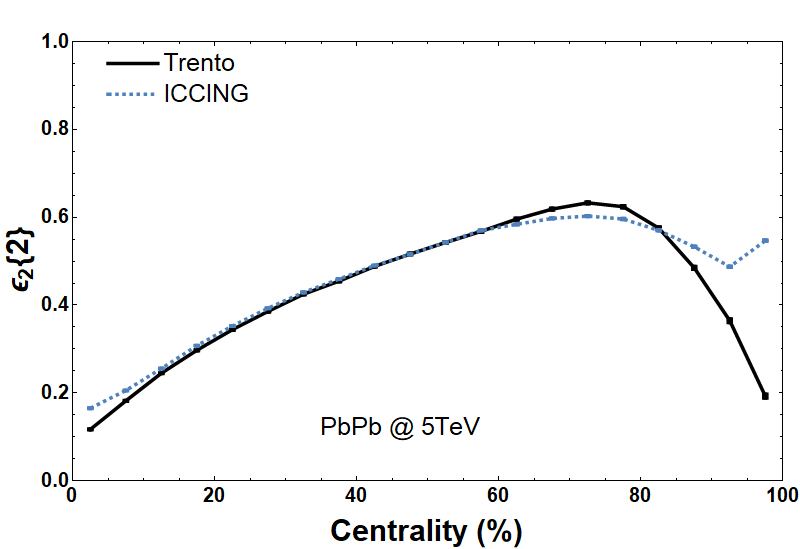} \,
    	\includegraphics[width=0.45 \textwidth]{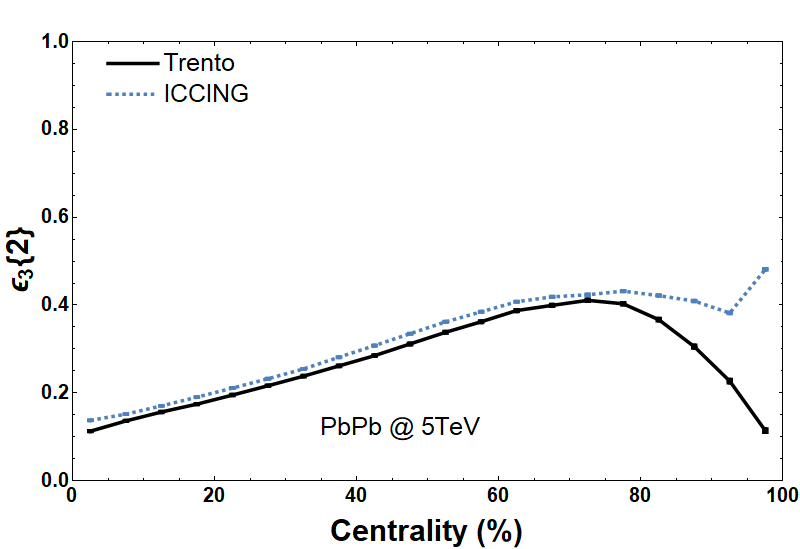}
    	\end{centering}
    	\caption{Eccentricities of the energy density as a function of centrality before (solid curve) and after running ICCING (dotted curve).}
    	\label{f:EccInOut}
    \end{figure}

The distribution of the RMS ellipticity $\varepsilon_2 \{2\} = \langle \varepsilon_2^2 \rangle^{1/2}$ and triangularity $\varepsilon_3 \{2\} = \langle \varepsilon_3^2 \rangle^{1/2}$ characterizing the energy density are shown in Fig.~\ref{f:EccInOut} as a function of centrality.  For $10-60\%$ -- most of the centrality range -- the effect of ICCING on the energy eccentricities is very small.  There is almost no modification to the ellipticity $\varepsilon_2 \{2\}$ and there is a small, roughly constant increase in the triangularity $\varepsilon_3 \{2\}$, which is consistent with the expectation that the additional pair splitting is a much smaller effect than the average elliptical geometry of the collision, but that it does introduce a small new source of event-by-event fluctuations.  Interestingly we see that ICCING does significantly modify the geometry of the energy density both in very central and in very peripheral collisions.  For the $0-10\%$ most central collisions, ICCING produces a nontrivial increase in the ellipticity; whereas the nucleon-level geometry in such collisions tends to be quite round, ICCING converts these round geometries into something more elliptical due to the back-to-back nature of the $g \rightarrow q \bar{q}$ splitting.  A similar effect appears in peripheral collisions $\sim 70-90\%$, where the peak of the ellipticity distribution shifts to the right.  As seen in previous work \cite{Sievert:2019zjr}, the location of this peak characterizes the transition between impact-parameter-driven geometry and finite-number-driven geometry characterized by $N_{part}$ at the level of Trento.  As such, the peak location reflects a resolution scale for the constituents of the medium, and a modification of this transition due to the introduction of sub-nucleonic degrees of freedom in ICCING is natural.  Lastly we note a sharp kink in the eccentricity distributions in the most peripheral centrality bin, $95-100\%$ characterized by significantly more eccentric events.  We attribute this feature to ICCING events in which only 1 or 2 quark pairs are produced, leading to a sharp increase in the eccentricities.  In any case, the final-state anisotropic flow in these most peripheral collisions is likely to be highly modified by effects like nonlinear response to the initial geometry  \cite{Noronha-Hostler:2015dbi,Sievert:2019zjr}.  Thus we conclude that the modifications to the bulk geometry induced by ICCING are generally small, with interesting systematic differences in very central and very peripheral events.

    %
   %

%
\subsection{Strangeness As a Distinct Probe of the Initial State}
%

    %
    \begin{figure}[h!]
        \begin{centering}
    	\includegraphics[width=0.6 \textwidth, trim = 2.25cm 0cm 3cm 1.15cm, clip]{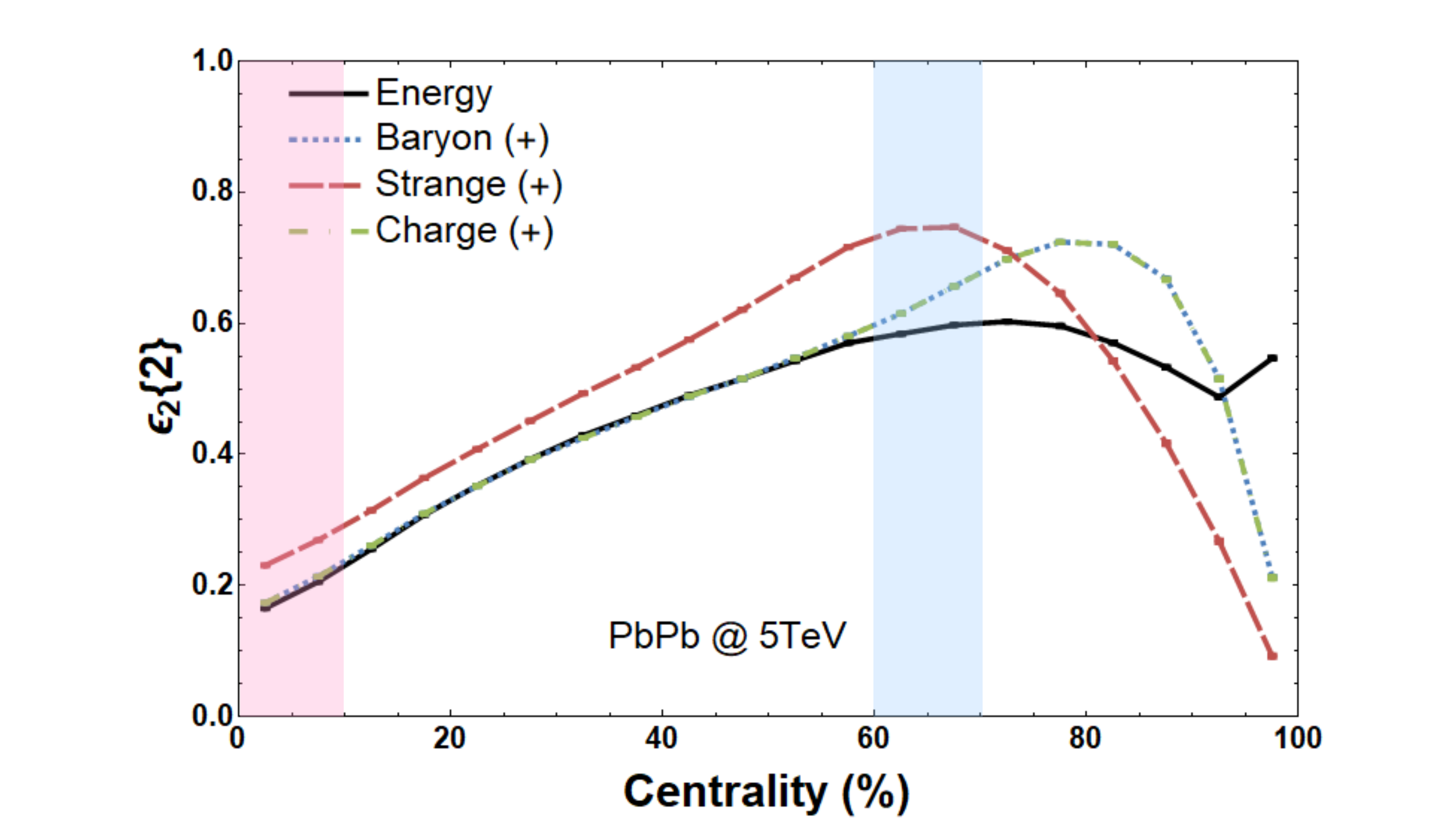} 
    	\\
    	\vspace{.5cm}
    	\includegraphics[width=0.45 \textwidth]{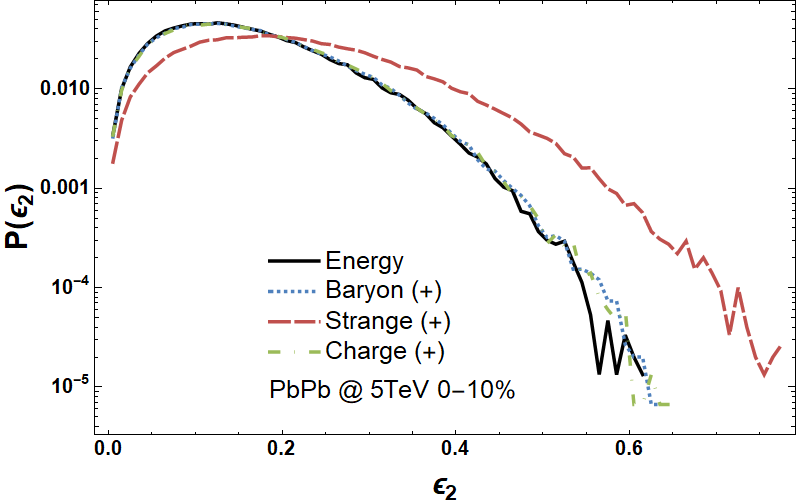} \,
    	\includegraphics[width=0.45 \textwidth]{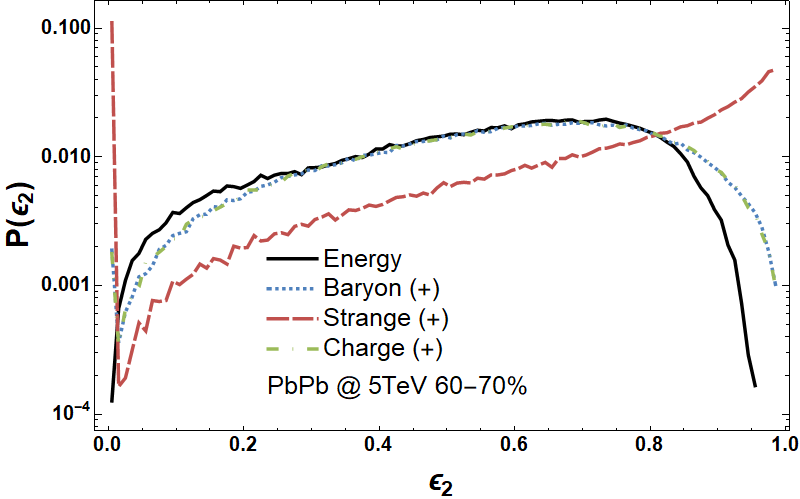}
    	\end{centering}
    	\caption{
    	Top:  RMS ellipticity $\varepsilon_2 \{2\}$ versus centrality.  Two centrality bins $(0-10\%)$ and $(60-70\%)$ are highlighted with the corresponding histograms shown below.
    	Bottom Left: Distribution of ellipticity in the $0-10\%$ bin.
    	Bottom Right: Distribution of ellipticity in the $60-70\%$ bin.
    	}
    	\label{f:IBSQCentrality}
    \end{figure}
    %

One of the primary results of this work is shown in Fig.~\ref{f:IBSQCentrality}, where we compare the RMS eccentricities $\varepsilon_2 \{2\}$ and $\varepsilon_3 \{2\}$ of the energy profile against the charge profiles of baryon number, strangeness, and electric charge.  For the charge distributions, we characterize the geometry of just the positive-charge distribution, with the corresponding distribution of negative charge having nearly-identical eccentricities in all cases.  From the plot of ellipticity $\varepsilon_2 \{2\}$ in the top left of Fig.~\ref{f:IBSQCentrality}, we note that, for $ < 60\%$ centrality, the baryon number and electric charge distributions track the energy distribution almost exactly, while the strangeness distribution is much more eccentric.  We examine this feature more differentially for the central $0-10\%$ centrality bin (pink band) by plotting the corresponding probability distribution in the bottom-left corner of Fig.~\ref{f:IBSQCentrality}.  This histogram shows that, on an event-by-event basis, the $B, Q$ distributions reproduce the geometry profile of the energy density; we attribute this feature to the huge abundances of $u, d$ quarks produced in these central events.  Because these nearly-massless quarks are copiously and nearly homogeneously produced in central collisions, their resulting charge densities mirror the original energy density profile, quantitatively.  In striking contrast, the strangeness distribution is far more eccentric than the bulk, even in the most central collisions.  This clearly indicates that the strange quark profiles do \textit{not} saturate the collision geometry, generating a lumpy, highly anisosotropic geometry even when the bulk geometry is quite round.

In more peripheral collisions starting at $60-70\%$ centrality, the $B, Q$ ellipticities begin to deviate significantly from the bulk energy density as the number of $u, d$ quarks being produced decreases and the transition to number-driven geometry sets in.  This centrality window, highlighted in the blue band, is shown in the histogram in the bottom-right panel of Fig.~\ref{f:IBSQCentrality}.  Here the $B, Q$ distributions track qualitatively the same shape as the energy density, reflecting the still-dominant role of impact-parameter-driven geometry, but nontrivial differences begin to appear.  There is a significant increase in events with $B, Q$ eccentricities which are larger than that of the energy distributions which produced them, indicating that the $u, d$ quark abundances have dropped low enough that they are no longer saturating the bulk geometry and are instead characterizing a different, lumpier distribution.  In contrast, in this centrality window the strangeness distribution exhibits a dramatic increase in events with strangeness ellipticity close to $1$, as well as a large spike around zero.  These simultaneous peaks at $0$ and $1$ likely reflect a set of events with few strange quarks, such as exactly one or two $s \bar{s}$ pairs, creating only one or two blobs of positive strangeness which are innately round (zero eccentricity) or innately elliptical (maximal eccentricity).  The appearance of this double-peak behavior in $\epsilon_2^{(S+)}$ also correlates with the location of the peak of the $\varepsilon_2^{(S+)} \{2\}$, which is again indicative of a transition to event geometries controlled by a small number of strange quarks being produced.

\begin{figure}
        \begin{centering}
    	\includegraphics[width=0.6 \textwidth]{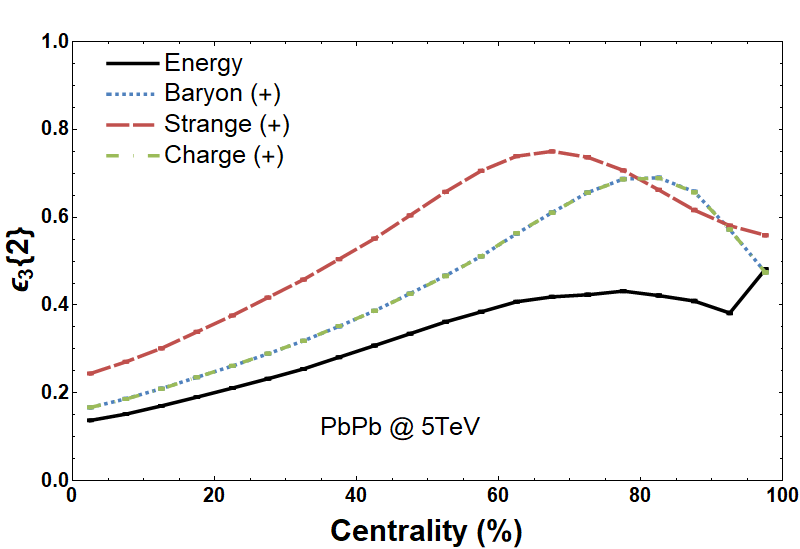}
    	\end{centering}
    	\caption{
         RMS triangularity $\varepsilon_3 \{2\}$ versus centrality.
    	}
    	\label{f:E3_BSQCentrality}
    \end{figure}

In Fig.~\ref{f:E3_BSQCentrality} we show the RMS triangularity distribution $\varepsilon_3 \{2\}$ of the bulk energy density and $BSQ$ charge distributions as a function of the centrality class.  Unlike the ellipticity $\varepsilon_2 \{2\}$ (top panel in Fig.~\ref{f:IBSQCentrality}) which for $\lesssim 60\%$ centrality is dominated by the mean-field elliptical geometry, the triangularity $\varepsilon_3 \{2\}$ arises entirely from fluctuations, without this mean-field background.  Up to very peripheral collisions $> 80\%$ centrality, a clear hierarchy $\varepsilon_3^{(S^+)} \{2\}  > \varepsilon_3^{(B^+ \, , \, Q^+)} \{2\} > \varepsilon_3^{(E)} \{2\}$.  This hierarchy is consistent with the introduction of new sources of sub-nucleonic fluctuations contributing to the $BSQ$ distributions, with the strange quarks producing the fewest number of particles and therefore fluctuating the most.  Because of its direct sensitivity to the fluctuating charge distributions, independent of the mean-field background, the triangularity $\varepsilon_3$ is the most sensitive probe we find here to the differences between the various charge and energy geometries.  We also note that, in all of these observables, the baryon number and electric charge distributions are essentially identical, because both quantum numbers are carried by the abundant $u, d$ quarks, with their mass difference being so small that the two distributions are indistinguishable

    %
    %

    %
    \begin{figure}[h!]
    	\includegraphics[width=0.6 \textwidth, trim = 2.25cm 0cm 3cm 1.15cm, clip]{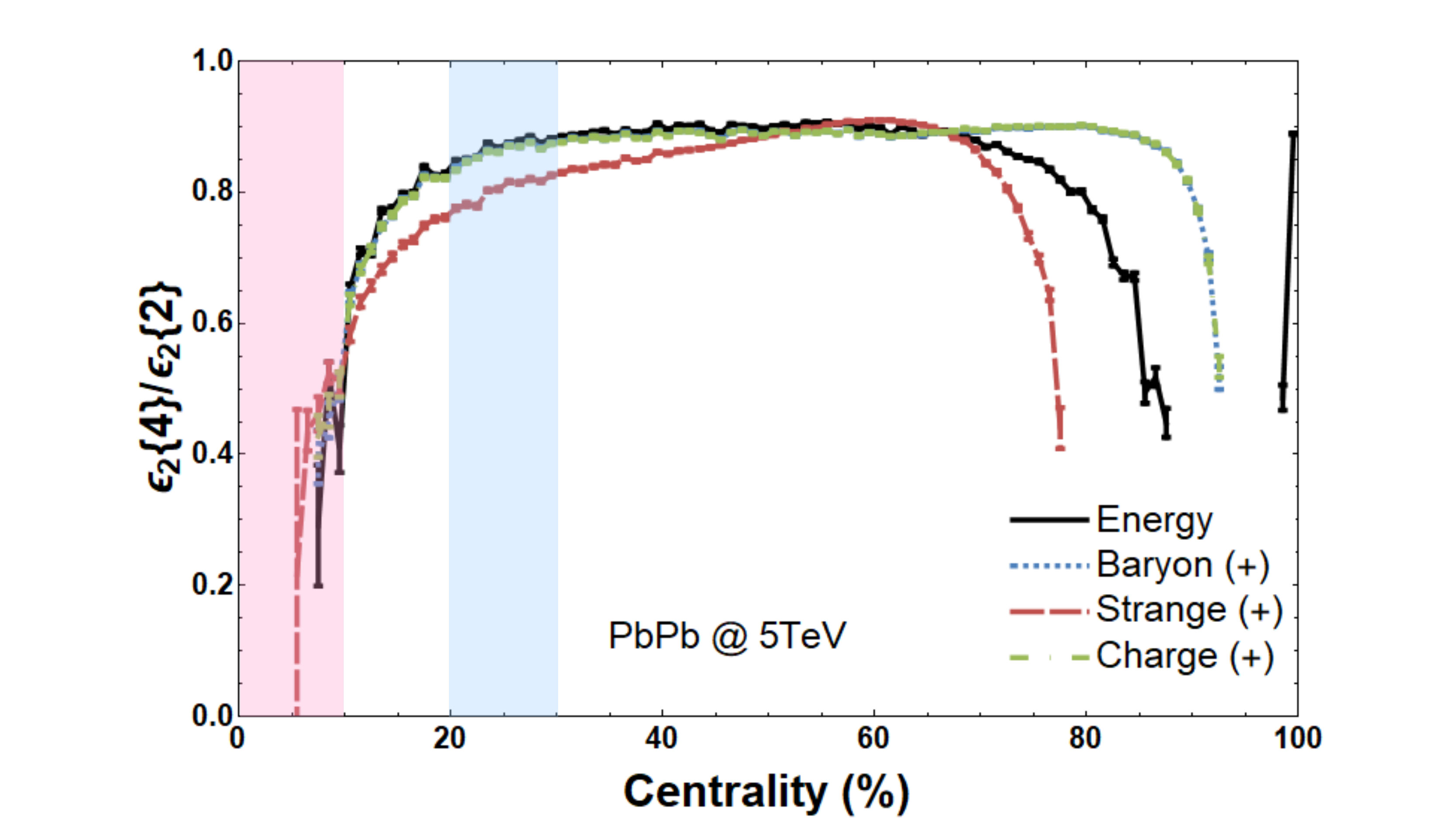}
    	\\
    	\vspace{.5cm}
    	\includegraphics[width=0.45 \textwidth]{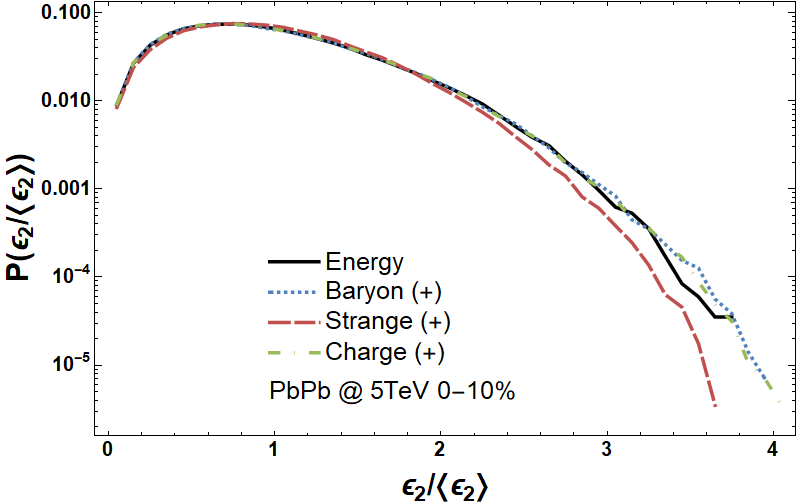}
    	\includegraphics[width=0.45 \textwidth]{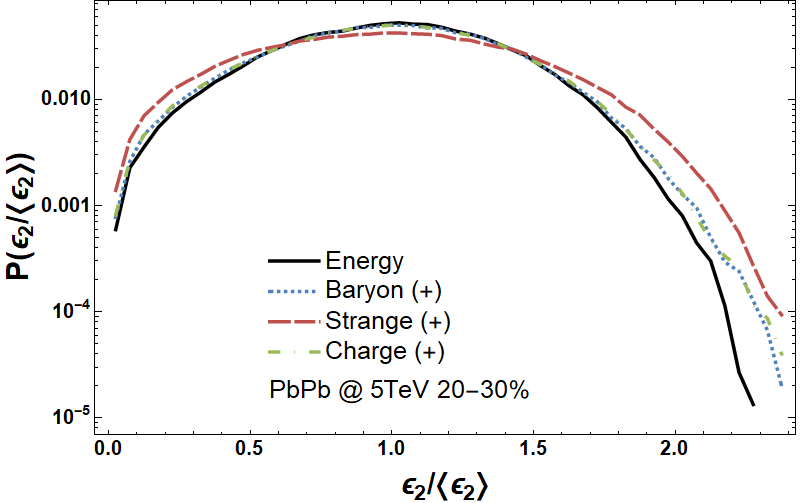}
    	\caption{
    	\textbf{Top:} Cumulant ratio $\varepsilon_2 \{4\} / \varepsilon_2 \{2\}$ versus centrality.  Two centrality bins ($0-10\%$ and $20-30\%$) are highlighted, corresponding to the histograms below.
    	\textbf{Bottom Left:} Probability distribution of the ellipticity $\varepsilon_2$ in the $0-10\%$ bin, scaled by the mean $\langle \varepsilon_2 \rangle$ to illustrate the width of the distribution.  Note that the strangeness distribution is \textit{narrower} than the bulk.
    	\textbf{Bottom Right:} Probability distribution of the ellipticity $\varepsilon_2$ in the $20-30\%$ bin, scaled by the mean to illustrate the width.  Note that the strangeness distribution is \textit{broader} than the bulk.
    	}
    	\label{f:IBSQ42Centrality100K}
    \end{figure}
    %

Another primary result of this work is the cumulant ratio $\varepsilon_n \{4\} / \varepsilon_n \{2\}$ for the various charge and energy distributions shown in Fig.~\ref{f:IBSQ42Centrality100K}.  As described in Sec.~\ref{app:Eccentricities}, the deviation of this ratio from unity is a measure of the variance of the $\varepsilon_n$ distribution, or equivalently the width of the $\varepsilon_n$ probability distribution normalized by the mean.  We see in the top panel of Fig.~\ref{f:IBSQ42Centrality100K} that for $< 70\%$ centrality, the baryon number and electric charge distributions track the energy density profile quite closely, while the strangeness distribution is significantly different.  Notably, there is a crossing of the curves at around $10\%$ centrality, where the strangeness distribution changes from having $\varepsilon_2 \{4\} / \varepsilon_2 \{4\}$ smaller than the bulk in mid-central collisions to a value larger than the bulk in central collisions.  This feature is explored from a different angle in the histograms for the $0-10\%$ central bin (pink band, bottom-left panel of Fig.~\ref{f:IBSQ42Centrality100K}) and the $20-30\%$ bin (blue band, bottom-right panel of Fig.~\ref{f:IBSQ42Centrality100K}).  These histograms are normalized by the mean to better reflect differences in the width of the distributions, corresponding to the deviation of the cumulant ratio from unity.  In the $20-30\%$ bin, we clearly see that the strangeness distribution is significantly \textit{broader} than the bulk, reflecting the increased number of event-by-event fluctuations from having a fairly small number of strange quarks being produced.  The event-by-event fluctuations of the energy density are comparatively narrower and more peaked around their mean value, which is dictated by the strong mean-field background of the bulk elliptical geometry.  In contrast, in the $0-10\%$ central bin, the strange quark distribution is \textit{narrower} than the bulk, reflecting a qualitative change of the strangeness distribution relative to the bulk.  As is clearly seen from the $\varepsilon_2$ cumulant ratio, in central collisions the fluctuations of the energy density greatly increase when going to central collisions, which can be understood from the disappearance of a mean-field elliptical geometry when the impact parameter goes to zero.  In comparison to this vanishing background, the event-by-event fluctuations are greatly magnified, resulting in a significant broadening of the width of the $\varepsilon_2$ histogram.  However, the strangeness distribution is broadened significantly less by this effect, resulting in the reduced slope of the $\varepsilon_2$ cumulant ratio and the crossing of the curves at $10\%$ centrality.  

We argue that this qualitative change in the strangeness distribution relative to the bulk with centrality indicates a dependence on more than just the number of quarks being produced; it further indicates that the underlying geometries which produce them are different and respond differently to changes in centrality.  If the only difference in the BSQ charge geometries with centrality came from the change in the number of particles produced, then we would expect to see the same hierarchy in $\varepsilon_2 \{4\} / \varepsilon_2 \{2\}$, that is, $E \approx B^+ \approx Q^+ > S^+$ maintained between mid-central and central collisions.  The fact that the disappearance of the bulk elliptical geometry in central collisions affects the strangeness distribution differently from the others reflects the fact that the geometry capable of pair producing strange quarks is not identical to the bulk geometry.  Rather, because the strange quarks have a nontrivial mass threshold of $2 m_s \approx 200 \, MeV$, they can only be produced by hot spots in the fireball with sufficient energy density to meet this mass threshold.  This feature can explain why the cumulant ratio of strangeness responds in a less singular manner than the bulk in central collisions: while the geometry of the bulk is becoming very round in central collisions, the geometry of the hot spots capable of pair produced strange quarks is innately lumpier.  As a result the width of the strange quark distribution is less sensitive to both the presence and the absence of a large elliptical geometry of the bulk.  

This effect on the strangeness distribution when going to central collisions can be compared with the behavior of the baryon number and electric charge distributions.  From the bottom-right panel of Fig.~\ref{f:IBSQ42Centrality100K}, we see that in $20-30\%$ collisions, the $B,Q$ distributions are in fact somewhat wider than the energy density distribution, although the difference in the cumulant ratio $\varepsilon_2 \{4\} / \varepsilon_2 \{2\}$ is harder to see.  It is interesting to note that the $B,Q$ distributions, too, have changed relative to the energy density when going from $20-30\%$ to $0-10\%$ centrality; while in the $20-30\%$ bin the $B,Q$ distributions are wider than the energy density, in the $0-10\%$ bin they have the same width.  Perhaps, with sufficient statistics in the ultracentral region, one may eventually see a similar crossing where the $B,Q$ distributions become narrower than the energy density as well.  If so, this may indicate that the hot spot coupling due to nonzero mass thresholds may extend to the light quarks to some degree as well.

    %
    \begin{figure}
    \begin{centering}
    \includegraphics[width=0.6 \textwidth]{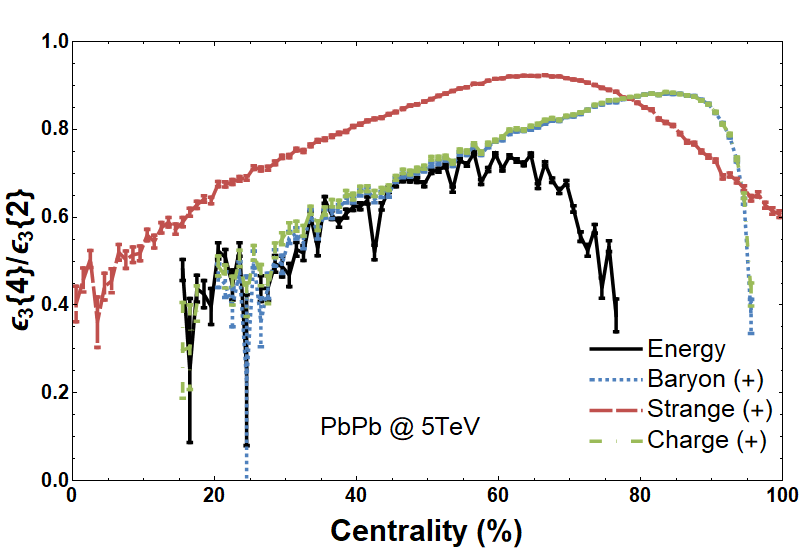}
    \end{centering}
    	\caption{
    	 Cumulant ratio $\varepsilon_3 \{4\} / \varepsilon_3 \{2\}$ versus centrality.
    	}
    	\label{f:E3_IBSQ42Centrality100K}
    \end{figure}
    %

It is also instructive to compare the behavior of the strangeness and bulk distributions for the cumulant ratio of triangularity shown in  Fig.~\ref{f:E3_IBSQ42Centrality100K}.  In this observable, the absence of a mean-field background geometry leads to an ordering hierarchy comparable to the one seen in the $0-10\%$ bin for $\varepsilon_2$: $S^+ > B^+ \approx Q^+ \approx E$.  This is the natural ordering in the absence of a strong geometrical effect, where the fluctuations are largest for the bulk quantities and smaller for the strangeness, which arises from an inherently lumpier hot spot distribution.  We argue that, taken together, these descriptors of the event-by-event fluctuations of the strangeness distribution and its centrality dependence provide systematic evidence for the coupling of strangeness to an independent underlying event geometry.

    %
    \begin{figure}
        \begin{centering}
        \includegraphics[width=0.8\textwidth]{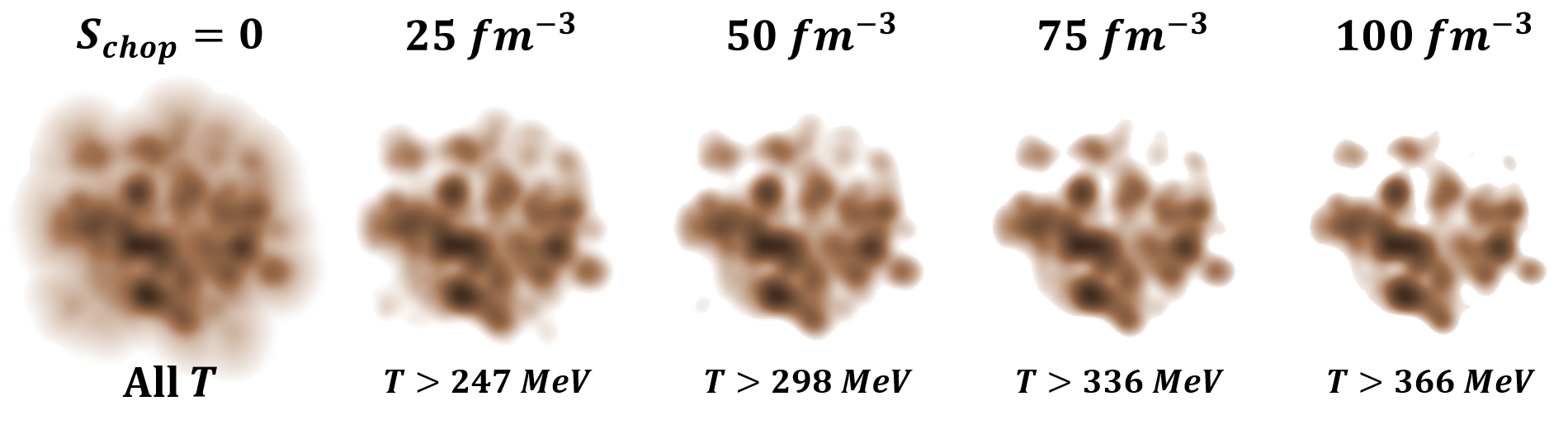}
    	\\
    	\includegraphics[width=0.45 \textwidth]{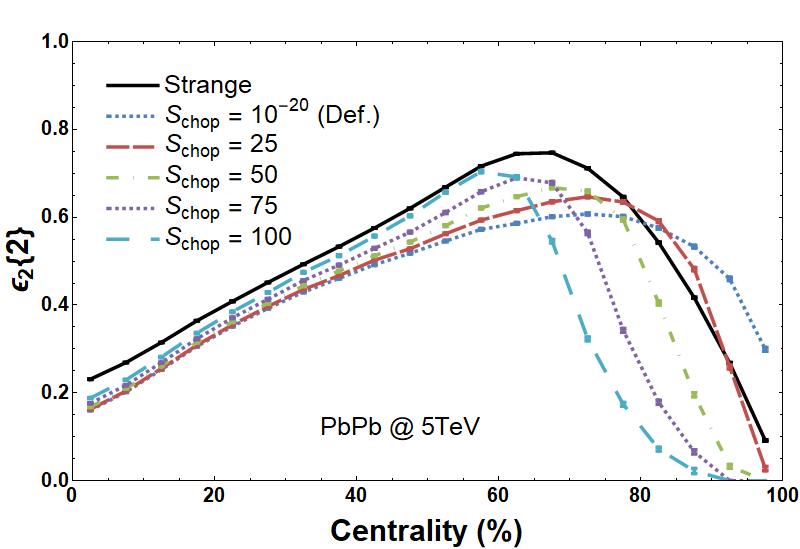} \,
    	\includegraphics[width=0.45 \textwidth]{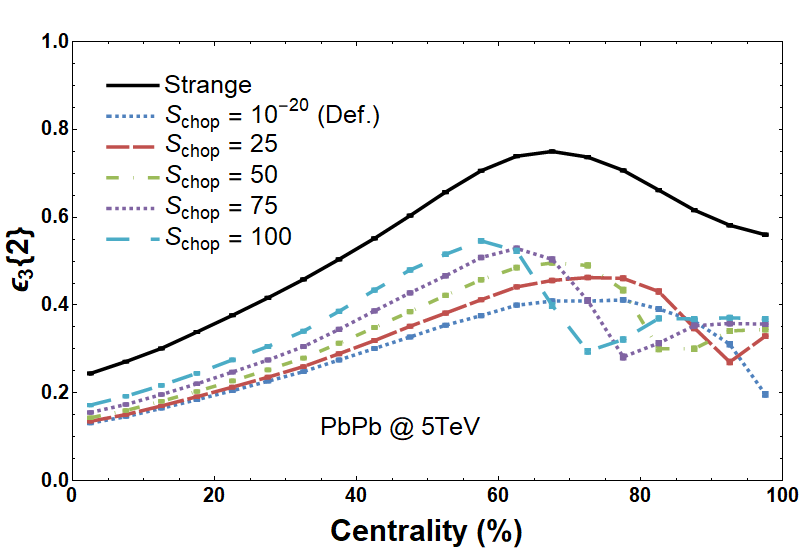}
    	\end{centering}
    	\caption{Comparison of the $S^+$ eccentricities (black) with the eccentricities of the energy distribution after applying a grooming cut $S_{\mathrm{chop}}$ between $0$ and $100 \, \mathrm{fm}^{-3}$ to the entropy density using the equation of state \cite{Alba:2017hhe}.  Top:  Increasing $S_\mathrm{chop}$ grooms the bulk geometry to select on the hot spots which dominate strangeness production.  Bottom: The geometry of the energy distribution after hot spot grooming converges toward the geometry responsible for the strangeness production.  
    	}
    	\label{f:Echop}
    \end{figure}
    %

To further investigate this idea of strangeness coupling to a hot spot geometry, we compare in Fig.~\ref{f:Echop} the RMS ellipticity and triangularity of the strangeness produced in ICCING (black curve) against the initial energy geometry (without modification by ICCING) as a function of the grooming parameter $S_\mathrm{chop}$ described in Sec.~\ref{subsec:Trento}.  By throwing out grid points with initial entropy densities below the cutoff, we can selectively impose tighter and tighter cuts on the hot spots present in the initial event upon which ICCING acts.  We see that, as the cutoff is tightened to select on progressively hotter parts of the initial geometry, the energy curve moves toward the strangeness distribution produced by ICCING.  The agreement is not perfect, but we do not expect it to be so since the original hot spot geometry is modified by ICCING along with the production of the strange particles.  Qualitatively, grooming the event to select on the hot spots increases the eccentricities and shifts the peak value to the left, just as seen in the strangeness distribution compared to the bulk energy density.  While further quantitative comparison to try to pin down the exact correlation between strangeness and hot spot geometry is warranted, this qualitative picture is consistent with the centrality dependence of the cumulant ratios seen in Figs.~\ref{f:IBSQ42Centrality100K} and~\ref{f:E3_IBSQ42Centrality100K}.

%
\subsection{Sensitivity Analysis to Model Parameters}
%

Let us now study the robustness of the above results by exploring their sensitivity to the various ICCING model parameters.  (We reiterate that unless otherwise specified, the default parameters are the ones indicated in Table~\ref{t:default}).  The discussion in this section is limited only to the set of parameters which affect the results in a sizeable way.  Parameters not explicitly discussed here, including the minimal splitting fraction $\alpha_{min}$ and the maximum $q\bar{q}$ separation $d_{max}$, do not significantly impact the observables. The insensitivity of our results to changes in details of the splitting function like $\alpha_{min}$ and $d_{max}$ indicate that the shape of the spatial distribution shown in Fig.~\ref{f:probabilities} plays a relatively small role in comparison to choices which affect the total chemistry of the initial state.  

%
%
\begin{figure}
    \begin{centering}
	\includegraphics[width=0.45 \textwidth]{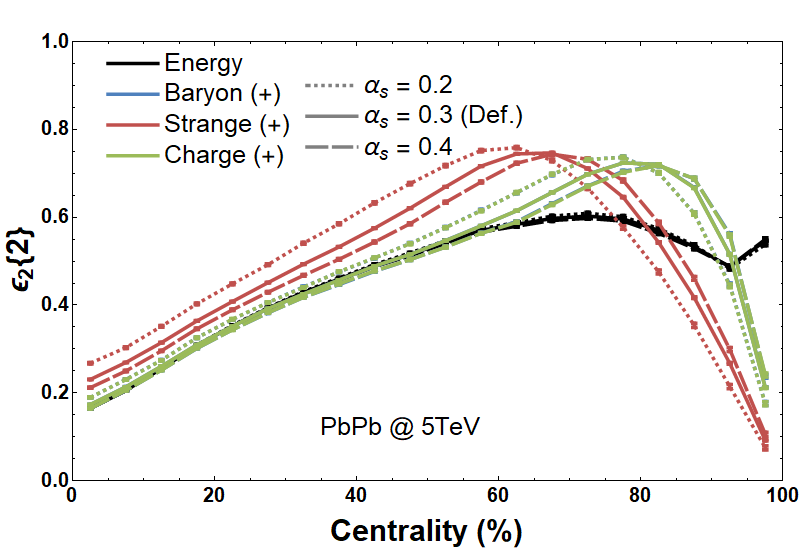} \,
	\includegraphics[width=0.45 \textwidth]{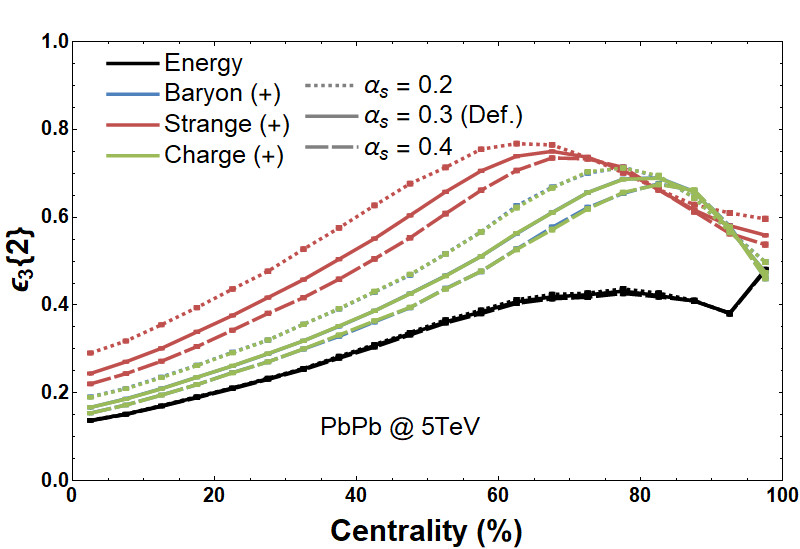}
	\end{centering}
	\caption{Comparison of the second cumulants $\varepsilon_n \{2\}$ for different values of the strong coupling $\alpha_s$.}
	\label{f:IBSQCentralityAlphaS}
\end{figure}
%
%

The simplest parameter affecting the chemistry of the initial state is the coupling constant $\alpha_s$, which affect the total rates of $q \barq$ pair production for all flavors but leave their ratios unchanged.  In Fig.~\ref{f:IBSQCentralityAlphaS} we show the RMS ellipticity $\varepsilon_2\{2\}$ and triangularity $\varepsilon_3\{2\}$ (left and right panels respectively) of the BSQ charge distributions and bulk energy density as a function of the centrality for different values of $\alpha_s$.  In both eccentricities we observe a fairly significant dependence on the values of the QCD coupling constant $\alpha_s$, leading to a flattening of the centrality dependence and a shift of the peak to higher centralities with increasing $\alpha_s$.  Both of these effects are a consequence of the increased probability to produce quarks of all flavors, resulting in $BSQ$ distributions which are overall less eccentric and closer to saturating the geometry of the bulk.  The rightward shift in the peak, whose location indicates the transition between a smooth geometry and a granular one, indicates that with larger $\alpha_s$ the onset of granularity of BSQ distributions has been pushed to higher centralities.  The triangularity $\varepsilon_3 \{2\}$, arising solely from fluctuations, is more sensitive to the changes in $\alpha_s$ compared with $\varepsilon_2 \{2\}$.  With $\varepsilon_3 \{2\}$ the convergence toward the bulk geometry is faster, and it again exhibits the same rightward shift of the peak.

%
\begin{figure}
    \begin{centering}
	\includegraphics[width=0.45 \textwidth]{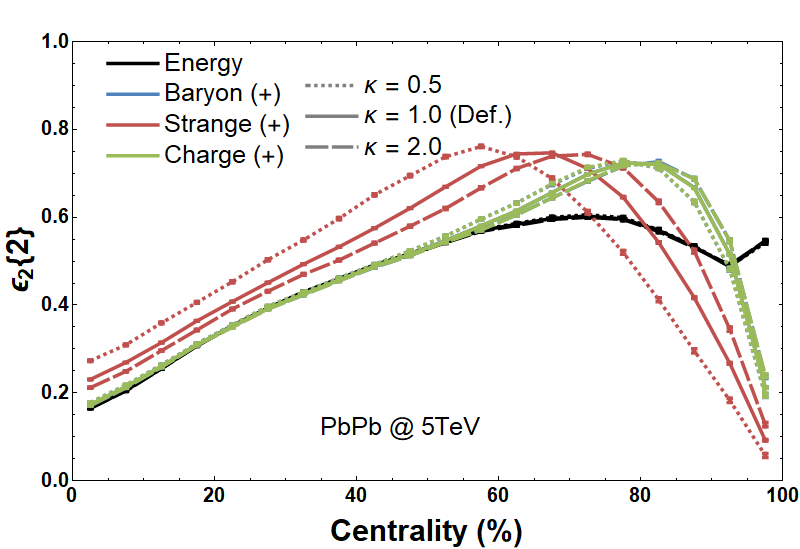} \,
	\includegraphics[width=0.45 \textwidth]{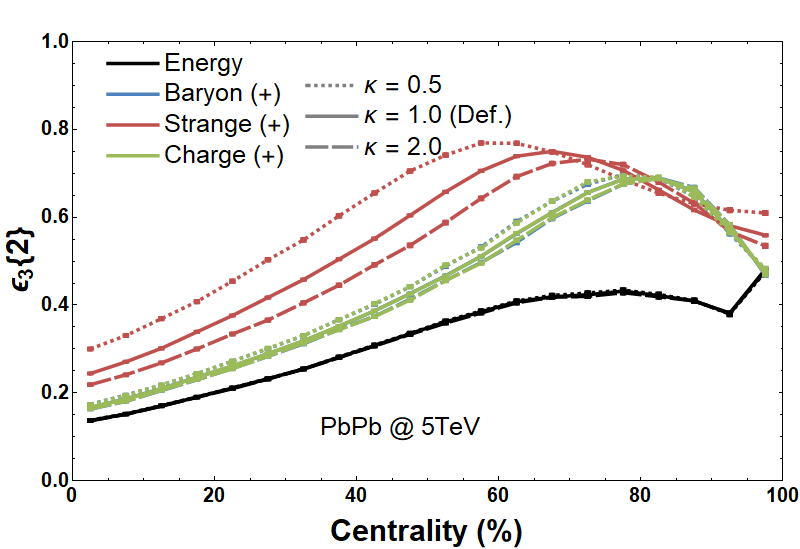}
	\end{centering}
	\caption{ Comparison of the second cumulants $\varepsilon_n \{2\}$ for different normalizations $\kappa$ of the saturation scale $Q_s$.}
	\label{f:IBSQCentralityQs}
\end{figure}
%

Another parameter which modifies the chemistry of the initial state is the scale-setting parameter $\kappa$ for the saturation scale (see Eq.~\eqref{e:kappadef}).  The value of $\kappa$ (and hence, $Q_s$) has little impact on the shape of the spatial correlation shown in Fig.~\ref{f:probabilities}, mostly affecting the quark chemistry as seen in Fig.~\ref{f:multratio}.  In Fig.~\ref{f:IBSQCentralityQs} we show the effect on $\varepsilon_2\{2\}$ and $\varepsilon_3\{2\}$ of varying $\kappa$ up and down by a factor of $2$ from its default value.  Both cumulants show a sizable dependence on $\kappa$, in a manner similar to the $\alpha_s$ dependence shown in Fig.~\ref{f:IBSQCentralityAlphaS}.  This indicates that the effect of varying $\kappa$ arises mostly through its impact on the total quark multiplicities for the different flavors.  However, there is an interesting difference from the $\alpha_s$ dependence seen previously: the change in the strangeness eccentricities is much larger than for baryon number and electric charge.  This is because an increase in $Q_s$ modifies the relative abundances of the quark flavors differently.  As seen in Fig.~\ref{f:multratio}, increasing the value of $Q_s$ from by a factor of 2 leads to a roughly equal increase in the total probabilities to produce all quark flavors.  But this constant increase in absolute terms corresponds to a greater \textit{percentage} increase in the abundance of strange quarks (~$+ 60\%$) versus up and down quark flavors (~$+ 15\%$).  This is due to the fact that the chemistry of the model we use here exhibits (exact or approximate) geometric scaling, depending only on the ratio $Q_s / m$ so that a change in $Q_s$ affects different flavors differently (see Appendix~\ref{app:Theory}).  Thus, when increasing $\kappa$ the geometry of strangeness gets significantly closer to the bulk, while the effect on the light flavors is much smaller.

%
\begin{figure}
    \begin{centering}
	\includegraphics[width=0.45 \textwidth]{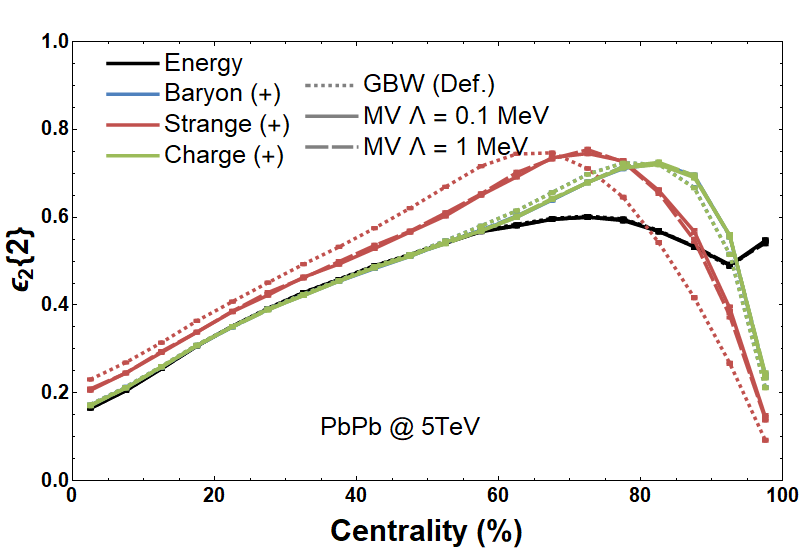} \,
	\includegraphics[width=0.45 \textwidth]{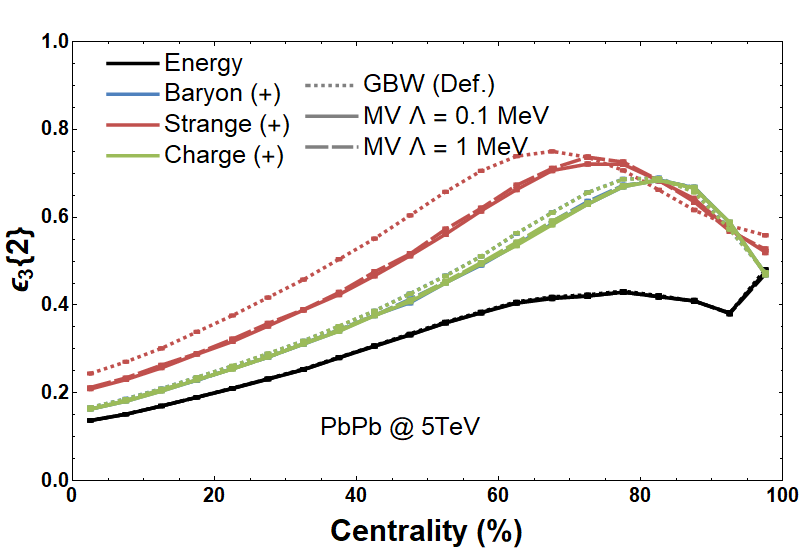}
	\end{centering}
	\caption{ Comparison of the second cumulants $\varepsilon_n \{2\}$ between the GBW and MV models, as well as for two different choices of the IR cutoff $\Lambda$ in the MV model.}
	\label{f:IBSQCentralityDipole}
\end{figure}
%
 
A third choice which influences the total quark multiplicities is the choice of GBW versus MV models (Eqs.~\eqref{e:multratio1_first} and \eqref{e:multratio2_first}, respectively).  As seen in Figs.~\ref{f:probabilities} and \ref{f:multratio}, the MV model has a higher probability to produce quarks at short distances, leading to a higher probability to produce quarks overall.  This results in the dependence seen in Fig.~\ref{f:IBSQCentralityDipole}: changing from the GBW to MV model leads to a flattening of the centrality dependence and a rightward shift of the eccentricity peaks.  As with the $\kappa$ dependence shown in Fig.~\ref{f:IBSQCentralityQs}, the roughly constant probability increase in absolute terms translates into a greater percentage increase for the strange quarks, leading to a larger shift in the strangeness distribution.  We also compare two different choices for the IR cutoff $\Lambda$ in the MV model, which has a negligible impact on the eccentricities. 

%
\begin{figure}
    \begin{centering}
	\includegraphics[width=0.45 \textwidth]{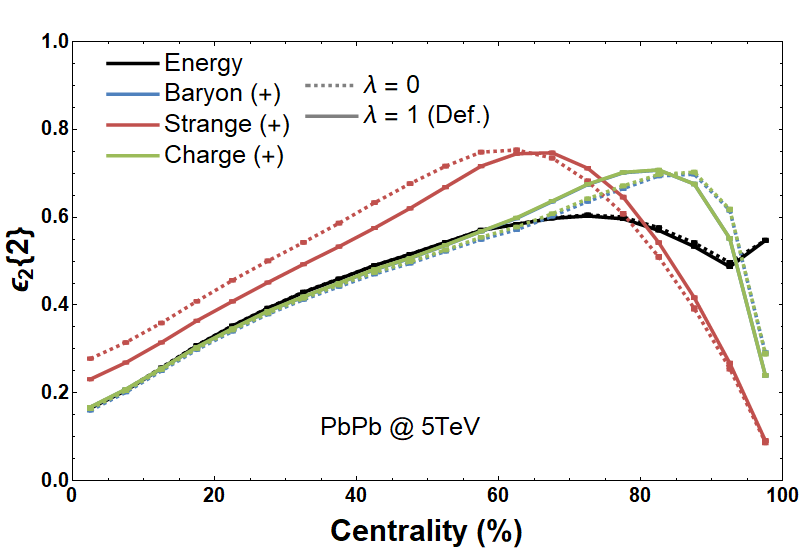} \,
	\includegraphics[width=0.45 \textwidth]{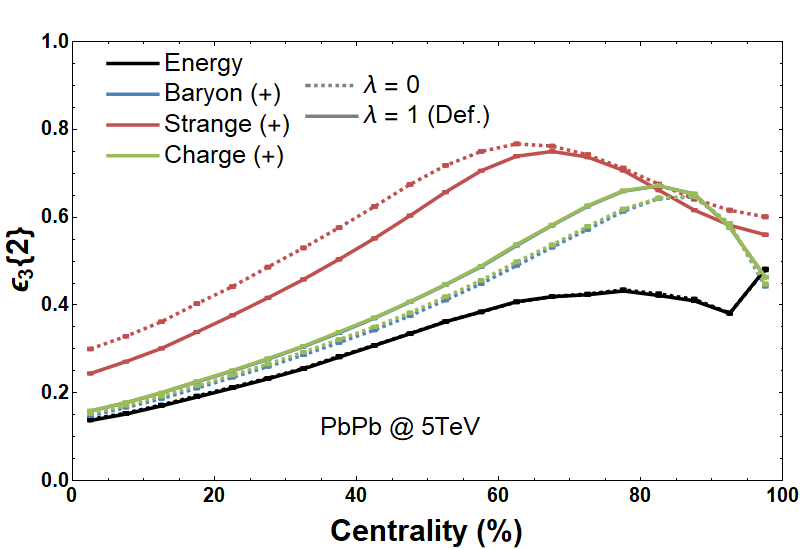}
	\end{centering}
	\caption{Comparison of the second cumulants $\varepsilon_n \{2\}$ for different distributions of gluon energy controlled by the parameter $\lambda$.}
	\label{f:IBSQCentralityLambda}
\end{figure}
%

The final ICCING parameter expected to have a significant impact on the initial-state chemistry is the exponent $\lambda$ from Eq.~\eqref{e:lambdaparam} which controls the energy dependence of the gluon spectrum.  As seen previously in Fig.~\ref{f:lambdaEffect}, changing from the boost-invariant spectrum $\lambda = 1$ to the constant spectrum $\lambda = 0$ leads to a significant \textit{decrease} in overall quark production due to apportioning the available energy into a smaller number of hard gluons.  We see in Fig.~\ref{f:IBSQCentralityLambda} that this change in quark multiplicities also leads to a corresponding change in the BSQ eccentricities.  The decrease in quark multiplicities in going from $\lambda = 1$ to $\lambda = 0$ results in an increase in the strangeness eccentricity and a leftward shift of its peak, as expected.  Interestingly, the shift in the baryon number and electric charge curves moves in the opposite direction, despite the fact that all quark multiplicities have decreased when going to $\lambda = 0$, as seen previously in Fig.~\ref{f:lambdaEffect}.

%
\begin{figure}
    \begin{centering}
	\includegraphics[width=0.45 \textwidth]{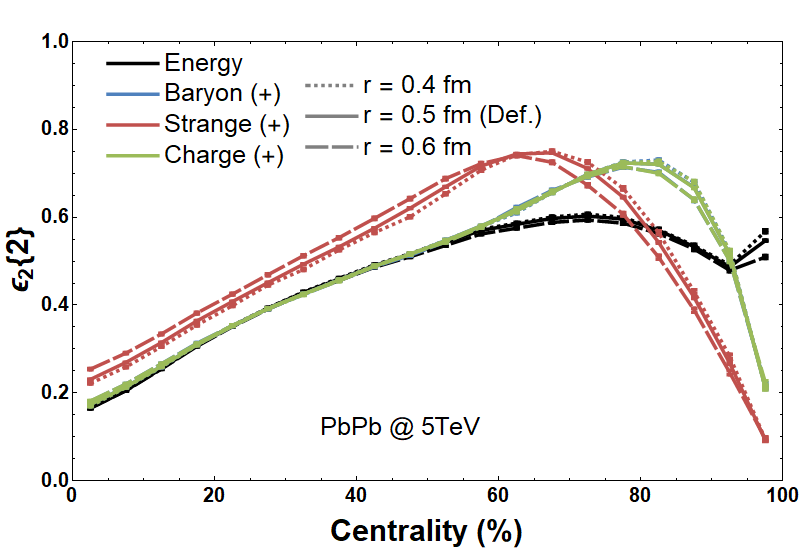} \,
	\includegraphics[width=0.45 \textwidth]{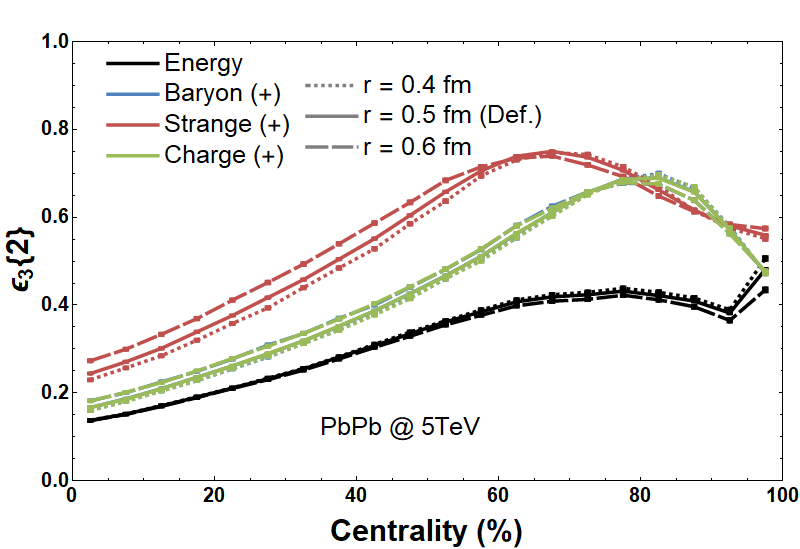}
	\end{centering}
	\caption{Comparison of the second cumulants $\varepsilon_n \{2\}$ for different radius parameters of the quarks and gluons.}
	\label{f:IBSQCentralityRad2}
\end{figure}
%

In studying the parameter dependence of ICCING's output, two other parameters are noteworthy.  The first is the radius $r$ of the quark and gluon blobs used in the redistribution algorithm described in Sec.~\ref{subsec:Redistribute}.  This parameter is perhaps the most \textit{ad hoc} choice in the model, but as seen in Fig.~\ref{f:IBSQCentralityRad2} it makes a relatively modest difference in the BSQ eccentricities.  These differences again reflect a small change in the quark multiplicities, in this case for a similar reason responsible for the $\lambda$ dependence of Fig.~\ref{f:IBSQCentralityLambda}.  If the circular blobs identified with gluons have a smaller radius, then it is possible to apportion the total available energy into more gluons, resulting in an increase in the number of quark pairs produced.

%
\begin{figure}
    \begin{centering}
	\includegraphics[width=0.45 \textwidth]{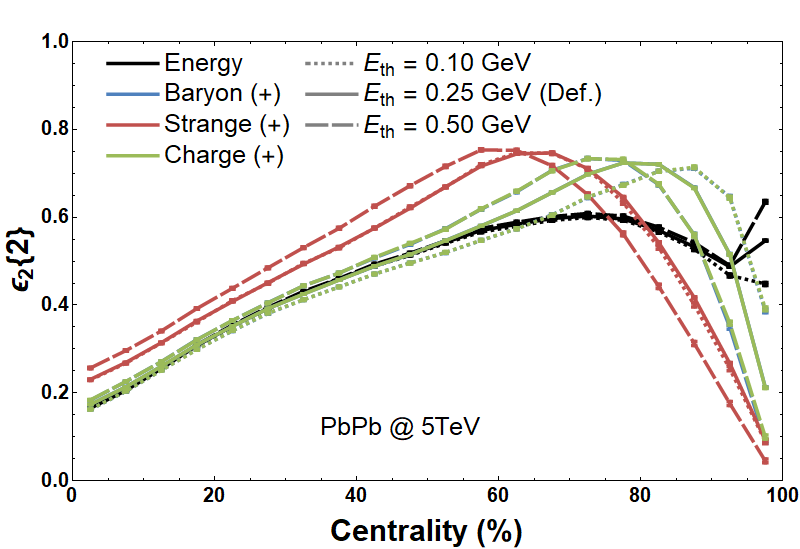} \,
	\includegraphics[width=0.45 \textwidth]{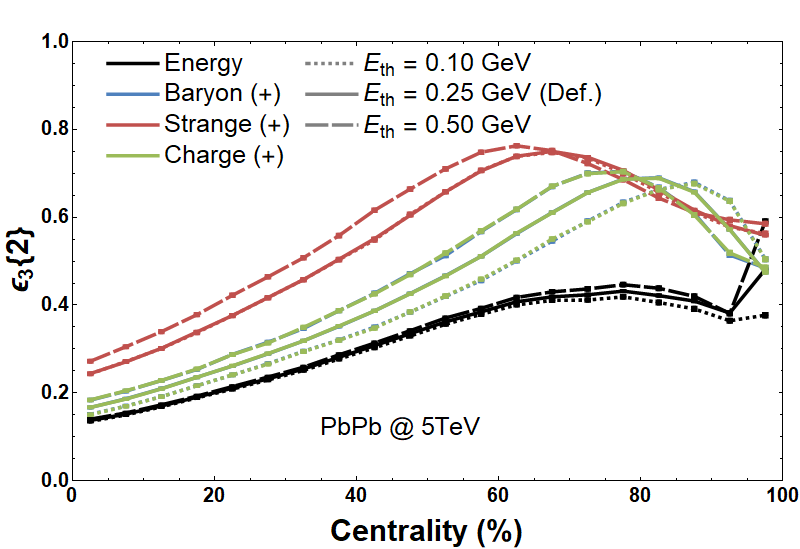}
	\end{centering}
	\caption{  Comparison of the second cumulants $\varepsilon_n \{2\}$ for different values of the threshold energy $E_{\mathrm{thresh}}$.}
	\label{f:IBSQCentralityEth}
\end{figure}
%

The last parameter with a significant impact on the output of ICCING is the threshold energy $E_\mathrm{thresh}$ determining when a gluon is eligible to undergo $g \rightarrow q \bar{q}$ splitting.  Because the algorithm of Sec.~\ref{subsec:Redistribute} deducts the input energy proportionately, the parameter $E_\mathrm{thresh}$ plays an important role in determining when the algorithm terminates.  It therefore is an important factor in the overall runtime of ICCING and should be chosen as large as possible without distorting the final results.  We see in Fig.~\ref{f:IBSQCentralityEth} that if $E_\mathrm{thresh}$ is chosen too high, it can artificially suppress the quark multiplicities and thereby impact the resulting eccentricities.  Consider, for instance, the impact on the strangeness distribution.  When $E_\mathrm{thresh}$ is below the $s \bar{s}$ mass threshold $2 m_s \approx 200 \, \mathrm{MeV}$, it has no effect on the strangeness distribution.  But the choice $E_\mathrm{thresh} = 500 \, \mathrm{MeV}$ discounts gluons which could have otherwise split into strange quarks, decreasing the strange quark multiplicity and shifting the strangeness eccentricity curve to the left.  For all three values considered, a similar effect is seen on the light quarks: the higher $E_\mathrm{thresh}$, the fewer up and down quarks which are produced.  The fact that a significant shift in the $B,Q$ eccentricities is still seen down to $E_\mathrm{thresh} = 100 \, \mathrm{MeV}$ indicates that this value is still too high; only if $E_\mathrm{thresh}$ is chosen below the up quark threshold $2 m_u \approx 5 \, \mathrm{MeV}$ can we be sure that it has not artificially prohibited the production of light quarks.  It is also interesting to note that varying $E_{thresh}$ has a significant effect on the sudden increase in the eccentricities of the bulk energy density in the most peripheral bin. This feature was noted previously in the context of Fig.~\ref{f:IBSQCentrality}, where it was associated with the accumulation of weight in bins corresponding to the production of only $1$ or $2$ $q \bar{q}$ pairs.  Here, as we decrease $E_\mathrm{thresh}$ from $0.5$ GeV to $0.1$ GeV, we see that the spike in the last bin decreases and disappears.  We interpret this as changing how ICCING resolves these most peripheral events: instead of tending to split a single blob of energy into one or two pairs, leading to the spike in eccentricity, with lower $E_\mathrm{thresh}$ ICCING is able to instead split it into several gluons and $q \bar{q}$ pairs.  As a result, the spike decreases and disappears as this discretization artifact of ICCING stops playing a dominant role in the last centrality bin.

In summary, we find that the greatest sensitivity of the results of ICCING is from parameters which affect the overall multiplicities of quark production.  These parameters include the coupling constant $\alpha_s$, saturation scale normalization factor $\kappa$, gluon energy parameter $\lambda$, and the choice of GBW versus MV models for the $g \rightarrow q \bar{q}$ splitting probabilities.  We find most of our default parameters to be reasonable starting points, with the exception of the splitting threshold energy $E_\mathrm{thresh}$ which should be set to less than $5 \, \mathrm{MeV}$ to avoid artificially suppressing light quark production.  Parameters which affect the shape of the quark/antiquark distribution (but not the overall chemistry), like $\alpha_\mathrm{min}$ and $d_\mathrm{max}$, have little effect on the eccentricities.  Finally, we note that the grooming parameter $S_\mathrm{chop}$ which was used in Fig.~\ref{f:Echop} to artificially groom an event to select on the hot spots should be carefully considered with respect to later hydrodynamic evolution.  For consistency, grid points below the freezeout temperature $T_{FO} \sim 150 \, \mathrm{MeV}$ should not be fed into the hydrodynamic equations of motion; in the equation of state we use here, this corresponds to $S_\mathrm{chop} \geq 2 \, \mathrm{fm}^{-3}$.

%
\section{Outlook and Conclusion}
\label{sec:Concl}
%

In this paper, we have introduced a new model which we denote ICCING -- Initial Conserved Charges in Nuclear Geometry -- which takes an arbitrary, externally-provided energy density profile $\epsilon(\vec{x}_\bot)$ and supplement it with new information about the distribution of conserved charges: baryon number $B$, strangeness $S$, and electric charge $Q$.  The primary result of this work is the distribution of BSQ charge eccentricities presented in Figs.~\ref{f:IBSQCentrality}-\ref{f:E3_IBSQ42Centrality100K}.  We find that for $<60\% $ centrality the eccentricities of $B$ and $Q$ closely track that of the bulk energy density, while the strangeness $S$ is significantly more eccentric.  We attribute this feature to the fact that the $B$ and $Q$ distributions are dominated by the production of $u$ and $d$ quarks, whose absolute mass thresholds $2m < 10 \, \mathrm{MeV}$ are negligible and are thus produced abundantly throughout the entire fireball geometry.  Strange quarks, on the other hand, having a nontrivial mass threshold $2m \approx 200 \, \mathrm{MeV}$, are produced less abundantly and only in regions which have sufficient energy to meet the threshold.  Both of these facts contribute to an increased eccentricity of the strangeness distribution across a wide range of centralities.  Although a large part of the enhanced strangeness eccentricities can be attributed to the smaller number of $s \bar{s}$ pairs being produced, we also see systematic dependencies that indicate that the strangeness distribution reflects a different underlying geometry than the bulk (see Figs.~\ref{f:IBSQ42Centrality100K} and \ref{f:Echop}).  Based on the nontrivial mass threshold of $s \bar{s}$ pair production, we attribute these differences to a coupling of strangeness to hot spots in the initial state.  Finally, while ICCING does not modify the energy eccentricities significantly in the 10 - 60\% centrality regime, it does produce a nontrivial enhancement in very central (0-10\%) and very peripheral (80-100\%) events due to the redistribution of energy in the $g \rightarrow q \bar{q}$ splitting (see Fig.~\ref{f:EccInOut}).

These results suggest that flow measurements of strange hadrons might provide new and different information than the bulk anisotropic flow even when there is a vanishing chemical potential. This effect could be further enhanced in hadrons containing multiple strange quarks, but estimating the flow of multi-strange hadrons from the initial state strays into murkier territory, which we leave to a future work.  A number of unresolved questions remain in the strange sector, for instance, the possibility of a flavor hierarchy \cite{Bellwied:2013cta, Noronha-Hostler:2016rpd, Bellwied:2018tkc, Bellwied:2019pxh} i.e. that strange particles freeze out at a higher temperature than light; mismatches when comparing particle spectra and flow harmonics \cite{Takeuchi:2015ana} at a fixed freeze-out temperature versus better fits with a higher strange freeze-out \cite{Devetak:2019lsk}; and difficulties capturing the strangeness multiplicities in small systems (although the core-corona approach appears promising \cite{Kanakubo:2019ogh}).  This work pushes us a step closer to a fully dynamical model that will be able to test the possibility of a flavor hierarchy. The next steps are to couple these initial conditions to BSQ hydrodynamics followed by a transport model. Since  the flavor hierarchy is predicted to be strongest at $\mu_B=0$, these initial conditions are precisely in the needed regime to test this hypothesis.

The algorithm and initial results presented here represent only the first layer of insights that can be gleaned from including the physics of $g \rightarrow q \barq$ splitting into the initial conditions of heavy ion collisions.  We can take advantage of the flexible implementation of ICCING by incorporating a range of theoretical inputs for the initial-state chemistry and spatial correlations, beyond the particular CGC-based model employed here.  In addition to incorporating more sophisticated CGC-based models, one could employ medium-modified splitting functions calculated in jet physics \cite{Sievert:2019cwq} or even take chemistry input from parton distribution functions as in Fig.~\ref{f:HERAPDF} and the spatial correlations from the vacuum light-front wave functions \cite{Gribov:1972ri, Altarelli:1977zs, Dokshitzer:1977sg}.  One can also incorporate more microscopic physics in the distribution of energy and charge around the partons in ICCING, for instance through the use of generalized parton distributions or Green's functions representing some pre-equilibrium evolution \cite{Kamata:2020mka}.  Future improvements may also be made to the algorithm itself, such as importance sampling when seeding the initial gluons.  Our framework can also be applied to a wealth of further charge-dependent observables, notably the charge correlations $\gamma_{112} , \gamma_{123}$ which are believed to be sensitive to the chiral magnetic effect \cite{Fukushima:2008xe, Kharzeev:2013ffa} and the charge balance functions \cite{Drijard:1979rz, Bass:2000az, Pratt:2018ebf}.  Additionally we can study various cross-correlations of charges with each other and with the bulk, as quantified by (e.g.) the symmetric cumulants.

In summary, the emergent physical picture from our new ICCING algorithm is that the distribution of different flavors and charges in the initial state resembles a rainbow cake, with different degrees of freedom forming distinct layers and geometries.  We have designed ICCING in a model agnostic way to maximize its utility and flexibility for the community; as such, the source code will be publicly released at the address \href{https://github.com/pcarzon/ICCING}{github.com/pcarzon/ICCING} upon publication.  It is our hope that this approach can provide a powerful new tool to the community which, when coupled with next-generation BSQ hydrodynamics, can be used to study the physics of charge transport and dissipation even at top collider energies.

%
\acknowledgements{The authors wish to thank G. Denicol, A. Dumitru, J. Jalilian-Marian, Y. Kovchegov, M. Luzum, J. Nagle, P. Romatschke, T. Schaefer, V. Skokov, P. Sorensen, and P. Tribedy for useful discussions.  The authors thank the anonymous referee for their excellent feedback which helped us to refine our original approach. M.M. acknowledges support from the US-DOE Nuclear Science Grant No. DE-FG02-03ER41260 and the BEST (Beam Energy Scan Theory) DOE Topical Collaboration.  M.D.S. acknowledges support from a startup grant from New Mexico State University.  D.E.W. acknowledges previous support from the Zuckerman STEM Leadership Program.  J.N.H. acknowledges the support from the US-DOE Nuclear Science Grant No. DE-SC0020633, and from the Illinois Campus Cluster, a computing resource that is operated by the Illinois Campus Cluster Program (ICCP) in conjunction with the National Center for Supercomputing Applications (NCSA), and which is supported by funds from the University of Illinois at Urbana-Champaign.}
%

\appendix

%
\section{Theoretical Framework}
\label{app:Theory}
%

Throughout this paper, we denote transverse vectors as $\tvec{v} = (v_x \, v_y)$ with magnitudes $v_\bot$.  We will also make use of the light-front momentum $p^+ \equiv \frac{1}{\sqrt{2}} (E + p_z)$ which we will largely consider to be synonymous with a particle's energy for an ultrarelativistic particle $\frac{p^+}{\sqrt{2}} = p_z = E$ moving primarily along the $z$ axis.  

%
\subsection{Sea Quark Multiplicity}
%

%
In Ref.~\cite{Martinez:2018ygo}, some of us computed the multiplicity of $q \barq$ pairs produced in the (semi)dilute / dense regime of the color glass condensate effective field theory~( see Ref.~\cite{Kovchegov:2012mbw} and references therein).  In Eq.~(68) of Ref.~\cite{Martinez:2018ygo}, we determined that the cross-section to produce $q \barq$ pairs with the quark (antiquark) at transverse position $\vec{B}_{1\bot} (\vec{B}_{2\bot})$ and with light-front momentum $k_1^+ (k_2^+)$, respectively, is given by
\begin{align} \label{e:quark0}
k_1^+ k_2^+ \frac{d\sigma^{q \bar q}}{d^2 B_{1\bot} dk_1^+ \, d^2 B_{2 \bot} dk_2^+} &=
\left( \frac{a \, \alpha_s^2 N_c}{4\pi^3} \ln\frac{1}{\Lambda} \right) \alpha (1-\alpha) m^2 \,
\left( 1 - e^{-\frac{1}{4} [\alpha^2 + (1-\alpha)^2] r_\bot^2 Q_s^2 (\vec{u}_\bot)} \right)
\notag \\ &\times
\left[ \left( \alpha^2 + (1-\alpha)^2 \right) K_1^2 (m r_\bot) + K_0^2 (m r_\bot) \right].
\end{align}
Here $\alpha = \frac{k_1^+}{k_1^+ + k_2^+}$ is the fraction of the $q \barq$ light-front momentum carried by the quark, $N_c$ is the number of quark colors, $m$ is the quark mass, and $\Lambda$ is an infrared cutoff.\footnote{
We note that the argument $\tfrac{1}{\Lambda}$ of the logarithm is dimensionful as written.   This is because the logarithm was obtained in a leading-logarithmic approximation with $\Lambda \rightarrow 0$, and the difference between different choices of the accompanying scale are beyond the precision of the approximation.  Rather than introduce another arbitrary scale to make the logarithm dimensionless, we simply write $\ln\tfrac{1}{\Lambda}$.  Ultimately this logarithm will cancel out in the ratios of interest to us.
}
The cross section \eqref{e:quark0} depends primarily on the quark/antiquark separation vector $\vec{r}_\bot \equiv \vec{B}_{1\bot} - \vec{B}_{2\bot}$, with the center-of-momentum coordinate $\vec{u}_\bot \equiv \alpha \vec{B}_{1\bot} + (1-\alpha) \vec{B}_{2\bot}$ entering only through the indirect dependence on the saturation scale $Q_s (\vec{u})$.

The expression \eqref{e:quark0} has been averaged over events in two ways.  The first is by averaging the scattering operators (Wilson lines) over the color configurations of both projectile and target.  The second is by averaging the overall impact parameter $\vec{B}_\bot$ between the colliding nuclei.  In this work, we wish to use the corresponding expression {\it{without}} averaging over all events, in order to account for fluctuating initial conditions.  We will do this by relaxing the averaging over impact parameter, so that the event-by-event fluctuations of collision geometry can be accounted for.  We will retain, however, the averaging over color configurations which in Eq.~\eqref{e:quark0} results in the factor $(1 - \exp(-\tfrac{1}{4} [\alpha^2 + (1-\alpha)^2] r_\bot^2 Q_s^2 (\vec{u})))$.  This choice results in the $q \barq$ pair production rates being a function of the collision geometry only, with the color field fluctuations being encoded entirely in the saturation scale $Q_s$ at a given transverse position.  In principle one can also account for event-by-event fluctuations of the local color fields as well by sampling color configurations as done in, e.g., Ref.~\cite{Mace:2018vwq}, but this extension is beyond the scope of the current framework and we leave it for future work.

The averaging over impact parameters is relaxed as follows.  The factor of $a$ (the mass number of the light projectile) in \eqref{e:quark0} arises from integrating the thickness function $T_a$ of the projectile over all impact parameters $\vec{B}_\bot$ (see Eq.~(17) of Ref.~\cite{Martinez:2018ygo}) in a ``valence quark model'' where each nucleon of the projectile is treated as a single quark.  Instead of integrating over $\vec{B}_\bot$, we keep it fixed in a given event so that the factor of $a$ is replaced by the thickness function $T_a$, and we can generalize beyond the valence quark model by converting from the thickness function to the momentum scale $\mu$ characterizing the width of color field fluctuations:
\begin{align} \label{e:mudict}
a \rightarrow T_a \rightarrow \frac{N_c}{2 \pi \alpha_s} \mu^2 .
\end{align}
(See also Eq. (A4) of Ref.~\cite{Martinez:2018tuf}).  We also further manipulate Eq.~\eqref{e:quark0} by converting from the cross section to the multiplicity and by making the change of variables
\begin{align}   \label{e:quark1prior}
\frac{dn^{q \barq}}{d^2 r_\bot \: d^2 u_\bot} &=
\frac{1}{\sigma_{\mathrm{inel}}} \int \frac{dk_1^+}{k_1^+} \frac{dk_2^+}{k_2^+}
\left[ k_1^+ k_2^+ \frac{d\sigma^{q \bar q}}{d^2 B_{1\bot} dk_1^+ \, d^2 B_{2\bot} dk_2^+} \right]
\notag \\ &=
\frac{1}{\sigma_{\mathrm{inel}}}
\int \frac{dq^+}{q^+} \int_0^1 \frac{d\alpha}{\alpha(1-\alpha)}
\left[ k_1^+ k_2^+ \frac{d\sigma^{q \bar q}}{d^2 B_{1\bot} dk_1^+ \, d^2 B_{2\bot} dk_2^+} \right] ,
\end{align}
where $\sigma_\mathrm{inel}$ is the total inelastic cross section, $q^+ = k_1^+ + k_2^+$ is the total light-front momentum of the $q \bar{q}$ pair (also equal to the light-front momentum of the gluon which produced them), and the Jacobian for the change of variables $d^2 B_{1\bot} \, d^2 B_{2\bot} = d^2 r_\bot \, d^2 u_\bot$ is $1$.

The distribution \eqref{e:quark1prior} is proportional to the longitudinal integral $\int\frac{dq^+}{q^+}$, reflecting its invariance under boosts along the $z$-axis.  If this were a distribution in momentum space, we could introduce the usual rapidity variable
\begin{align} \label{e:rap}
    y \equiv \frac{1}{2} \ln\frac{q^+}{q^-} = \ln\frac{\sqrt{2} q^+}{\sqrt{q_\bot^2 + m^2}}
    = \ln\frac{\sqrt{q_\bot^2 + m^2}}{\sqrt{2} q^-} ,
\end{align}
which, for fixed transverse momentum $\tvec{q}$, leads to $dy = \frac{dq^+}{q^+}$.  Then, moving the rapidity infinitesimal $dy$ to the left-hand side of \eqref{e:quark1prior}, we would conclude that the distribution is independent of rapidity $y$ and therefore boost invariant.  The situation in \eqref{e:quark1prior} is more subtle, however, because we have already integrated over the transverse momentum $\tvec{q}$ in order to perform the Fourier transform to fixed transverse position $\tvec{u}$.  This Fourier transform over $\tvec{q}$ at fixed $q^+$ therefore integrates over the rapidity \eqref{e:rap} through its dependence on $q_\bot$.  Nevertheless, the light-front momentum $q^+$ is held fixed during the Fourier transform, and the boost invariance of the distribution \eqref{e:quark1prior} is still present in the longitudinal integral $\int\frac{dq^+}{q^+}$.  We can make the boost invariance explicit, even in coordinate space, by defining a modified or ``quasi-rapidity''
\begin{align}   \label{e:rap2}
    \tilde{y} \equiv \ln\frac{{\sqrt{2}} q^+}{m}
\end{align}
with $m$ the particle mass.  We could use any fixed scale in the denominator, not just the mass (for instance on could use $\Lambda_{QCD}$ for a massless gluon).  Although this does not correspond to ``rapidity'' in the usual sense, it plays the same role in describing a boost-invariant distribution: $\frac{dq^+}{q^+} = dy = d\tilde{y}$.  Therefore we will describe the longitudinal distribution using the light-front momenta $\frac{dq^+}{q^+} \equiv d\tilde{y}$ using the language of ``rapidity,'' even though the expressions are in coordinate space.  The two definitions \eqref{e:rap} and \eqref{e:rap2} coincide in that they both transform additively under boosts, so that independence of either $y$ or $\tilde{y}$ reflects invariance under boosts.  The definitions differ in their assignment of the ``zero rapidity direction''; the true rapidity $y$ is zero for a particle whose velocity along the $z$-axis is zero (but which may be moving in the transverse plane), while the quasi-rapidity $\tilde{y}$ is zero for a particle at rest.  With this caveat, we change variables to the multiplicity per unit ``quasi-rapidity'' $d\tilde{y}$, giving
\begin{align} \label{e:quark1}
\frac{dn^{q \barq}}{d^2 r_\bot \: d^2 u_\bot \: d\tilde{y}} &=
\frac{\alpha_s N_c^2}{8 \pi^4}
\frac{\mu^2(\vec{u}_\bot) m^2 }{\sigma_{inel}} \ln\frac{1}{\Lambda}
\int_0^1 d\alpha \,
\Big[ 1 - \exp\Big( - \tfrac{1}{4} [\alpha^2 + (1-\alpha)^2] \, r_\bot^2 Q_s^2 (\vec{u}_\bot) \Big) \Big]
\notag \\ &\hspace{1cm}\times
\Big[ \big(\alpha^2 + (1-\alpha)^2 \big) K_1^2 ( m r_\bot) + K_0^2 (m r_\bot) \Big] .
\end{align}

Eqs.~\eqref{e:quark0} and \eqref{e:quark1} were obtained under a series of approximations.  The full structure of the color-averaged scattering operators, shown in Eq.~(37) of Ref.~\cite{Martinez:2018ygo}, has been simplified using the large-$N_c$ approximation.  This expresses the cross section \eqref{e:quark0} in terms of the fundamental dipole amplitude $D_2 (x,y)$ describing the high-energy scattering of a quark at position $\vec{x}_\bot$ and an antiquark at position $\vec{y}_\bot$.  This dipole amplitude, in turn, has been evaluated using the Golec-Biernat-Wusthoff (GBW) model \cite{GolecBiernat:1998js}: $D_2 (\tvec{x}, \tvec{y}) = \exp(-\tfrac{1}{4} |\tvec{x}-\tvec{y}|^2 Q_s^2)$ which introduces dependence on the target saturation scale $Q_s$.  The saturation scale $Q_s$ and the analogous scale $\mu$ in the projectile are both dependent on transverse position through their respective nuclear thickness functions.  In Eq.~\eqref{e:quark1} we have evaluated both of these scales at the $q \barq$ transverse center of momentum $\vec{u}_\bot$; in doing so, we have neglected small perturbative shifts in the local densities, corresponding to a gradient expansion of the thickness functions of the projectile and target.  Finally, we note that the factor $\ln\frac{1}{\Lambda}$ appearing in \eqref{e:quark0} arises from a leading-logarithmic approximation to the integral over the positions of the projectile color sources $\mu$ (see Eqs.~(67-68) of Ref.~\cite{Martinez:2018ygo}).  These approximations are the simplest nontrivial ones which can be made to evaluate the $q \barq$ multiplicity \eqref{e:quark1}; all of them can be relaxed in future work, and in this paper we explore in greater detail the model dependence of the dipole amplitude $D_2 (\tvec{x}, \tvec{y})$ by comparing the GBW model with the more complete McLerran-Venugopalan (MV) model \cite{McLerran:1993ni, McLerran:1993ka, McLerran:1994vd} in Secs.~\ref{subsec:MV} and \ref{sec:Results}.

%
\subsection{Gluon Multiplicity and Sea Quark Multiplicity Ratio}
%

We compare the $q \barq$ multiplicity from Eq.~\eqref{e:quark1} with the analogous inclusive gluon multiplicity calculated under the same approximations.  The momentum-space cross section for gluon production in the CGC framework is textbook material \cite{Kovchegov:2012mbw}, but our use of it here is unorthodox in that we need to employ it in coordinate space.  As such, we will briefly repeat the derivation in the format we need here.  In coordinate space, the amplitude\footnote{Here $A$ is a scaled amplitude, related to the usual scattering amplitude $M$ by $A = M / 2 s$ with $s$ the center-of-mass energy squared.} to produce a low-$x$ gluon of color $a$ and spin $\lambda$ at transverse position $\vec{u}_\bot$ and light-front momentum $q^+$ is
\begin{align}
A^a_\lambda (\vec{u}_\bot , x) = \frac{i}{\pi} \int d^2 b_\bot \, \rho^b (\vec{b}_\bot)
\, \frac{\vec{\epsilon}_\lambda^* \cdot (\vec{u}_\bot - \vec{b}_\bot) }{\left| \tvec{u} - \tvec{b} \right|^2}
\, \left[ U^{a b}_{\vec{u}_\bot} - U^{a b}_{\vec{b}_\bot} , \right] ,
\end{align}
where $U^{a b}_{\vec{u}_\bot}$ is a Wilson line \cite{Kovchegov:2012mbw} in the adjoint representation and $\rho$ is a classical source charge.  Squaring the amplitude, we sum over the gluon spins using $\sum_\lambda (\epsilon_\lambda^*)_\bot^i (\epsilon_\lambda)_\bot^j = \delta^{i j}$ and average over the source charges using (c.f. Eq.~\eqref{e:mudict})
\begin{align}
\langle \rho^a (\vec{x}_\bot) \, \rho^{b \, *} (\vec{y}_\bot) \rangle =
\delta^{a b} \, \delta^2 (\vec{x}_\bot - \vec{y}_\bot) \, \mu^2 (\vec{x}_\bot) ,
\end{align}
we obtain the cross section
\begin{align}   \label{e:gluon_boost}
\frac{d\sigma^G}{d^2 u_\bot  \, dq^+} = \frac{N_c^2 - 1}{2\pi^3 q^+}  \int d^2 b_\bot \,
\frac{ \mu^2 (\vec{b}_\bot) }{\left| \tvec{u} - \tvec{b} \right|^2} \,
\left[ 1 - D_2^{adj} (\vec{u}_\bot, \vec{b}_\bot) \right] .
\end{align}
Here the the global impact parameter $\vec{B}_\bot$ determining the geometry of the event has been held constant and is not explicitly denoted, and $D_2^{adj} (\vec{u}_\bot, \vec{b}_\bot) \equiv \frac{1}{N_c^2 - 1} \langle \tr[ U_{\vec{u}_\bot} U_{\vec{b}_\bot}^\dagger ] \rangle$ is the high-energy dipole scattering amplitude in the adjoint representation.

For an apples-to-apples comparison with Eq.~\eqref{e:quark1}, we make the same approximations.  We take the large-$N_c$ limit, for which $D_2^{adj} = |D_2|^2$, and we employ the GBW model $D_2 (\vec{u}_\bot, \vec{b}_\bot) = \exp[-\tfrac{1}{4} |\tvec{u}-\tvec{b}|^2 \, Q_s^2]$.  We also change variables from light-cone momentum $q^+$ to ``quasi-rapidity'' $d\tilde{y} = \frac{dq^+}{q^+}$, and we evaluate both $\mu^2$ and $Q_s^2$ at the external position $\vec{u}_\bot$ up to gradient corrections.  This gives the gluon multiplicity as
\begin{align} \label{e:gluon0}
\frac{dn^G}{d^2 u_\bot \: d\tilde{y}} &= \frac{N_c^2}{2\pi^3 \sigma_{\mathrm{inel}}} \mu^2 (\vec{u}_\bot) \int d^2 b_\bot \,
\frac{1}{\left| \tvec{u} - \tvec{b} \right|^2} \, 
\left[ 1 - \exp\left(- \thalf |\tvec{u}-\tvec{b}|^2 Q_s^2 (\vec{u}_\bot) \right) \right]
\notag \\ &\hspace{2cm}\times
\theta(1 - | \tvec{u} - \tvec{b} | \Lambda)
\notag \\ &=
\frac{N_c^2}{\pi^2 \sigma_{\mathrm{inel}}} \mu^2 (\vec{u}_\bot) \,\int\limits_0^{1/\Lambda} \frac{ d w_\bot}{w_\bot} \, \left[ 1 - \exp\left(- \thalf w_\bot^2 Q_s^2 (\vec{u}_\bot) \right) \right].
\end{align}
Note that the integral over $d^2 b$ is convergent in the UV due to the saturation exponential (color transparency), but logarithmically divergent in the IR.  For this reason it must be regulated by the explicit cutoff $\Lambda$.  Keeping only the log-divergent part of the $d^2 w_\bot$ integral (that is, employing the leading-logarithmic approximation) we obtain
\begin{align} \label{e:gluon1}
\frac{dn^G}{d^2 u_\bot \: d\tilde{y}} = \frac{N_c^2}{\pi^2 \sigma_{\mathrm{inel}}} \, \mu^2 (\vec{u}_\bot) \, \ln\frac{1}{\Lambda} .
\end{align}

By taking the ratio of the $q \barq$ multiplicity \eqref{e:quark1} to the gluon multiplicity \eqref{e:gluon1} at the same position, we can compare the number of quark pairs versus gluons produced in a given event:
\begin{align} \label{e:multratio0}
\frac{n^{q \barq} (\vec{u}_\bot, \vec{r}_\bot, \alpha)}{n^G (\vec{u}_\bot)} &=
\frac{\alpha_s}{8 \pi^2}  m^2
\int d^2 r_\bot \int\limits_0^1 d\alpha \,
\Big[ 1 - \exp\Big( - \tfrac{1}{4} [\alpha^2 + (1-\alpha)^2] \, r_\bot^2 Q_s^2 (\vec{u}_\bot) \Big) \Big]
\notag \\ &\hspace{1cm} \times
\Big[ \big(\alpha^2 + (1-\alpha)^2 \big) K_1^2 ( m r_\bot) + K_0^2 (m r_\bot) \Big] ,
\end{align}
where we have used the shorthand $n(\vec{u}_\bot) \equiv \frac{dn}{d^2 u_\bot \: d\tilde{y}}$ for the quark and gluon densities.  The multiplicities are also independent of the ``quasi-rapidity'' $\tilde{y}$, which we omit from here on; this is a consequence of the high-energy asymptotics which are boost-invariant at the semi-classical level.  We note that Eq.~\eqref{e:multratio0} is exactly the same as the correlation function $\mathcal{C}$ calculated in Eq. (75) of Ref.~\cite{Martinez:2018ygo}, except that now the saturation scale $Q_s^2$ is evaluated locally at the position $\vec{u}_\bot$.  Note also that the linear dependence of both Eqs.~\eqref{e:quark1} and \eqref{e:gluon1} on $\mu^2$ has dropped out entirely due to the (semi-)dilute / dense approximation used.

Interpreting the total $q \barq$ to gluon multiplicity ratio as the probability for an individual splitting, we can express the differential probability density to split into a $q \barq$ pair separated by transverse distance $r_\bot$ and with momentum fraction $\alpha$ carried by the quark as
\begin{align} \label{e:prob1}
\frac{dP}{dr_\bot d\alpha} &= 2\pi r_\bot \: \frac{n^{q \barq} (\vec{u}_\bot, \vec{r}_\bot, \alpha)}{n^G (\vec{u}_\bot)}
\notag \\ &=
\frac{\alpha_s}{4 \pi}  m^2 r_\bot
\Big[ 1 - \exp\Big( - \tfrac{1}{4} [\alpha^2 + (1-\alpha)^2] \, r_\bot^2 Q_s^2 (\vec{u}_\bot) \Big) \Big]
\notag \\ &\hspace{1cm} \times
\Big[ \big(\alpha^2 + (1-\alpha)^2 \big) K_1^2 ( m r_\bot) + K_0^2 (m r_\bot) \Big] .
\end{align}
It is also interesting to observe that the total $q \barq$ splitting probabilities exhibit a kind of geometric scaling, depending only on the ratio of scales $Q_s / m$.  This scaling can be made explicit through the change of variables $\zeta \equiv m r_\bot$, giving
\begin{align} \label{e:multratio1}
\frac{n^{q \barq} (\vec{u}_\bot)}{n^G (\vec{u}_\bot)} &=
\frac{\alpha_s}{4 \pi}
\int\limits_0^1 d\alpha \, \int\limits_0^\infty d\zeta \, \zeta \,
\Big[ 1 - \exp\Big( - \tfrac{1}{4} [\alpha^2 + (1-\alpha)^2] \, \tfrac{Q_s^2 (\vec{u}_\bot)}{m^2} \, \zeta^2 \Big) \Big]
\notag \\ &\hspace{1cm} \times
\Big[ \big(\alpha^2 + (1-\alpha)^2 \big) K_1^2 (\zeta) + K_0^2 (\zeta) \Big] .
\end{align}
Eqs.~\eqref{e:prob1} and \eqref{e:multratio1} are plotted in Figs.~\ref{f:probabilities} and \ref{f:multratio} and further discussed in Sec.~\ref{subsec:MV}.

Let us emphasize that the calculations presented here are not evaluations of the real-time dynamics of the quarks and gluons in the early stages of a heavy-ion collision \cite{Gelis:2004jp, Gelfand:2016prm, Tanji:2017xiw, Tanji:2017suk}.  Rather, the multiplicities computed here correspond to quarks and gluons produced as asymptotic ``out'' states in scattering amplitudes.  While we will apply these spatial correlations and multiplicity ratios to the initial conditions of a heavy-ion collision on a fixed-proper-time hypersurface, this is a model assumption which can be further explored in future work.  For our present purposes, we use these quantities to effectively fix the chemistry and correlations arising in the initial state, before subsequent modification by the strong final-state hydrodynamic evolution in the quark-gluon plasma.

%
\subsection{MV Model and Geometric Scaling}
\label{subsec:MV}
%

The quark/antiquark \eqref{e:quark1} and gluon \eqref{e:gluon0} multiplicities above have been computed using the GBW model for the dipole scattering amplitude.  This Gaussian model correctly captures the nonlinear effects in the deep saturation regime at large dipole sizes, but it misses the transition to power-law behavior in the dilute regime of small dipoles.  This latter feature is more properly captured by the McLerran-Venugopalan (MV) model \cite{McLerran:1993ni, McLerran:1993ka, McLerran:1994vd}, which modifies the Gaussian exponent to include a logarithm:
\begin{align} \label{e:dipoles}
D_2 (\tvec{x}, \tvec{y}) =
    \begin{cases}
    \exp\left(-\tfrac{1}{4} |\tvec{x}-\tvec{y}|^2 Q_s^2\right) & \mathrm{GBW} \\
    \exp\left(-\tfrac{1}{4} |\tvec{x}-\tvec{y}|^2 Q_s^2 \ln\tfrac{1}{|\tvec{x}-\tvec{y}| \Lambda} \right) & \mathrm{MV}
    \end{cases}
\end{align}
with $\Lambda$ an infrared cutoff in the MV model.  Using the MV model dipole amplitude in the $q \barq$ multiplicity \eqref{e:quark1} modifies the saturation exponent accordingly; the same is true of the gluon multiplicity \eqref{e:gluon0} initially, but since it is dominated by the IR this does not affect the expression \eqref{e:gluon1} at leading-logarithmic accuracy.  As such, we can immediately write down the differential and integrated splitting probabilities as
\begin{align} \label{e:prob2}
\left.\frac{dP}{dr_\bot d\alpha}\right|_{MV} &=
\frac{\alpha_s}{4 \pi}  m^2 r_\bot
\Big[ 1 - \exp\Big(
- \tfrac{1}{4} \alpha^2 \, r_\bot^2 Q_s^2 \ln\tfrac{1}{\alpha r_\bot \Lambda}
- \tfrac{1}{4} (1-\alpha)^2 \, r_\bot^2 Q_s^2 \ln\tfrac{1}{(1-\alpha) r_\bot \Lambda}
\Big) \Big]
\notag \\ &\hspace{1cm} \times
\Big[ \big(\alpha^2 + (1-\alpha)^2 \big) K_1^2 ( m r_\bot) + K_0^2 (m r_\bot) \Big]
\end{align}
and
\begin{align} \label{e:multratio2}
\left.\frac{n^{q \barq}}{n^G}\right|_{MV} &=
\frac{\alpha_s}{4 \pi}
\int\limits_0^1 d\alpha \, \int\limits_0^\infty d\zeta \, \zeta \,
\Big[ 1 - \exp\Big(
- \tfrac{1}{4} \alpha^2 \, \zeta^2 \tfrac{Q_s^2}{m^2}  \,  \ln\tfrac{1}{\alpha \zeta \Lambda/m}
- \tfrac{1}{4} (1-\alpha)^2 \, \zeta^2 \tfrac{Q_s^2}{m^2} \,  \ln\tfrac{1}{(1-\alpha) \zeta \Lambda/m}
\Big) \Big]
\notag \\ &\hspace{1cm} \times
\Big[ \big(\alpha^2 + (1-\alpha)^2 \big) K_1^2 (\zeta) + K_0^2 (\zeta) \Big] .
\end{align}

We note that there are other phenomenologically relevant parameterizations of the dipole amplitude which can be considered, in particular the AAMQS fit \cite{Albacete:2010sy} which allows for flexibility in the exponent of $(|\tvec{x}-\tvec{y}| Q_s)$ and in the argument of the logarithm.  We find, however, that the differences between the optimal values of these parameters within AAMQS and the standard MV model is far smaller than the difference between MV and GBW.  We thus conclude that the most important effect for us to consider in the dipole model is the treatment of the short-distance UV region, which the MV model handles differently than the GBW model.  For this reason, we restrict our discussion in this paper to the GBW and MV models.

The differential splitting probabilities for the GBW \eqref{e:prob1} and MV \eqref{e:prob2} models are shown in Fig.~\ref{f:probabilities} as a function of the $q \barq$ distance.  Both models exhibit the same IR behavior for large dipole sizes, but differ for small dipoles in the UV.  While the dipole amplitude \eqref{e:dipoles} in the  GBW model dies off strongly as a Gaussian at short distances, the MV model transitions to a much milder power-law behavior.  As a result, the MV model leads to an enhancement of quark production at short distances relative to the GBW model.  We also see that the quark mass $m$ leads to modest changes in the splitting function, as reflected in the differences between the curves for the up and strange quarks.

The integrated quark/gluon multiplicity ratio is shown in Fig.~\ref{f:multratio} for both the GBW \eqref{e:multratio1} and MV \eqref{e:multratio2} models for up, down, strange, and charm quarks.  In all cases, the quark production rates increase with increasing $Q_s$, with lighter quarks being produced more abundantly for a given $Q_s$ than heavy quarks.  As pointed out above, the integrated ratio \eqref{e:multratio1} in the GBW approximation exhibits exact geometric scaling, depending only on the ratio $Q_s/m$.  As such, the four curves in Fig.~\ref{f:multratio} all collapse into a single universal curve when scaled by the mass.  Because of this scaling property, we note that the production rates for up and down quarks are noticeably different.  This is because, although the up $(2.3~\mathrm{MeV})$ and down $(4.8~\mathrm{MeV})$ quark masses are so light as to be practically massless in absolute scales, they differ by a factor of $2$.  Consequently, up and down quarks ``see'' an effective saturation scale $Q_s$ which differs by a factor of $2$, leading to a $\sim 20\%$ difference in their abundances at the same $Q_s$.  In the MV model \eqref{e:multratio2}, the introduction of a dimensionful cutoff scale $\Lambda$ breaks explicit geometric scaling, albeit only logarithmically.  As seen explicitly in \eqref{e:multratio2}, if one (artificially) fixes the ratio of the MV cutoff $\Lambda / m$ compared to the quark masses, then the integrated chemistry again possesses exact geometric scaling.  At fixed cutoff $\Lambda$ in absolute terms, this leads to a quantitatively small breaking of geometric scaling since different quark masses ``see'' different values of the scaled cutoff $\Lambda / m$.

%
\section{Eccentricities, Cumulants, and Anisotropic Flow}
\label{app:Eccentricities}
%

%
\subsection{Standard Definitions of the Eccentricities}
%

The standard definition of the complex eccentricity vector $\bm{\mathcal{E}_n}$ is usually given as
\begin{align} \label{e:ecc1}
\bm{\mathcal{E}_n} \equiv \varepsilon_n \, e^{i n \psi_n} \equiv -
\frac{\int r dr d\phi \, r^n e^{i n \phi} \, f(r, \phi)}
{\int r dr d\phi \, r^n \, f(r, \phi)} ,
\end{align}
where $f(r,\phi)$ is some initial state distribution like the energy density or entropy density which specifies the initial state.  Here and throughout this paper we denote the magnitude of the eccentricity by $\varepsilon_n$ and its complex (event-plane) angle by $\psi_n$.  It is convenient to re-express this quantity in terms of the complex position vector $\bm{r} \equiv x + i y$ through $r^n e^{i n \phi} = \bm{r}^n$:
\begin{align} \label{e:ecc2}
\bm{\mathcal{E}}_n \equiv -
\frac{\int d^2 \bm{r} \, \bm{r}^n \, f(\bm{r})}
{\int d^2 \bm{r} \, |\bm{r}|^n \, f(\bm{r})} \: ,
\end{align}
where we use boldface to denote the complex vector.  Usually the definition \eqref{e:ecc1} or \eqref{e:ecc2} is specified as applying only in the center of mass frame (or the central frame of whatever the observable $f$ is).  In a general coordinate system, this is
\begin{align} \label{e:ecc3}
\bm{\mathcal{E}}_n \equiv -
\frac{\int d^2 \bm{r} \, (\bm{r} - \bm{r}_{CMS})^n \, f(\bm{r})}
{\int d^2 \bm{r} \, |\bm{r} - \bm{r}_{CMS}|^n \, f(\bm{r})}
\end{align}
with the center-of-mass vector
\begin{align}
\bm{r}_{CMS} \equiv
\frac{\int d^2 r \, \bm{r} \, f(\bm{r})}
{\int d^2 r \, f(\bm{r})}
=
\frac{1}{f_{tot}} \, \int d^2 r \, \bm{r} \, f(\bm{r}) .
\end{align}
One consequence of this definition is that the directed eccentricity $\bm{\mathcal{E}}_1$ vanishes identically, since
\begin{align}
\bm{\mathcal{E}}_1 &\propto
\int d^2 r \, (\bm{r} - \bm{r}_{CMS}) \, f(\bm{r})
\notag \\ &=
\int d^2 r \, \bm{r} \, f(\bm{r})
-
\bm{r}_{CMS} \, \int d^2 r \, f(\bm{r})
\notag \\ &=
f_{tot} \: \bm{r}_{CMS} - f_{tot} \: \bm{r}_{CMS} = 0.
\end{align}

These properties and the preference for the definition \eqref{e:ecc1} or \eqref{e:ecc2} is not accidental; they are necessary conditions for the eccentricities to serve as candidate estimators for the final-state anisotropic flow vectors
\begin{align} \label{e:Vdef2}
	\bm{V_n} \equiv v_n \, e^{i n \phi_n} = \frac{1}{N_{tot}}
	\int d^2 k_\bot \, e^{i n \phi} \, \frac{dN}{d^2 k_\bot} .
\end{align}
Considerations such as translational invariance, rotational invariance, and other discrete symmetries strongly constrain which intial-state functions transform in the same way as the flow vectors $\bm{V_n}$, and the hydrodynamic principle of long-wavelength dominance leads to a natural power counting of candidate estimators to the flow harmonics \cite{Gardim:2011xv,Gardim:2014tya}.  For this reason, the eccentricities we use to quantify the initial state in Sec.~\ref{sec:Results} are the magnitudes $\varepsilon_2$ and $\varepsilon_3$.  We thank Matt Luzum for notes and helpful discussions clarifying subtleties of these issues.

%
\subsection{Cumulants}
%

The two- and four-particle flow cumulants are defined as
\begin{subequations}
\begin{align}
v_n \{2\} &\equiv \sqrt{
\left\langle \frac{1}{N_2} \int_{p_1 p_2} e^{i n (\phi_1 - \phi_2)} \,
\frac{dN_2}{d^2 p_1 d^2 p_2}
\right\rangle
}
\\
v_n \{4\} &\equiv \sqrt[4]{
2 \left\langle \frac{1}{N_2} \int_{p_1 p_2} e^{i n (\phi_1 - \phi_2)} \,
\frac{dN_2}{d^2 p_1 d^2 p_2} \right\rangle^2 -
\left\langle \frac{1}{N_4} \int_{p_1 \cdots p_4} e^{i n (\phi_1 + \phi_2 - \phi_3 - \phi_4)} \,
\frac{dN_4}{d^2 p_1 \cdots d^2 p_4} \right\rangle
} ,
\end{align}
\end{subequations}
where $N_2$ and $N_4$ are the number of particle pairs and quadruplets, respectively \cite{Luzum:2013yya}.  If these multiparticle correlations arise entirely from independent particle emission coupled to a collective flow, then the multiparticle distributions factorize on an event-by-event basis, and the cumulants can be written entirely in terms of the statistical distribution of flow harmonics $v_n$:
\begin{subequations} \label{e:Vcumdefs}
\begin{align}
v_n \{2\} &= \sqrt{ \left\langle v_n^2 \right\rangle }
\\ \notag \\
v_n \{4\} &= \sqrt[4]{
2 \left\langle v_n^2 \right\rangle^2 -
\left\langle v_n^4 \right\rangle
}
\notag \\ &=
v_n \{2\} \,
\sqrt[4]{
1  - \frac{\mathrm{Var}(v_n^2)}{\langle v_n^2 \rangle^2}
}.
\end{align}
\end{subequations}
Thus, in a flow picture, $v_n \{2\}$ measures the RMS of the $n^\mathrm{th}$ harmonic flow, and $v_n \{4\}$ is a measure of its fluctuations: the greater its suppression below unity, the greater the fluctuations (variance) of $v_n^2$.  To quantify the initial state geometry, we will use the analogous cumulants to \eqref{e:Vcumdefs} for the initial eccentricities $\varepsilon_n$:
\begin{subequations} \label{e:Ecumdefs}
\begin{align}
\varepsilon_n \{2\} &= \sqrt{ \left\langle \varepsilon_n^2 \right\rangle }
\\ \notag \\
\varepsilon_n \{4\} &= \sqrt[4]{
2 \left\langle \varepsilon_n^2 \right\rangle^2 -
\left\langle \varepsilon_n^4 \right\rangle
}
\notag \\ &=
\varepsilon_n \{2\} \,
\sqrt[4]{
1  - \frac{\mathrm{Var}(\varepsilon_n^2)}{\langle \varepsilon_n^2 \rangle^2}
} .
\end{align}
\end{subequations}

%
\subsection{Subtleties of Eccentricities for Conserved Charges}
%

The above discussion is well established when the quantity $f(\bm{r})$ being described is something like the energy density or entropy density which is positive definite.  But if $f(\bm{r}) = \rho(\bm{r})$ is a charge density, in particular one in which the total net charge is zero, then the situation becomes much more subtle.  To see this, define the net charge and net dipole moment (as a complex vector) as
\begin{subequations}
	\begin{align}
	q_{tot} &= \int d^2 \bm{r} \, \rho(\bm{r}) \\
	\bm{d} &= \int d^2 \bm{r} \, \bm{r} \, \rho(\bm{r}).
	\end{align}
\end{subequations}
If $q_{tot} = 0$, as in the case at top collider energies where net baryon stopping is suppressed, then the center of charge
\begin{align}
\bm{r}_{COC} \equiv
\frac{\int d^2 r \, \bm{r} \, \rho(\bm{r})}
{\int d^2 r \, \rho(\bm{r})}
= \frac{1}{q_{tot}} \bm{d}
\end{align}
becomes undefined.  One consequence of vanishing $q_{tot}$ is that the dipole moment is the same with respect to any origin of coordinates.  If the distribution is shifted to an arbitrary origin at $\bm{R}$,
\begin{align}
\int d^2 r (\bm{r} - \bm{R}) \rho(\bm{r}) =
\bm{d} - \underbrace{q_{tot}}_{= \: 0} \bm{R} = \bm{d}
\end{align}
then the dipole moment is unchanged.  For this reason, it is impossible to define a corresponding frame such that $\bm{\mathcal{E}_1} = 0$ when the total charge vanishes.  Instead, there is always a nonzero directed eccentricity  proportional to the dipole moment.  Because of the inability to construct a center-of-charge frame and ensure the vanishing of $\bm{\mathcal{E}_1}$ for a conserved charge with $q_{tot} = 0$, the usual definitions \eqref{e:ecc1} or \eqref{e:ecc2} must be modified.

For any of the three conserved charge densities $\rho_{\mathcal{X}}(\bm{r})$, where $\mathcal{X}$ denotes any of baryon number $B$, strangeness $S$, or electric charge $Q$,
 we will treat separately the regions of positive charge with $\rho_{\mathcal{X}}(\bm{r}) > 0$ and negative charge with $\rho_{\mathcal{X}}(\bm{r}) < 0$ by decomposing
\begin{align}
\rho_{\mathcal{X}} \equiv \rho^{(\mathcal{X}^+)} \: \theta(\rho_{\mathcal{X}}) +
\rho^{(\mathcal{X}^-)} \: \theta(-\rho_{\mathcal{X}}) ,
\end{align}
where we have suppressed the position argument $\bm{r}$ for brevity.  Then the corresponding eccentricities of the positive and negative charge densities are
\begin{align}
\varepsilon_n^{(\mathcal{X}^{\pm})} &\equiv
\left|
\frac{
    \int d^2 \bm{r} \, \left( \bm{r} - \bm{r}_{COC}^{(\mathcal{X}^{\pm})} \right)^n \,
    \rho^{(\mathcal{X}^{\pm})}(\bm{r})
    }{
    \int d^2 \bm{r} \, \left| \bm{r} - \bm{r}_{COC}^{(\mathcal{X}^{\pm})} \right|^n \,
    \rho^{(\mathcal{X}^{\pm})}(\bm{r})
    }
\right| ,
\end{align}
where
\begin{align}
\bm{r}_{COC}^{(\mathcal{X}^{\pm})} \equiv \frac{\int d^2 \bm{r} \, \bm{r} \, \rho^{(\mathcal{X}^{\pm})}(\bm{r})}{\int d^2 \bm{r} \, \rho^{(\mathcal{X}^{\pm})}(\bm{r})}
\end{align}
is the center of charge.  For all of these quantities, we will consider the event-by-event distribution of eccentricities, and the second and fourth cumulants of that distribution with the standard definitions.

%
\subsection{Initial-State Quantifiers vs. Final-State Estimators}
%

These various eccentricities are all measurements of the shape of the initial state geometry.  While they can quantify in detail the modification of the initial state by the ICCING algorithm and the relationships between the geometries of energy and of the conserved charges, these eccentricities are not directly observable themselves.  Rather, they serve as a method of quantifying  initial conditions for subsequent hydrodynamic evolution and freezeout into the final measured distribution of particles in momentum space.  However, while some important steps have been made in this direction \cite{Demir:2008tr,Denicol:2013nua,Kadam:2014cua,Rougemont:2015ona,Rougemont:2017tlu,Monnai:2016kud,Greif:2017byw,Denicol:2018wdp,Martinez:2019bsn,Du:2019obx}, a full hydrodynamics code which incorporates charge diffusion and dissipative effects with all relevant interplay between the three conserved charges does not yet exist.  In the absence of such a complete hydrodynamic model of the charge dynamics in the QGP, quantifying the initial state through the eccentricities defined above is the best first step we can take.

As progress in the field continues toward the development of a full hydrodynamic picture of $B$, $S$, and $Q$, we will be able to directly compare these quantifiers of the initial state with the corresponding anisotropic flow observables in the final state.  For bulk particle production from the eccentricities of the initial energy (or entropy) density, strong evidence already exists that linear or quasi-linear response dominates the hydrodynamic mapping from initial to final state for central PbPb collisions at the LHC  \cite{Teaney:2010vd, Gardim:2011xv, Niemi:2012aj, Teaney:2012ke, Qiu:2011iv, Gardim:2014tya, Betz:2016ayq, Noronha-Hostler:2015dbi}.  Cubic deviations from linear response have been seen to play a more important role when moving toward more peripheral collisions and when considering smaller collision systems \cite{Niemi:2015qia, Noronha-Hostler:2015dbi, Sievert:2019zjr,Rao:2019vgy,Wei:2019wdt}.  Using the initial charge densities provided by ICCING, we will in future work be able to explore various candidate estimators for the flow of conserved charges in the final state.  For now, though, we will use the eccentricities to quantify in detail the structure of the initial state geometry produced by the ICCING algorithm.

%
\section{Note on the Density Scales}
\label{app:Densities}
%

The absolute scale for the charge densities is set both by the quantum numbers of the various quark flavors and by the way in which our model deposits those quantum numbers as density distributions in space.  As an example, consider the charge density profile of a quark which carries total baryon number $Q = +1/3$.  An infinitesimal charge $dq$ is related to the three-dimensional charge density $\rho$ by
\begin{align}
dq = \rho \, d^3 x = \rho \, d^2 x_\bot \, \tau_0 \, d\eta
\end{align}
such that
\begin{align}
\rho = \frac{1}{\tau_0} \frac{dq}{d^2 x_\bot \, d\eta}
\end{align}
is the charge density across the initial hypersurface at proper time $\tau_0$.  Because our framework is explicitly boost-invariant, all charges are expressed per unit rapidity $dq/d\eta$.

In our model, the two-dimensional charge density $dq/d^2x_\bot \, d\eta$ is deposited with a Gaussian profile with radius $r$, with default values  $r=0.5 \, \mathrm{fm}$ and $\tau_0 = 0.6\,\mathrm{fm}$.  If that 2D Gaussian profile were distributed continuously across all space, then the 3D charge density would be
\begin{align} \label{e:continfty}
\rho(x,y) = \frac{Q}{2\pi r^2 \tau_0} \, \exp\left[ -\frac{x^2 + y^2}{2 r^2} \right]
\end{align}
such that
\begin{align} \label{e:chargecons}
\frac{dq}{d\eta} = \tau_0 \int d^2 x_\bot \, \rho = Q
\end{align}
recovers the total charge of the quark (per unit rapidity $d\eta$).  For a baryon number $Q = +1/3$ distributed across the continuous distribution \eqref{e:continfty} over all space using default parameters, the peak charge density at the origin would be
\begin{align}
\rho(0,0) = \frac{1/3}{2\pi (0.50 \, \mathrm{fm})^2 (0.60 \, \mathrm{fm})} = 0.35 \, \mathrm{fm}^{-3} .
\end{align}

However, rather than using the Gaussian distribution normalized to unity across all space in Eq.~\eqref{e:continfty}, we instead cut off the density at a finite radius $r$.  This changes the normalization of the density profile by squeezing the entire charge into a smaller area in order to preserve the normalization \eqref{e:chargecons}, leading instead to
\begin{align} \label{e:contfinite}
\rho(x,y) = \frac{Q}{2\pi r^2 \tau_0 \, \mathrm{Erf}^2 \left( 1/\sqrt{2} \right)} \, \exp\left[ -\frac{x^2 + y^2}{2 r^2} \right],
\end{align}
where $\mathrm{Erf} \left( 1/\sqrt{2} \right) \approx 0.683$ is the error function.  This choice enhances the overall density by a factor of $\approx 2$ such that the charge density at the center now becomes
\begin{align}
\rho(0,0) = \frac{1/3}{2\pi (0.50 \, \mathrm{fm})^2 (0.60 \, \mathrm{fm}) (0.683)^2} = 0.76 \, \mathrm{fm}^{-3} .
\end{align}

This normalization is further modified by the fact that the charge is distributed over a finite grid rather than continuously.  One consequence is that the radius must be measured in integer lattice units, which we have taken to have a spacing of $\Delta x = \Delta y = 0.06 \, \mathrm{fm}$.  Thus when the nominal radius of $r = 0.5\,\mathrm{fm}$ is rounded to lattice units, the actual radius used in practice is $r \approx 0.48 \, \mathrm{fm}$ which further enhances the overall density scale by $\sim 10\%$:
\begin{align}
\rho(0) = \frac{1/3}{2\pi (0.48 \, \mathrm{fm})^2 (0.60 \, \mathrm{fm}) (0.683)^2} = 0.82 \, \mathrm{fm}^{-3} .
\end{align}
The other consequence of discretization is that the normalization \eqref{e:chargecons} is enforced across only the enclosed grid points $(x_i , y_i)$ within a circle of radius $r$, giving
\begin{align} \label{e:gridfinite}
\rho_i = \frac{Q}{\Delta x \, \Delta y \, \tau_0} \,
    \frac{
    \exp\left[ -\frac{x_i^2 + y_i^2}{2 r^2} \right]
    }{
    \sum_{j\in\mathrm{circle}} \exp\left[ -\frac{x_j^2 + y_j^2}{2 r^2} \right]
    },
\end{align}
which satisfies the discretely-normalized version of Eq.~\eqref{e:chargecons}
\begin{align}
\frac{dq}{d\eta} = \tau_0 \, \Delta x \, \Delta y \, \sum_i \, \rho_i = Q .
\end{align}
For the default parameters $r = 0.5 \, \mathrm{fm} \approx 0.48 \, \mathrm{fm}$ and $\Delta x = \Delta y = 0.06 \, \mathrm{fm}$, the lattice sum gives
\begin{align}
\sum_{j\in\mathrm{circle}} \exp\left[ -\frac{x_j^2 + y_j^2}{2 r^2} \right] \approx 165.3 ,
\end{align}
corresponding to a density at the origin of
\begin{align}
\rho(0) = \frac{1/3}{(0.06 \, \mathrm{fm})^2 (0.60 \, \mathrm{fm}) (165.3)} = 0.93 \, \mathrm{fm}^{-3} .
\end{align}

Thus we see that in our model implementation, the baryon density at the center of a single quark is close to $\sim 1 \, \mathrm{fm}^{-3}$.  For a central event such as the one shown in Fig.~\ref{f:Events}, an overlap of 3 quarks is not uncommon, leading to baryon densities on the order of $\sim 3 \, \mathrm{fm}^{-3}$ at certain grid points.  The corresponding scale of the electric charge density can be further enhanced by a factor of 2 because of the electric charge $Q=+2/3$ of the up quark.  Similarly, the scale of the strangeness density can be increased even further because of the convention $Q = -1$ for the strange quark, although there tend to be fewer overlapping strange quarks.  Two highly overlapping strange quarks would lead to a strangeness density of $\sim -6 \, \mathrm{fm}^{-3}$.  This analysis explains the origin of the absolute charge density scales seen in events like the one shown in Fig.~\ref{f:Events}.  It also suggests that these absolute scales can be substantially model-dependent; we will therefore explore the effect of different charge deposition profiles in future work.

%
\section{Estimate of the Parameter $a$ from Eq.~\eqref{e:Trentoa}}
\label{app:TrentoEstimate}
%

As a motivation for where the typical value of the parameter $a$ from Eq.~\eqref{e:Trentoa} arises, consider the following heuristic argument.  If we consider the final hadronic state of a heavy-ion collision to be composed of an ideal gas of $\pi^+, \pi^-, \pi^0$, then by the equipartition theorem the energy density is given by $\epsilon = \tfrac{9}{2} n T$ in units where $k_B = 1$.  Similarly, the ideal gas law gives $p = n T$, and together with the first law of thermodynamics $\epsilon = T s - p$, this yields the relation $s = \tfrac{11}{2} n$ between the entropy density and number density.  Then, assuming a matching of the final pion entropy at freezeout to the fluid entropy, together with a nearly ideal (isentropic) hydrodynamic state, the total entropy of the initial state is directly related to the number of pions produced in the final state by
\begin{align}
\int d^3 x \, s_0 = \frac{11}{2} N_\mathrm{final} .
\end{align}
Assuming by isospin symmetry that the charged $\pi^\pm$ states account for $2/3$ of $N_\mathrm{final}$, and changing to Milne coordinates $d^3 x = \tau_0 \, d^2 x_\bot \, d\eta$, we have
\begin{align}
\tau_0 \int d^2 x_\bot \, s_0 (\vec{x}_\bot) = \frac{33}{4} \frac{dN_{ch}}{d\eta}
\end{align}
Then assuming the model \eqref{e:Trentomodel} and taking $\frac{dN_{ch}}{d\eta} \approx \frac{dN_{ch}}{dy} \sim \ord{1000}$, $\tau_0 \approx 0.6 \, \mathrm{fm}$, and $\int d^2 x_\bot \, T_R (\vec{x}_\bot) \approx 140$ for a typical central in Trento with $p=0$, this gives
\begin{align}
a = \frac{
\tfrac{33}{4} \frac{dN_{ch}}{dy}
}{
\tau_0 \, \int d^2 x_\bot \, T_R (\vec{x}_\bot)
}
\approx
\frac{
\tfrac{33}{4} (1000)
}{
(0.6 \, \mathrm{fm}) \, (140)
}
\sim \ord{100 \, \mathrm{fm}^{-1}},
\end{align}
roughly consistent with \eqref{e:Trentoa}.  We thank Gabriel Denicol for elucidating this useful order-of-magnitude estimate.

%
%

\end{document}